\documentclass[nofootinbib,showpacs,showkeys]{revtex4}
\usepackage{amssymb}
\usepackage{latexsym}
\usepackage{epsfig}

\usepackage[utf8]{inputenc}
\usepackage{amsfonts}
\usepackage{amsmath}
\usepackage{amsthm}
\usepackage{amscd}
\usepackage{epsfig}
\usepackage{epstopdf}
\usepackage{graphicx}
  \graphicspath{{/}{figures/}}
  \DeclareGraphicsExtensions{.pdf,.png}
\usepackage{comment}
\usepackage{enumitem}

\begin{document}

\title{The Chern-Simons Current in Time Series of  Knots and Links in Proteins}

\author{Salvatore Capozziello $^{1,2,3}$}
\email{capozziello@na.infn.it} \affiliation{$^{1}$Dipartimento di
	Fisica "E. Pancini",   Universit\`a di Napoli "Federico II", I-80126, Napoli,
	Italy}
\affiliation{$^{2}$INFN Sez. di Napoli, Compl. Univ. di Monte S. Angelo,
	Edificio G, I-80126, Napoli, Italy }
\affiliation{$^{3}$Gran Sasso Science Institute, Viale F. Crispi, 7, I-67100,
L' Aquila, Italy}

\author{Richard Pincak $^{4,5}$}\email{pincak@saske.sk}
\affiliation{ $^{4}$Institute of Experimental Physics, Slovak Academy of Sciences,
Watsonova 47,043 53 Kosice, Slovak Republic}
\affiliation{ $^{5}$ Bogoliubov Laboratory of Theoretical Physics, Joint
Institute for Nuclear Research, 141980 Dubna, Moscow region, Russia}

\begin{abstract}
  
A superspace model of  knots and links for  DNA time series data is proposed to take into account   the 
 feedback loop from docking to undocking state of protein-protein interactions. 
 In particular, the   direction of interactions between the 8 hidden 
states of DNA is considered. 
It is a $E_{8}\times E_{8}$ unified spin model    where the  genotype, from active  
and  inactive side of DNA  time data series, can be considered  for any living organism.
 The mathematical model is borrowed from  loop-quantum  gravity and adapted to
 biology. It is  used to derive  equations for gene expression describing  transitions  from
 ground  to excited states, and  for the 8 coupling states between geneon and anti-geneon
 transposon and retrotransposon in trash DNA.   Specifically,  
we adopt a  modified Grothendieck cohomology  and a modified Khovanov cohomology
 for biology. The result is  a 
Chern-Simons current  in  $(8+3)$ extradimensions of a given unoriented supermanifold 
 with ghost fields of protein structures.
 The $8$ dimensions come from the 8 hidden  states of
 spinor field of genetic code. The extradimensions come from the 3 types of principle fiber  bundle in the
 secondary protein.

 \end{abstract}

 
 \date{\today}

%%%%%%%%%%%%%%%%%%%%%%%%%%%%%%%%%%%%%%%%%%%%%%%%%%%%%%%%%%%%%%%%%%%%%%%%%%%%%%%%%%%%%%%%%%%%%%%%%%%%%%%%%%
%%%%%%%%%%%%%%%%%%%%%%%%%%%%%%%%%%%%%%%%% SECTION I %%%%%%%%%%%%%%%%%%%%%%%%%%%%%%%%%%%%%%%%%%%%%%%%%%%%%
\maketitle

%-------------------------------------------------------------------------------
\section{Introduction}\label{sec:intro}
%-------------------------------------------------------------------------------

The  attempt  to explain living organisms
 by mathematical model is a long story  that dates back to Turing \cite{machine}. It is well known that all living organisms  have their 
own character deriving from the gene expression of their genome. Up to now, there is  no   general mathematical expression for the coordinate changes in gene expression of  genetic code for  active  and 
inactive areas of  DNA, for RNA and for proteins.
Recent  discoveries  in biology are  the knotted DNA, the
 knotted protein \cite{knot3}, 
and the unknotted RNA folding with the role of knotted protein in codon correction of RNA in
 methyl transfer \cite{knot1}. Knots are  four dimensional   topological objects, embedded in 3 dimensions,  used for long time in loop-quantum  gravity and superconductor theory. 
 One can also  
set up a theoretical route \cite{vortex} searching for the general equation to solve 
knotted protein folding \cite{folding} by using  the Wilson loop operator \cite{loop2} in loop-quantum  gravity \cite{wilson,phase}
 for gene expression with a boundary phase\cite{phase2, foam} condition.
The theory should support the problem of adaptive changing of
docking curvature of knotted protein folding.
 The definition of curvature from parallel transport of gauge field of genetic code along fiber of protein while docking to each other
 is still have no precise definition. One of the most suitable equation to explain this phenomenon is the Seiberg-Witten   equation that can be adapted to biology. 
 Here, we use a  new approach, based on  algebraic topology,  to define  biological structures:   specifically the
Grothendieck topology is the  mathematical structure that we are going to adopt this work.

A sheaf cohomology of retrotransposon replication cycle  is defined as an exact 
short   sequence in loop space of underlying moduli state of genetic code.
 This sequence is missing of knotted property \cite{knot2} and folding behavior of 
protein and RNA\cite{knot4}. We use the category approach  for a
 modified algebraic construction coming from 
 loop-quantum gravity where Khovanov cohomology\cite{khovanov} and Grothendick 
topology are used in order to  describe  biological properties of knots and  links 
 into time series of  protein, DNA and RNA. 
 The key of protein folding behavior are the curvature and volume  of hyperbolic knot of underlying 
time series of link in protein. We modify the Khovanov cohomology\cite{witten2} to define 
a new mathematical object in the  framework of modified Grothendieck cohomology \cite{topology, topology2} for the  definition of time series \cite{super} of 
knots and links in proteins.  The source of folding diversity  from the primary structure of 
protein is considered to
 be a source of knot hidden super state of link \cite{knot5} between 
interaction of all species.  This happens in the space
 of superstates  over principal fiber bundle of secondary protein with curvature of docking. The process occurs
 in the modified Wilson loop of gauge field over the invariant property of spinor field in genetic code
 of a  living organism. It can be defined as a  Witten invariant \cite{witten} for
 biology. Knots and links in protein folding might be the resulting  structures 
of optimized  free energy over the partition function of co-states between 2 pairs of geneon  and anti-geneon states with 
transposon and retrotransposon in the wave function of genotype. The result of secondary protein optimized complex surface  is
 given by the transition of hidden states among adaptive feedback loops that 
  change the curvature of protein docking state in viruses and host cells in replication cycle represented as hyperbolic knots.
 The time series model of knotted protein can be  defined by a modified Khovanov  sheaf cohomology
 for  living organisms. It  can be useful to
 understand the behavior of hidden states in the parasitism of viral attach to host cell with 
co-cycle elements defined by the  curvature  of protein docking state in spin space as transitions 
between partition function  in the cell structure \cite{current}. The underlying context is  the  algebraic topology of knots and links.

A new approach for solving knotted protein folding problem is based on the quantum loop 
invariant of  the link number with 
homotopic path in ribbon graph of time series of knotted protein folding.  Results are expressed in hyperbolic volume and hyperbola  graph as relations between the
  parameters  gene sequence  evolution. 
 With this model of  Chern-Simons current in biology \cite{preprint}, we can give a  new definition of Ramanujan-Jones-Laurent polynomials \cite{fusion,loop,superspace2}.
We use the homotopy class of hyperbolic knotted  fundamental group \cite{hyperbolic,hyperbolic2} to define
 the obstruction curvature components of viral 
glycoprotein as transponson and retrotransposon \cite{retrotransposon,retrotransposon2,retrotransposon3,retrotransposon4}. 
The Chern-Simons supercurrent \cite{preprint2} is a potential field of life free energy
 in the canonical form of genotype. It is a new definition of gene partition function \cite{partition}
 $Z_{t}$ in the form of modified Wilson loop of the gene behavior field.
We derive curvature by the normalized curvature of unit circle. The blend
 radius is  centered at  the cell nucleus. 
The curve of equivalent class of current, the    
single strand nucleotide homotopy path of protein primary structure,  before folding into 
modified Grothendieck cohomology class of equivalent curvature, is a moduli state space with 
Hopf fibration as principle fiber for the classification of secondary protein folding parameters.

The biological functions of  proteins depend on the intramolecular docking processes 
that can transmit the free energy as Chern-Simons supercurrent between the substrate binding sides
 with a sum of curvature which is  zero at  equilibrium state.
 It is an  adjoint left and  right group action over the Hopf fibration of gene expression. With this new definition, we can explain the inactive area of trash DNA by using 
 the feedback  loop in extra-dimensions. The process gives rise to   a 
change of protein docking state from non-equilibrium to equilibrium.  The successful docking state of drug with 
receptor protein is represented as an icosahedral  
 viral glycoprotein  induced from knotted DNA with underlying 8 hidden states in
 octomer of histone complex \cite{epi}. An example is  the chromatin structure of human nucleosomes inside the chromosome.
 The extra $ (8+3)$ supersymmetry dimensions of human genome
  come from the so called Reidemeister moves  in histone modification with 
 knots and links structure. The  move represent the docking mechanism between feedback loop 
of underlying  central dogma between coupling state of 
  DNA, RNA, and protein structure.

The superpace of data time series  in DNA and their hidden states of genetic code  
can be model with a loop space plus extra-properties of  adjoint representation as category of
sieve in extend central dogma of  Grothendieck topology.
 The observation of gene expression can appear only on one 
side of supersymmetry of chiral molecule in living organism.
The other side  is the inactive part of hidden state in trash DNA  which can be explain by the  Laurent polynomial with negative 
degree as hidden state in left and right hand chiral molecular supersymmetry space of
  DNA, RNA and protein.

Another   problem in biology  is 
how  the  genome of viral DNA is knotted and
why TrmD bacterial knot protein \cite{epi1} is  involved with RNA methyl transfer and genetic code error correction.  
 In other words, the problem consists in how  we can  set up  a general  equation with initial conditions 
for  knotted DNA, RNA and proteins  in
 living organism.   The coupling states between 
 2 biological systems can be modeled  by 
a  distribution over  a statistical theory with  2 Laurent polynomials 
of knotted states. It implies an underlying vortex operator  with the convolution of 2 quantum wave
 functions of  2 geneotype.
 Jones \cite{jone} defines relations of hyperbolic $4_{1}$ knot 
invariants by statistical mechanics and von Neumann algebras.
 Witten \cite{witten} was the first who explained the meaning of Jones polynomials  as
 Chern-Simons currents and in relation to Khovanov cohomology \cite{khovanov}. The description is achieved  in 
loop-quantum  gravity   appearing in a system of partial differential equation over connection, a
 Seiberg-Witten equation for  instantons and monopoles.
 
  We can also interprete Jones polynomials, Laurent polynomials and Khovanov cohomology as coordinates
 of Grothendieck topology in  the categories of  living organisms with the geneotype which is the  Yang-Mill field acting 
 in replication cycles  over the principle bundle of protein structure.
 The interaction between their adaptation behavior in ecosystem are coordinates changing under 
their morphisms with co-adjoint 
 co-continous functors as parasitism state or mutual state 
between viral replication cycle coupling with host cell or mother cell fertilized by  father cell. 
We  solve these complicated self-dual supersymmety  
problem, based on the genotype  of capsid protein in 
icosahedral virus and knotted protein
 in methyl transfer of  inactive gene,  resetting methylation state in
 DNA, RNA  and in histone modification.
 
 It is an effective  supersymmetry  of  spinor field hidden states  of time series data of knotted protein which represents a 8 hyperbolic knotted 
loop gravity. Specifically, the  atomic units of gene can be defined in term of
 Jones polynomial over the  Seiberg-Witten equation.
 The duality of unoriented supermanifold  over living organism with extradimensions can 
 break a chiral supersymmety over central dogma adopting the dynamics  of  
 string and d-brane theory. 
 
 The paper is organized as follows. In Section II we summarize the basic definitions of 
  modified Grothendieck topology for biology.   Modified  Khovanov cohomology for time series of knotted protein is discussed in Section III.
  The  Seiberg-Witten invariant  for  knots and links   in proteins  is considered in Section IV.
  In Section V, a computation in synthetic time series of knots and links in proteins is presented. 
  Another computation of Chern-Simons current in viral capsid glycoprotein is reported in Section VI.
  Results of data analysis are in Section VII.  Discussion and conclusions are in Section VIII.

%===========================================
\newtheorem{Definition}{Definition}
\section{Grothendieck Topology  for Biology}
%============================================

We use existence axioms of Grothendieck topology as a main tool to define a 
general coordinate frame for every protein 
 in metabolism of living organisms by the open set of space of secondary protein
 folding  structure.
In the  new approach of Grothendieck topology, we  define 
sheaves on a category of protein structure and their modified Khovanov cohomology for biology.

 The open set,  as the coordinates
 of transmission signal intracellar superspace and intercellular superspace, is
 the  co-adjoint co-continuous functor of the Grothendieck open set with more analytic extra properties
 of the Laurent polynomial of knot and complex surface of protein with curvature. 
 
 We need to modify the  Atiyah axioms for Topological Quantum Field Theory  for all  parameterized 
 components, i.e. organelles, proteins, DNA, RNA and cell membrane. The choice of solutions of higher 
algebraic topology and differential geometrical object is a sheaf cohomology theory  
intuitively derived from the axioms of Grothendieck topology with some extra property of Atiyah and Hitchin systems.
Let $D$ be the space of DNA, R be the space of RNA, P be the space of protein. They have a mirror symmetry with their partner superspace, mitDNA $D^{\ast}$, mitRNA, $R^{\ast}$, $P^{\ast}$, structural (i.e. histone protein) inactive enzyme state of protein.
A free Abelian group over sheaf sequence of coordinates defines a genetic code as 
a link or knotted quantum observable states.  For these 4 particle-like in 
genotype, the   shift and drift are transition hidden states
 in inactive area of trash DN.  This analogy  works also  for DNA, RNA methylation and histone 
modification, for open and closed geneon states, for active gene states to anti-geneon states, for  inactive gene state transition to exited state by methyl transfer life energy group. It is worth noticing that  methyle group consists of a  tetrahedral geometry  with the shape of water molecule
 with high energy in respiration without oxygen,  i.e. fermentation in Krebs cycle. The methyl transfer is in analogy with some property of the 
transfer of proton, $H^{+}$ and electron in Krebs  cycle, where it is  known  that stem cells have a resetting of all methyl states. The methyl transfer states hold some memory in form of entaglement states from mother and father genes for their methyl transfer purpose. 
 
In this section, we discuss the  role of RNA as an enzyme to control protein docking state in reversed direction
of central dogma $\cdots \rightarrow \mathcal{O}_{R^{\ast}}\rightarrow \mathcal{O}_{R}\rightarrow \cdots $
compared with enzyme property of induced passive and active structure
 of protein in $\cdots \rightarrow\mathcal{O}_{P^{\ast}}\rightarrow \mathcal{O}_{P}\rightarrow \cdots $.
The example of  Reidemeister move in knot theory is a gene splicing of pre-mRNAs 
with the consensus sequence at the exon-intron junction of eukareotic cell.
 The phenomenon of the so called  self-splicing of Tetrahymena group I intron
 is in analogy with the new definition of knot differential operator and of  link operator
 to generate closed loops of cyclized intron and spliced exons. It is a well known
 result that some groups of knotted RNA form by introns giving the  so called tetraloops. 
The closed loop is formed by knotted RNA and generate a new type of RNA called group I
 ribozymes. In this section, we give an example of time series of knotted RNA,
  called exon-intron self-dual-splicing time series of knotted RNA.
 The existence of hidden states of proteins in nature is induced from hidden states as a genotype: it is the 
 so called  lac operon state and  trp operon state in noncoding   area of trash DNA.
The active state of protein interaction, as starting and ending  unknotted state in protein 
partner in this example,  is an active repressor state of RNA polymerase complex. It is a ghost field of protein state $\Phi^{+}(A_{\mu})$ and inactive
 repressor state, as antighost field,  $\Phi^{-}(A_{\mu}).$  The threshold for the change state from ghost field to anti-ghost field is well knon in biology as an
  amount of tryprophan in attenuation of trp operon. 
  
Another example of application of 
Khovanov cohomology in  biology is a methyl transfer control in transcription encoding mechanism.
The process  is well known  as DNA metylation. It is an interaction of protein to DNA directly  called
 DNA metyltransferase (DNA MTase). This mechanism can turn on and off gene. 
It is involved with gene imprinting and process of life energy production in mitochondria and in some bacteria.
Some recent  experiments  revealed that without  metylation, all cells  die. The theory to explain this process is based on  a 
quantum field approach for biology. This gene expression process consumes  life energy and is related to the  Krebs cycle. The enzymes in  Krebs cycle
 in all living organism share 8 common  states of Grothendieck topology in a common equation to produce the life energy for controlling the gene expression.
Recently, some  experiment revealed  that stem cells induce,  from the resetting state of DNA-Methylatition,
 a mechanism to clean all methyl-transfer group from mother and father genome. The result of reset state cells can be a differentiation to other
 types of cells and also to nerve cell. This happens also when a retrotransposon gene in trash DNA is contaminated by the  methyl transfer group. There will be an exception
 for reset states in trash DNA of methyl group in retrotransposon. Researchers  still cannot 
explain this phenomena and the biophysical mechanism by which the  cell can reset methyl group only in active area
 but not in inactive area. 
 
\begin{figure}[!t]
 \centering
\epsfig{file=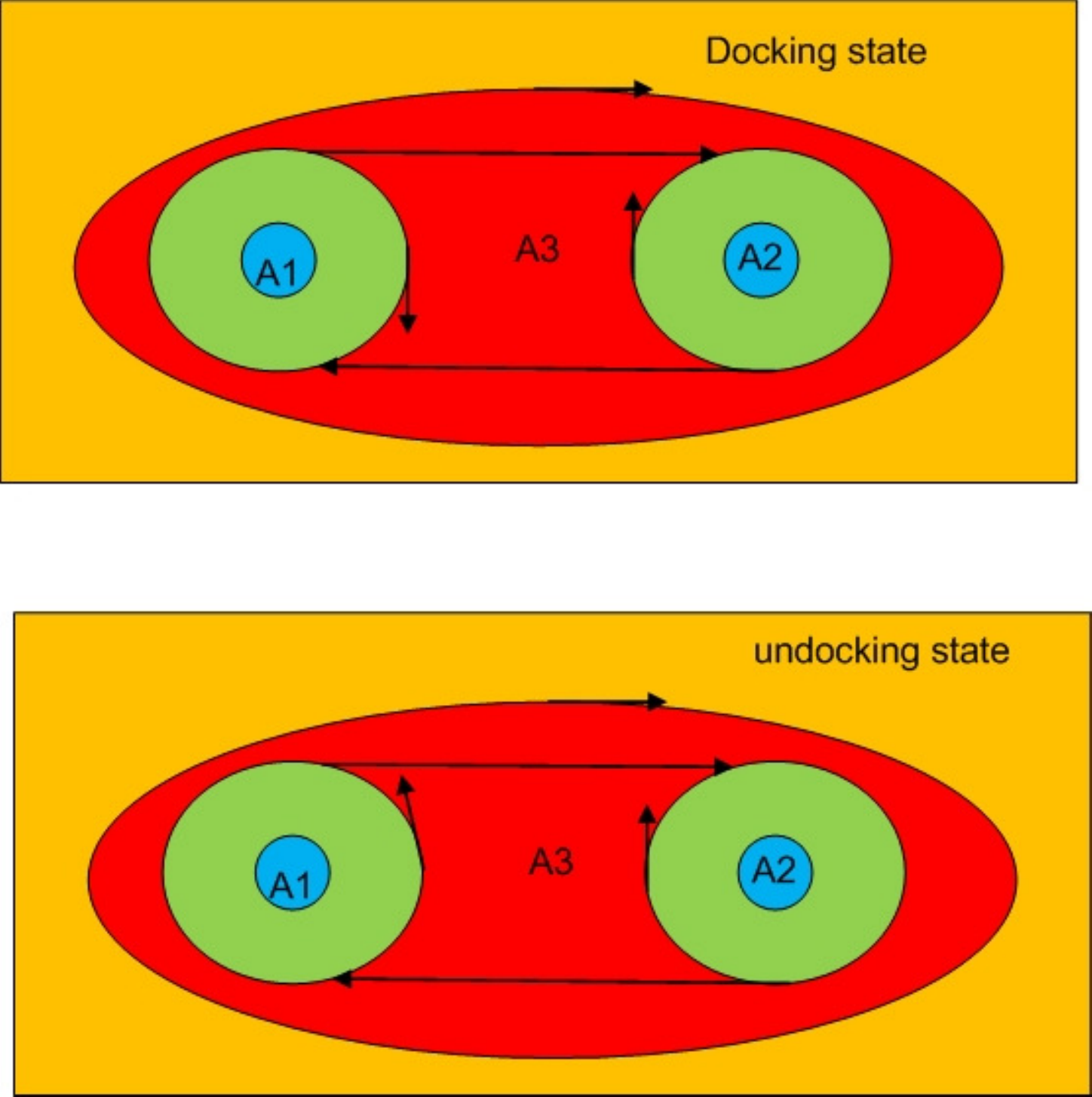,width=5cm}
 \epsfig{file=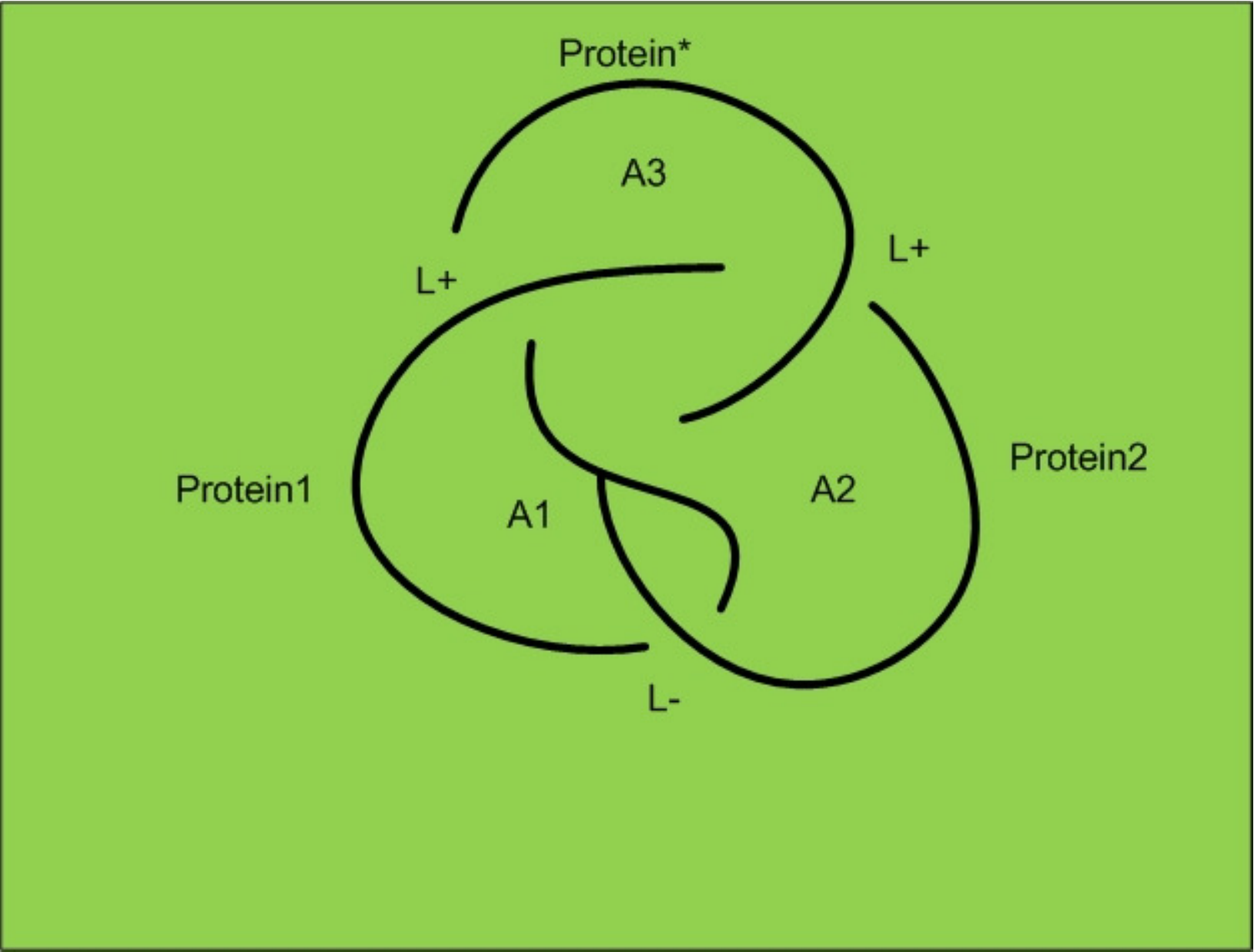,width=5cm}
 \epsfig{file=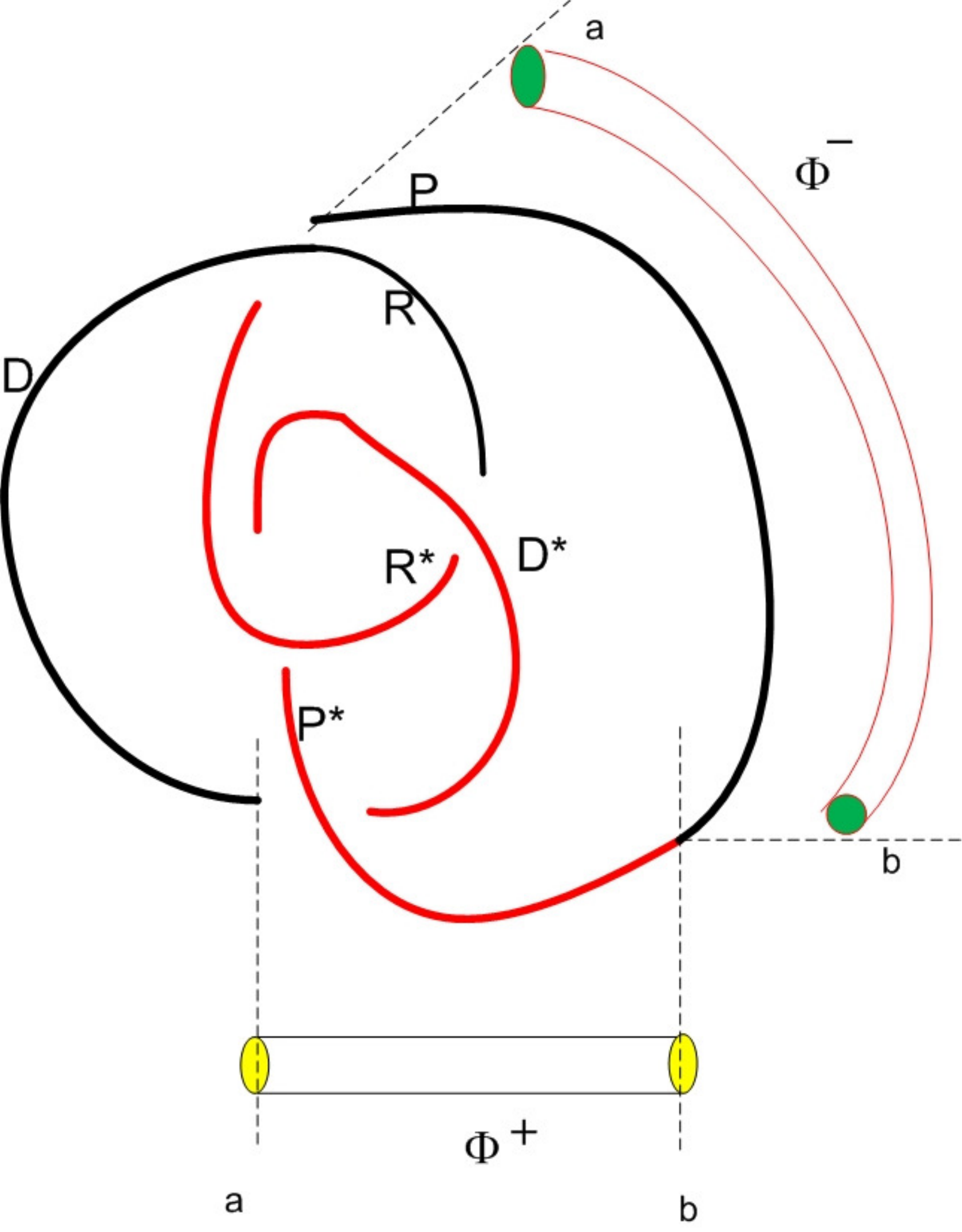,width=5cm}
\caption{On the left, a diagram for the  interaction of 2 proteins with 3  gauge fields is represented  as the connection 
$A_{1},A_{2}$ and $A_{3}$. The docking state is induced from  opposite direction of spinor field in $A_{1}$ and $A_{2}$ 
and is canceled  to each other to produce an equilibrium state embedded into the  connection $A_{3}$.
For undocking state,  the coupling between $A_{1}$ and $A_{2}$ does not cancel  each other.
 In  the picture shown in the middle, a Chern-Simons current, represented as a knot $3_{1}$ with 3 types of connections over 3 types of principle 
fiber bundle of secondary protein structure is reported.\label{knot}.  The picture shown on the right is 
a 8 knot of short exact sequence of extended central dogma. We glue the starting point 
of a short exact sequence $0_{dbarne}\simeq 0_{anti-dbarne}$ like a hyperbolic $4_{1}$ knot
  $0_{dbrane}   \rightarrow \mathcal{O}_{D}\stackrel{\phi^{d_{1}}_{\pm}}{  \rightarrow }\mathcal{O}_{R} 
\stackrel{\phi^{d_{2}}_{\pm}} {  \rightarrow }\mathcal{O}_{R^{\ast}} 
    \rightarrow   \mathcal{O}_{P}
 \stackrel{\phi^{d_{3}}_{\pm}}{  \rightarrow }\mathcal{O}_{P^{\ast}}
\stackrel{\phi^{d_{4}}_{\pm}} {  \rightarrow }\mathcal{O}_{D^{\ast}} 
\rightarrow  0_{anti-dbrane}$.The trajectory in $4_{1}$ 
 induces an infinite  modified Khovanov cohomology sequence in 
higher dimensions in categories of biological systems.  
We label $4_{1}$ knot with 2 different color from coupling between 2 cycles, i.e. host cell and viral particle. This $4_{1}$
 knot can also be used to understand parasitism states of a virus attack to the host cell 
as hyperbolic manifold with hyperbolic volume.} \end{figure}

 Here, we use supersymmetry with knotted time series data in RNA to explain this situation by adopting the Khovanov 
cohomology for knotted RNA. The methylation can appear in both RNA, DNA and histone protein to  switch the 8 hidden  states in underlying geneotype
 to a transition to each other among the  64 hidden states.
The approach is based on the experimental fact  that protein structure comes from 3 layers 
in transcription process with 3 types of connection. According to this fact, the active protein states, $P$ have their partner in passive protein $P^{\ast}$ in the docking process (see Fig. \ref{knot} for a detailed   explanation). This 3 types of sequence are involved with
 structure of protein as 1-1 maps of intrinsic homogenous coordinates  from the 8 hidden  spinor states in DNA molecule. 
We classify the biopolymer molecules involved with the protein folding structure with co-continuos co-adjoint Grothendieck topology as an open set for measuring the degree of protein folding in the higher dimensions space of loop space of cell.
 We identify by their own  symmetry and chemical property by using invariant properties of curvature in closed surface of protein pair docking state.
 
Let $\mathcal{O}_{D},\mathcal{O}_{R}$ and $\mathcal{O}_{P}$ be
   active objects in a categories of DNA, RNA, active state of protein.
 Let $\mathcal{O}_{D^{\ast}},\mathcal{O}_{R^{\ast}}$ and $\mathcal{O}_{P^{\ast}}$
 be passive states of DNA, RNA and protein in hidden transition state over
 inactive area of trash DNA.
Let the open set of Grothendieck topology be a continuous adjoint and co-adjoint function 
over 2 categories of active objects and inactive objects with an extend sheaf sequence of
 short exact sequences of  central dogma. The continuous functor between these 4 objects from different 4 categories  play a role of open set in the  topology of cell.
Let an adjoint continuous functor be
$C_{\ast}: 0   \rightarrow \mathcal{O}_{D}   \rightarrow  \mathcal{O}_{R} 
  \rightarrow  \mathcal{O}_{R^{\ast}} 
    \rightarrow   \mathcal{O}_{P}
    \rightarrow  \mathcal{O}_{P^{\ast}}
   \rightarrow  \mathcal{O}_{D^{\ast}} 
\rightarrow  0
$ and co-adjoint co-continuous functor be $C^{\ast}:0\leftarrow  \mathcal{O}_{D}\leftarrow \mathcal{O}_{R}
 \leftarrow\mathcal{O}_{R^{\ast}}\leftarrow  \mathcal{O}_{P }\leftarrow  
   \mathcal{O}_{P^{\ast}}    \leftarrow     \mathcal{O}_{D^{\ast}} \leftarrow     0 $.
 If these co-adjoint continuous  functor over 2 superspace  exists  in 2 layers of dbrane and anti-dbrane of living organism $X$ and $Y$, then the chain and the co-chain complex of 
short exact sequences of extended central dogma are co-adjoint and co-continous as replication cycle, that is 
$  C_{\ast}(X)\circ C^{\ast}(Y)= C_{\ast}(Y)\circ C^{\ast}(X)=0$.
 We can define a genotype as a grade over genetic shift and drift operator from their evolution of their own species with analogy with differential and co-differential 
operator in a  chain sequence of cohomology theory, that is $ \phi^{d_{i}}_{\pm} = 0 \rightarrow  \mathbb{Z} \rightarrow  0,i=1,2,3,4$.
 This operator has 4 types of shift in hidden transition states of  inactive area of genetic code. It 
 represents 4 hidden directions of gene expression represented  as 4 wave function 
  of a particle-like  in a quantum field theory for biology. They are the
 geneon, anti-geneon, transposon and retrotransposon over categories of
$\mathcal{O}_{D},\mathcal{O}_{R},\mathcal{O}_{P}$ and $\mathcal{O}_{D^{\ast}},\mathcal{O}_{R^{\ast}}$ and $\mathcal{O}_{P^{\ast}}$.
 Transposon is represented as knotted DNA in inactive area of trash DNA (the unknotted DNA also classified as a special case of knotted DNA with unknotted state). The particle-like are the geneon, anti-geneon, transpopson and retrotransposon.
The knots and links in  protein are represented by time series over transition coupling states of these 4 particles in genotype.
 We have enzyme properties of some protein with active and 
inactive states  which imply that geneons have their partner passive states, the  anti-geneon states of genotype.
The transposon is defined as a cut and glue protein complex hypersurface into 2 parts.
 The Atiyah-Segal and Hitchin-system axioms   define the sheaf sequence of transposon, retrotransposon, geneon and anti-geneon 
 states in protein complex manifold. This approach   opens the possibility to link under the same standard  quantum field theory and  biophysics.
\begin{Definition}
Let $G$ be the gauge group of genotype. Let $[s_{i}]=[e^{i\beta}],[s_{i}^{\ast}]=[e^{i\beta}]^{\ast}=[e^{-i\beta}]$ be a spinor field of genotype over DNA.
Let $\Phi^{\pm}: (\mathcal{A},s)\mapsto \{1,-1\}$ be ghost and anti-ghost field
 in protein secondary structure with supersymmetry in genetic code. Let
 $\mathcal{A}$ be  a supermanifold of living organism and $s$ be
 the  Hamiltonian system of gene expression mechanism in  living organism DNA, RNA and protein 
over underlying gauge group of genotype as lie group, $G$.   
Let $U\subset \mathcal{K}$ be an open set in a manifold  of protein secondary structure with
  smooth embedded homotopy path of genetic code lift a fix alphabet as based point to equivalent class of fiber with parallel transport in covering space  by 
$[\gamma (t)]\in [\mathcal{K},E]$ . It contain a starting point at 
$\gamma(t_{1})=a\in U$ and an ending point of gene expression at $\gamma(t_{n})=b\in U$.
Let $[A_{\mu}]=[\Gamma_{a}^{b\mu}]$ be an equivalent class of representation of genetic code  independent of the chosen alphabet.
It is a  connection over principle fiber of underlying  protein encoding,
 $P_{\mathcal{K}}$ (for more detail of definition see \cite{preprint}.)
The connection used for parallel transport from path of state in  the genetic code of DNA
$[A_{\mu}]=\{(  [s_{i}(\gamma(t_{1}))],[s_{i}(\gamma(t_{2}))]), 
([s_{i}(\gamma(t_{2}))],[s_{i}(\gamma(t_{3}))]),(
[s_{i}(\gamma(t_{3}))],[s_{i}(\gamma(t_{4}))]),
\cdots ,([s_{i}(\gamma(t_{n-1}))],[s_{i}(\gamma(t_{n}))])      \}$
 starts at the position $\gamma(t_{1})=a$ to
 $\gamma(t_{n})=b$  in fiber (with the Hopf fibration) of  $U\times F \rightarrow E=P_{\mathcal{K}}$ 
with  $F_{a}  \ni\Phi^{+}_{a}=\pi^{-1}(a)$
 and $F_{b}  \ni\Phi^{+}_{b}=\pi^{-1}(b)$ with the size of gene defined by 
$rank F_{b}=n=rank F_{a}$, a ghost field of gauge field over genetic code in fiber of protein secondary structure.
 The connection $[A_{\mu}]$  represents the gauge field of genetic code as the parallel transport with gauge group 
operation as translation and reversed translation in gene expression from protein fiber $F_{a}$  to $F_{b}$ in the tangent of manifold of
 living organisms for their genotype. We have that the spinor field  of genetic code 
spans in fiber $F_{a}$
and transports to $F_{b}$ by $A_{\mu}\Phi^{+}_{a}:=\Phi^{+}_{b}(A_{\mu})$, where 
$A_{\mu}=\Gamma_{a}^{b\mu}(\Phi^{+}_{a})=\Phi^{+}_{b}(A_{\mu}),\Phi^{+}_{a},\Phi^{+}_{b}=<<[s_{1}],[s_{2}],\cdots ,[s_{n}]>> $
and $\Phi^{-}_{a},\Phi^{-}_{b}$ is the anti-ghost field. It comes from translation in reversed direction as reversed transcription with hidden state
in dual fiber $\Phi^{-}_{a}\in F_{a}^{\ast},\Phi^{-}_{a}=<<[s_{n}^{\ast}],[s_{n-1}^{\ast}],\cdots ,[s_{1}^{\ast}]>>.$
\end{Definition}
The definition of connection in biology is used for two reasons:  firstly, it is a source of gauge potential field
 over genetic code where the  original transmission comes from our ancestors in the past in which 
we do not know yet where DNA come from. Secondly, it defines coordinates over the fiber of secondary protein in  
 knots and links of time series data of proteins where the  protein structure wave function, the  geneon  denoted by $\Phi^{\pm}(A_{\mu})$ of underlying genotype,  is defined independent 
of the chosen coordinate representation but it is intrinsically defined  by $[A_{\mu}]$ over a principal fiber bundle of protein $P_{K}$  which satisfy   7  of Atiyah axioms for biological quantum field.

There exists several ways to define connections $A_{\mu}$ with different purposes of applications in both
differential geometry  and  quantum field theory.
The reasons is related  to the connection to be used to define the  parallel transport of principal fiber along the tangent of 
a Riemannian manifold as vertical differentiation perpendicular to the horizontal axis.

 \begin{Definition}[Parallel transport along gene sequence]
Let a path of sequence of alphabet in genetic code of underlying genotype be parametrized by underlying smooth curve $\gamma(t)$ with path 
ordering. We can define an equilibrium state of genotype without evolution in specific species  where different species can be obtained by the  parallel transport along the curve of expression of secondary structure protein.
Let $X$ be a space of protein structure. 
Let $E$ be a covering space of underling secondary protein.
We have a connection of path in genetic code of underlying genotype with  start from position $a$ and stop
 translation in position $b$ defined by $\Gamma (\gamma)_{a}^{b}$,
\begin{equation}
\Gamma (\gamma)_{a}^{b}: E_{\gamma_{a}} \rightarrow E_{\gamma_{b}}
\end{equation}
such that
\begin{itemize}
\item  $\Gamma (\gamma)_{a}^{a}=Id$ the identity transformation  of fiber of secondary protein $E_{\gamma(a)}$.
\item  $\Gamma (\gamma)_{a}^{b}\circ \Gamma(\gamma)_{c}^{a}=\Gamma (\gamma)_{c}^{b}. $
\item $\Gamma$ is smooth  transport curve with preserved direction of gauge field and connecting  all fibers together along $\gamma_a,\gamma_a $and $\gamma_c$.
\end{itemize}
The gauge field of genetic code $A_{\mu}=\Gamma_{a}^{b\mu}$ is a parallel transport along  geodesic curve of protein (the minimum distance along protein folding)
 $\gamma : I\rightarrow X$ and $\beta=\frac{d\gamma}{dt}$ be tangent smooth curve embedded in underlying protein surface if and only if
\begin{equation}
\bigtriangledown_{\beta(t)}\beta(t)=0.
\end{equation}
\end{Definition}
\begin{Definition}
Let $D^{+}_{\Phi}$ be an active docking operator of 2 proteins in cell membrane. We use the 
same notation of Dirac operator for transition state in gene, $D^{+}$ and attach operator for docking process in protein $D^{+}_{\Phi}$ in different category of biological object. The first operator acts over category of DNA $cat(D,D^{\ast})$, the second operator acts over category of protein $Cat(P,P^{\ast})$, we can define also with attachment 
of protein to DNA in $cat(P^{\ast},D)$, also process of miRNA to RNA in $cat(R,R^{\ast})$ with the similar definitions of Grothendieck topology as  coordinate of co-adjoint functor between them. It is a modified Dirac operator which 
 brings the  attachment of spinor field to the ghost field  of viral glycoprotein  $\Phi^{+}(A_{\mu})$:
 It  translates,  by the  parallel transport of spinor field of genetic code to  dock with host
 cell receptor protein $\Phi^{-}(A_{\mu})$ as anti-ghost field with fixes the  site embedded in host cell membrane.  
 The definition is analogue to the  covariant derivative of differential geometry with some  extra-properties of  genetic code gauge field
 over connection. We have a docking between active protein and inactive (passive) protein  by docking operator defined as
\begin{equation}
D^{+}_{\Phi^{+}(A_{\mu})}| \Phi^{-}(A_{\mu})>= 
\lim_{t_{a}\rightarrow t_{b}}\frac{ \Gamma_{b}^{a\mu}\Phi_{b}^{+}(\gamma(t_{b})) -\Phi_{a}^{-}(\gamma(t_{a}))  )}{t_{b}-t_{a}}.
\end{equation}
Let $D^{-}$ be a passive docking operator of 2 proteins in cell membrane. It is a docking of antibody to viral glycoprotein in anti-parallel  spin, that is  in reversed direction of docking.
We have 
\begin{equation}
D^{-}_{\Phi^{-}(A_{\mu})}| \Phi^{+}(A_{\mu})>= 
\lim_{t_{b}\rightarrow t_{a}}\frac{ \Gamma_{a\mu}^{b}\Phi_{a}^{-}(\gamma(t_{a})) -\Phi_{b}^{+}(\gamma(t_{b}))  )}{t_{a}-t_{b}}.
\end{equation}
\end{Definition}
\begin{Definition}
(Principle fiber bundle of secondary protein structure)
Let $X_{t}=S^{3}$ be a Kolmogorov space \cite{kolmogorov} in spinor field of  time data series.
Let  $\rho$ be the representation of gauge group $G=SU(2)$.
Let $G$ be a Lie group of spinor field of geneon-anti-geneon  pair (analogue with gene in chromosome pairs
 between dominant and recessive genotype) 
  in  supermanifold of living organism $\mathcal{M}$
 with fundamental group of complement in hyperbolic knot
 $\kappa\in K$ representation;  $\pi_{1}(S^{3}-K)=<x,y,z>,x,y$ are alphabets  for
 geneon  pairs in hidden state of DNA sequence of genotype and $z$ is the alphabet for 
a supersymmetry of transposon and retrotransposon hidden in trash DNA.
We choose $\kappa$ be a hyperbolic knot $4_{1}$, a figure 8 with  $x,y,z$. 
 Let $\rho$ be a representation of knotted group over genetic 3-alphabet of supercodon.
 It is a group action to finite vector space $G\times X_{t}\rightarrow X_{t}$ ,
 with dimension $dim V=n$.
Let $\beta$ be the co-cycle of genotype ,$\beta(x_{0}):X_{t} \rightarrow X_{t}/G$. Let the projection $\pi: P_{X_{t}} \rightarrow X_{t}$ 
be the canonical projection with 
$p\in \pi^{-1}(x_{0}),P_{X_{t}}$, a principal  bundle.  Let $A_{\mu}$ be a connection
 with underlying equivalent class of gauge field in genetic code $[A_{\mu}]$ (see \cite{preprint} for definition of genetic code as connection in detail) on covering space $P_{X_{t}}$,
 then the parallel transport along  co-cycle $\beta_{A_{\mu}}$ maps the fiber bundle 
$\pi^{-1}(x_{0})  \rightarrow \pi^{-1}(x_{0})$. Because group action $G$ is a trivialization on principal fiber bundle, 
so there exists 
$g_{A_{\mu}} \in G$ such that co-cycle of particle-like wave function of 
quantum biological observable over gauge field
 (connection or gauge potential of genetic code $[A_{\mu}]$) is the holonomy of connection $A_{\mu},\mu=1,2,3 $ represented, for folding type of fibers, in beta sheet, alpha helix and
 loop region of secondary structure of protein folding according to the Axioms of Atiyah for biology. 
 
We have a group action as folding types of translation of adjoint co-continuous functor in context of 
Grothendieck topolgy for biology,  at $p$, i.e. $\beta_{A_{\mu}}(p)=pg_{A_{\mu}}\in X_{t}/G$.
 Let  $\Phi$ be a quantum observable of coupling state between geneon and anti-geoneon pairs and transposon-retrotransposon pair.
Let $W_{\rho ,\beta,\kappa}$ be the modified Wilson loop over a link states between retrotransponson (RNA) to geneon and retrotransposon
 and folding fiber type as parallel transport of geneon along connection $A_{\mu}$ (gauge potential);
 then, we have to define a quantum biological observable of  protein structure as a  ghost field by 
\begin{equation}
\Phi(A_{\mu}):=W_{\rho,\beta}(A_{\mu})= Tr[\rho (g_{A_{\mu}})] ,\hspace{0.5cm} \forall A_{\mu} \in \mathcal{A}_{X_{t}}
\end{equation}
We call $W_{\rho,\beta,\kappa}(A_{\mu})$ the modified Wilson loop associated with 
the representation $\rho$ and co-cycle $\beta$. 
A special case is realized when $\rho =Ad$, the adjoint representation of $G$ and $\kappa $ is a knot and link of underlying DNA, RNA and protein structure.
 \end{Definition}

 \begin{Definition}
Let $\kappa$ will be a hidden knotted state in the   trajectory of metabolism of protein in underlying genotye of gene expression. Let 
$\beta$ will be cocycle of genotype in living cell (e.g. viral replication cycle or cell circle).
A modified Wilson loop with Wick rotation of path ordering gene by gauge group operation for transcription process over moduli state space of  real line, ($\mathbb{R}/\sigma(s[i])
\in \mathbb{R}/G,\sigma([s_{i}]\in G)$ real line with  extra properties of spin permutation operator attach as extradimension to each continuous point in line) for gene expression   for real path ordering  with genotype in form of connection $[A_{\mu}]$ is defined by
\begin{equation}
W_{\beta,\rho,\kappa}(A_{\mu})=Tr_{\rho}Pe^{\oint A_{\mu}  }=0.
\end{equation}
where  $Tr$ means the  trace of gauge group representation. $P$ is a solution path ordering  of genetic code with solutions ordered by permutations with wedge product property of spinor fields that satisfy the Pauli exclusion principle for  all solution states and $\rho$ is the adjoint representation of underlying gauge group operation for gene translation in gene  expression.
\end{Definition}
If we use the Wigner ray $W(z)=\frac{1}{z}$ projection to project state $[s_{1}],[s_{2}]$ 
with path ordering  (i.e. $[s_{1}],[s_{2}] \neq [s_{2}],[s_{1}]$) of solution 
of gene expression state in unit cycle of fiber with parallel transport 
along connection $A_{\mu}=\Gamma_{i}^{j\mu}$,  the equation for gene expression for one single state of gene is
\begin{equation}
W_{\beta,\rho,\kappa}(A_{\mu})=Tr_{\rho}Pe^{\oint \frac{1}{s-[s_{1}]}ds   }Tr_{\rho}Pe^{\oint \frac{1}{s-[s_{2}]}ds   }=0.
\end{equation}
so, we have the result of gene expression along direct transcription process along 
space of DNA, i.e. $X=D, G=P^{\ast},$
 $G\times X\rightarrow X, gx_{t}=x_{t+1}$ with  character $Tr\rho(G)=1$.
We have solution of gene expression by integrating over circle in fiber of genetic code. We get
\begin{equation}
|s-[s_{1}]||s-[s_{2}]|=0,
\end{equation}
\begin{equation}
s=\pm [s_{1}],\pm [s_{2}].
\end{equation}
The solution states with path ordering mean that we cannot change an order of solutions. The solution have their mirror supersymmetry 
in hidden state for transition. We will later call $+[s_{i}]$ a geneon state and a mirror state in which inactive state in trash DNA
 is the  anti-geneon state of repeated inactive gene  and denoted by $-[s_{i}]$ (the state in genotype here is a sequence
 of alphabet in biology i.e. $[s_{i}],i=1,2,3,\cdots 8=\{ATCGTTTAAAA\}$). 
 
For every genotypes, it  contains 8 hidden  states. We will extend this solution to a Laurent polynomial and a Jones polynomial by using the Seiberg-Witten invariant.
The first 4 states are geneon, anti-geneon, transposon and retrotransposon in dsDNA and
the other 4 states are in hidden DNA or passive state of DNA (it might be
 some form of moving gene in mitDNA, mitochrondrian DNA analogy with miRNA), $D^{\ast}$.
 We have $<[s_{i}]|[s_{i}^{\ast}]>=0,<[s_{i}]|[s_{i}]>=1, D^{+}|s_{i}>=|[s_{i}^{\ast}]>,D^{-}|[s_{i}^{\ast}]>=|s_{i}>$
 for all $i$ for a stable pair of orbital of geneon and anti-geneon with modified Dirac operator  
 with transition energy of ground state sets to be zeros as closed shell
 of spinor field for genotype. This pair is analogy of pair of gene in pair of chromosome in which 
biologists use for phenotype gene expression with dominant state and recessive state of genotype.
 Later, we will use the  link operator to map this sequence to Khovanov cohomology for time series of 
knots and links in protein,
$\mathcal{F}\{ 0   \rightarrow \mathcal{O}_{D}\stackrel{\phi^{d_{1}}_{\pm}}{  \rightarrow }\mathcal{O}_{R} 
\stackrel{\phi^{d_{2}}_{\pm}} {  \rightarrow }\mathcal{O}_{R^{\ast}} 
    \rightarrow   \mathcal{O}_{P}
 \stackrel{\phi^{d_{3}}_{\pm}}{  \rightarrow }\mathcal{O}_{P^{\ast}}
\stackrel{\phi^{d_{4}}_{\pm}} {  \rightarrow }\mathcal{O}_{D^{\ast}} 
\rightarrow  0\}   $. 
This algebraic construction  is analogue to  a  Khovanov cohomology in  loop-quantum  gravity context but with different meaning.
It is a discretized reaction which is the  quantization of a differential map in forward and reversed direction of the transcription, $\phi^{d_{i}}_{\pm}$ , of the genotype. It can be represented  as the  covering space of sieve and site over the supermanifold coordinates of living cell with sieve and site assuming the  Grothendieck topology. The modified Atiyah axioms for quantum biology are used to define the coordinates in the loop space of cell
 by a pullback functor from the categories of states in DNA or the 3-types genotype wave function as a morphism between states in these categories to other categories of the fiber
 of secondary protein folding structure. The observed categories of a protein $Cat(P,P^{\ast})$ are assumed to be isomorphic to  a compact oriented smooth n-dimensional manifold.
The manifold of adaptive behavior of gene expression in a living organism with geneon states in genotype  are assumed
to be isomorphic to the   categories of $(n+1)$ dimensional complex vector space of  hidden states in cell superspace, $\mathcal{M}\in Obj(Cat(\mathcal{M}))$.  The morphism is a system transmission  of a given  cohomology  sequence of 
protein docking as cell signal between intercellular transmission state of geneon, retroposon and transposon in form of  
DNA, enzyme or siRNA, miRNA , tRNA and noncoding RNA in which arise from intron. The histone structural protein  contains an octomer with histone modification process capable of  changing the transition state of spinor  fields in time series of genetic code according to the homogeneous coordinates 
of 8 states in $E_{8}\times E_{8}$ grand unify theory  of cell.

 \begin{Definition}[Atiyah-Segal and Hitchin System axioms for quantum field in biology]
Let $\mathcal{H}$ be a modified Khovanov cohomology-Grothendieck open set-co-continuous functor  map between the category of  $Cat(P,P^{\ast})$ to the category of $Cat(\mathcal{M})$ which satisfy the following axioms:
\begin{itemize}
\item {\em Axiom 1. (Orientation) } Let $Pro_{\Phi^{+}(A_{\mu})}$ be  the co-bordism of superspace of protein folding or the  boundary of  
protein complex surface during the  docking process 
with underlying connection $A_{\mu=1,2,3}$ for $\beta$ sheep, $\alpha$ helix and loop region folding in secondary structure.  The 
 curvature of the structure is derived  by the variation of  connection in parallel transport along the surface of the  protein.   We denote the  curvature inside this object as an extra-property of the adaptive behavior.  This curvature emerges as  a change docking in evolution of the  
feedback loop of retrotransposon related to the 
 strength of the gauge field $A_{\mu}$ with $F^{\mu\nu}=\partial_{\mu} A_{\nu}-
\partial_{\nu} A_{\mu}+[A_{\mu}, A_{\nu}]:=[\bigtriangledown_{\mu},\bigtriangledown_{\nu}] $
where $\bigtriangledown_{\mu}$ is a covariant derivative of  folding structure, with $\mu=1,2,3$ ($\beta$ sheet, $\alpha$ helix, loop),
interacting  with protein partner $\nu=1,2,3 $, in the active state of docking 
 with the opposite orientation of central dogma (also analog with the opposite direction of docking in protein receptor in cell membrane)
 $Pro^{\ast}_{\Phi^{-}(A_{\mu})}$
 of protein  in passive undocking state or in the hidden state of protein layer in feedback adaptive loop of evolution.

 Let $\mathcal{M}^{\ast}$ be the dual vector space of superstate in cell where  $\mathcal{M}\in Cat(\mathcal{M})$. Then
$\mathcal{H}(Pro^{\ast}_{\Phi^{-}(A_{\mu})})=(\mathcal{H}(Pro_{\Phi^{+}(A_{\mu})}))^{\ast},
\forall Pro_{\Phi^{+}(A_{\mu})}\in Cat(P,P^{\ast}) $.
\item {\em Axiom 2. (Multiplicativity)}  Let $\coprod$ be a disjoint union then 
\begin{equation}
\mathcal{H}({Pro_{\Phi^{+}(A_{\mu})}}_{1}  \coprod {Pro_{\Phi^{+}(A_{\mu})}}_{2} )=
\mathcal{H}({Pro_{\Phi^{+}(A_{\mu})}}_{1}) \otimes  \mathcal{H}({Pro_{\Phi^{+}(A_{\mu})}}_{2}),\hspace{0.5cm}
\forall {Pro_{\Phi^{+}(A_{\mu})}}_{1},{Pro_{\Phi^{+}(A_{\mu})}}_{2}\in
 Cat(P,P^{\ast}).
\end{equation}
\item {\em Axiom 3. (Transitivity)} Let $  F_{i}: {Pro_{\Phi^{+}(A_{\mu})}}_{i}  \rightarrow  {Pro_{\Phi^{+}(A_{\mu})}}_{i+1}  $
be a morphism. Then
\begin{equation}
\mathcal{H}(F_{1} F_{2} )  =   \mathcal{H}(F_{2}) \mathcal{H}(F_{1})\in 
      Hom(\mathcal{H}( {Pro_{\Phi^{+}(A_{\mu})}}_{1}) ,  \mathcal{H}(  {Pro_{\Phi^{+}(A_{\mu})}}_{3} )),
\end{equation}
where $F_{1},F_{2}$ denote the co-continuous Grothendieck topology functor as an open set in coordinates of cell metabolism.
The docking mechanism of enzyme catalytic interaction is given by joining fusion surface between protein complex
 surface of principle bundle
 ${Pro_{\Phi^{+}(A_{\mu})}}_{2} $ in the sense of co-bordism,   $  F_{1}\coprod_{{Pro_{\Phi^{+}(A_{\mu})}}_{2} } F_{2}$ in  analogy with quantum field theory.

\item {\em Axiom 4. (Identity)} Let $\phi_{k}$ be the empty of $n-$ dimensional supermanifold of living organism. Then
\begin{equation}
     \mathcal{H}(\phi_{k})  =\mathbb{C}.
\end{equation}
\item {\em Axiom 5. (Homotopy of protein folding)}  For each $ Pro_{\Phi^{\pm}(A_{\mu})}\in  Cat(P,P^{\ast}), $ we have a homotopy path of protein folding in secondary structure given by,
\begin{equation}
      \mathcal{H}( Pro_{\Phi^{\pm}(A_{\mu})} \times [0,1] ) :  \mathcal{H}(  Pro_{\Phi^{\pm}(A_{\mu})}    )\rightarrow 
\mathcal{H}(  Pro_{\Phi^{\pm}(A_{\mu})}  ).
\end{equation}
It is an identity map. This map induce  a moduli-state space model in  protein folding.
We have

\item {\em Axiom 6. (Partition of Primary and Secondary Structure )}
 Let $\mathcal{K}=C_{n}(P)$ be a $n-$manifold of chain of primary structure of protein with $dim P=n$  a number of states  represented as a compact oriented vector space in protein primary structure.
The   differential $d: Primary(P)\rightarrow Secondary(P)$ maps (we use $C_{n}(\cdot)$ for 
primary structure Grothendieck topology and $C_{n-1}(\cdot)$ for secondary structure) a primary protein structure Grothendieck topology to a
secondary protein folding structure defined by the protein folding in the sense of homotopic map (cohomology functor is the homotopic invariant functor of the protein folding by this new definition)
 of modified Khovanov cohomology decomposed space into kernel and boundary $d: C_{n}(P)\rightarrow C_{n-1}(P)$
with $ \partial C_{n}(P)=ker(d(C_{n}(P))+Boundary(d(C_{n-1}(P)))$. If morphism is injective then $Ker(d(C_{n}))=0$. We have
a modified co-bordism for global surface of protein docking as the only boundary state of amino-acid sequence of secondary structure 
without the inner structure of amino-acid sequence. We have
\begin{equation}
\partial \mathcal{K}= Pro_{\Phi^{+}(A_{\mu})} \coprod Pro^{\ast}_{\Phi^{-}(A_{\mu})}
\end{equation}
with a linear transformation as evolution of feedback loop between docking and undocking state of  curvature in 
secondary protein structure, 
\begin{equation}
 \mathcal{H}(  \mathcal{K}): \mathcal{H}({ Pro_{\Phi^{+}(A_{\mu})}}_{1}) \rightarrow  \mathcal{H}({ Pro_{\Phi^{+}(A_{\mu})}}_{2})
\end{equation}
with co-adjoint co-continuous functor.

\item {\em Axiom 7. (Hitchin system  in protein docking )}
In this axiom, we define the self dual morphism of fiber bundle of protein  $P_{\mathcal{K}}$  as $End(P_{\mathcal{K}})$. The notion of tensor product 
for secondary protein folding is the effect of evolutional adaptive behavior feedback loop gauge field. It  expresses as
 knots and links in proteins with changing curvature  defined in this extension from  Atiyah to 
Hitchin system axioms for biology. Let $P_{\mathcal{K}}$ be the principle bundle over manifold of   folding structure as covering space.
The change of internal curvature by evolutional path $[\beta]$ over co-cycle of  fiber bundle of secondary 
protein structure is defined by endomorphism of co-adjoint functor map between fiber, $[\beta]\in End(P_{\mathcal{K}})=Hom(P_{\mathcal{K}}, P_{\mathcal{K}})$.
Let  $ H^{1}(End(P_{\mathcal{K}})):=[End(P_{\mathcal{K}}),S^{1}] $ be the first cohomology group..
Let $L$ be the space of links induced by  external influence epigenetic factors of environment  which can produce an  feedback path evolution. 
The tensor  product $End(P_{\mathcal{K}})\otimes L$ means that
 the result of  mathematical complex structure, induced   from  the right hand side of multilinear algebras,
 is  $End(P_{\mathcal{K}})\otimes L=Hom ( End(P_{\mathcal{K}}),L^{\ast} ),  
L^{\ast}=Hom(L,\mathbb{Z})$.

The time series of knots and links in proteins  is parameterized by 3 types of link states in quantum loop crossing, 
as a projection from 3 dimension, the plane defined by the cohomology sequence of links 
 with the involution of a group operation of rank $rank(G)=n$, where $G$ is the translation group in amino-acide sequence along the peptide chain
(i.e. the protein has  a length with $k=1,2,3,\cdots n$ number of amino-acid )  
 $L^{\otimes k},k=1,2,3\cdots n$. We  define a more exact sequence of link operators for the folding  of secondary protein structure by,
\begin{equation}
<L>: 0\rightarrow <L_{\pm,0}>(t_{1})\rightarrow <L_{\pm,0}>(t_{1})
\cdots \rightarrow <L_{\pm,0}>(t_{n})\rightarrow 0. 
\end{equation}
where $<L_{\pm,0}>$ mean $<L_{0}>$ or $<L_{+}>$ or $<L_{-}>$
are  link states  which satisfy the modified skein relation parametrized as 3nd exact sequence layers in  knots and links in proteins  folding.
All protein structure exists in pair $(\Phi_{i}^{\pm}(A_{\mu}) ,P_{\mathcal{K}}))$  independently  of  changing curvature in 
protein folding. It   is defined by
\begin{equation}
End(P_{\mathcal{K}})\otimes L, \phi^{\pm}(A_{\mu})\in H^{0}( End(P_{\mathcal{K}})\otimes L )
\end{equation}

An element of evolutional gauge field is defined by $H^{0}(L^{\otimes k})$ where $k=1,2,3,\cdots rank(G)$, $G$ is a group
 operation over fiber of protein. It can be   represented  as a 
change of coordinates in genetic code as the effect of genetic variation in principal fiber  bundle  of protein structure. The Yang-Mill field
  and the Seiberg-Witten invariant  are  solutions of the Hitchin pair for
 biology, that is  $(\Phi_{i}^{\pm}(A_{\mu} ),A_{\mu}(P_{\mathcal{K}}))  )$. 
\end{itemize}
\end{Definition}

\begin{Definition}
Knots and links in proteins  $\kappa$ are a piecewise smooth embedding of primary structure aminoacid peptide sequence, homotopic to
 the circle $S^{1}$ in 3-space of protein supermanifold. They  satisfy the axioms of quantum field for biology. 
A unknotted protein is a special case  with link number zero.
A collection of knots and links in proteins  is called a link of proteins $[L]$. For a 2-knot link, we  define the linking number
 as the number of times where one the knot wraps around satisfying the skein relation in a Jones  polynomial.
\end{Definition} 

This protein folding structure with knots is actually the $U(1)$ Chern-Simons theory.
The characteristic class of co-bordism of secondary protein structure can be identified  with ghost  and anti-ghost fields
 for link $[L]_{\mu}^{[s_{i}]}$ of knotted protein with connection $A_{\mu}$ and underlying hidden state in genetic code $[s_{i}]$ in this 
 modified Khovanov cohomology for biology.
 This axioms  of secondary protein structure is  analogue with axiom of quantum field theory. It will be 
indentified with Jones-Ramanujan-Witten-Laurent
 polynomial for the ghost field $\Phi_{i}^{\pm}([L]^{[s_{i}]}_{A_{\mu}})$ of link of knotted protein in 
 modified Khovanov cohomology for loop quantum  gravity for biology.

\section{Modified Khovanov Cohomology for Time series of knotted protein}
\begin{figure}[!t]
 \centering
 \epsfig{file=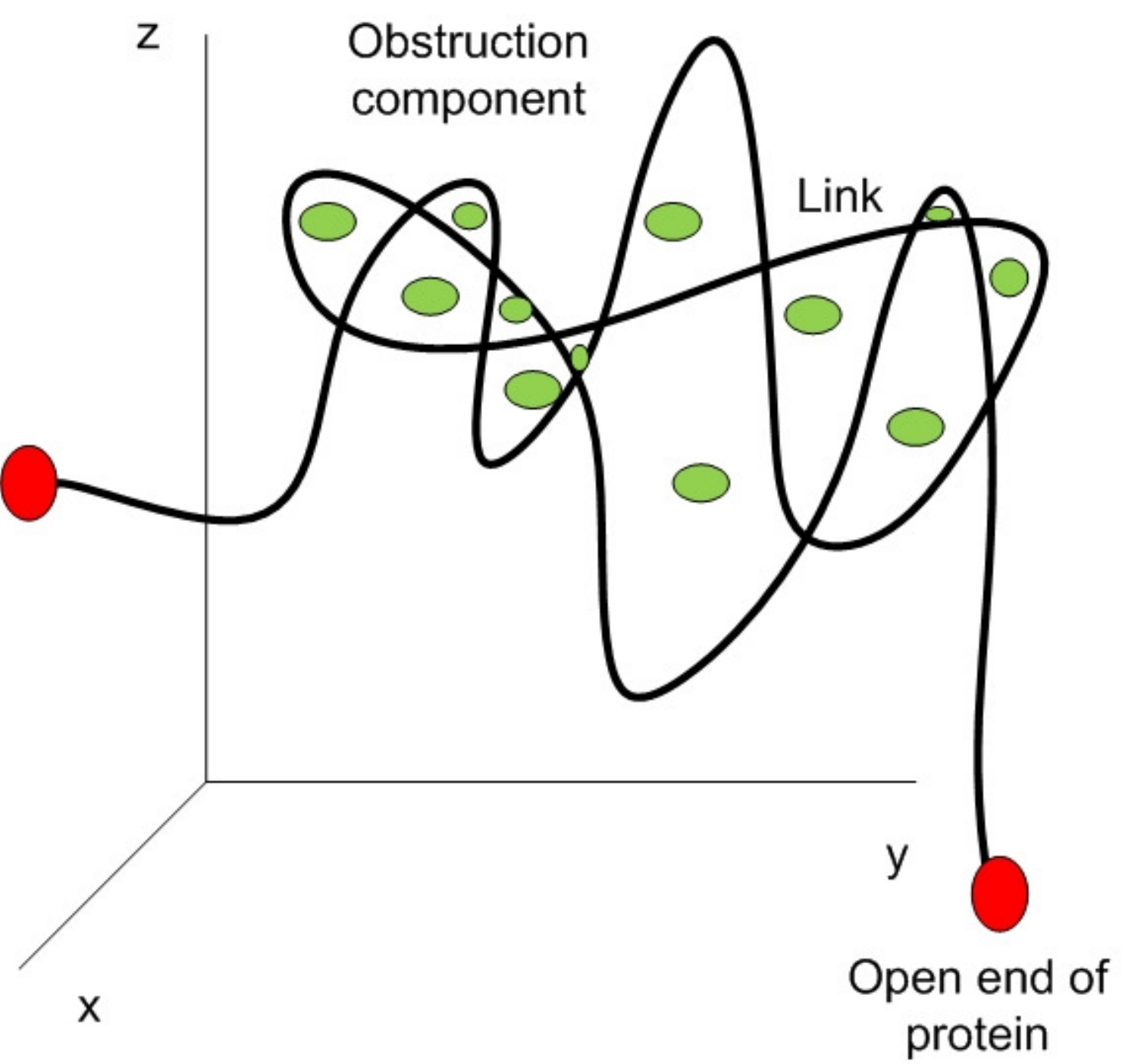,width=8cm}
 \epsfig{file=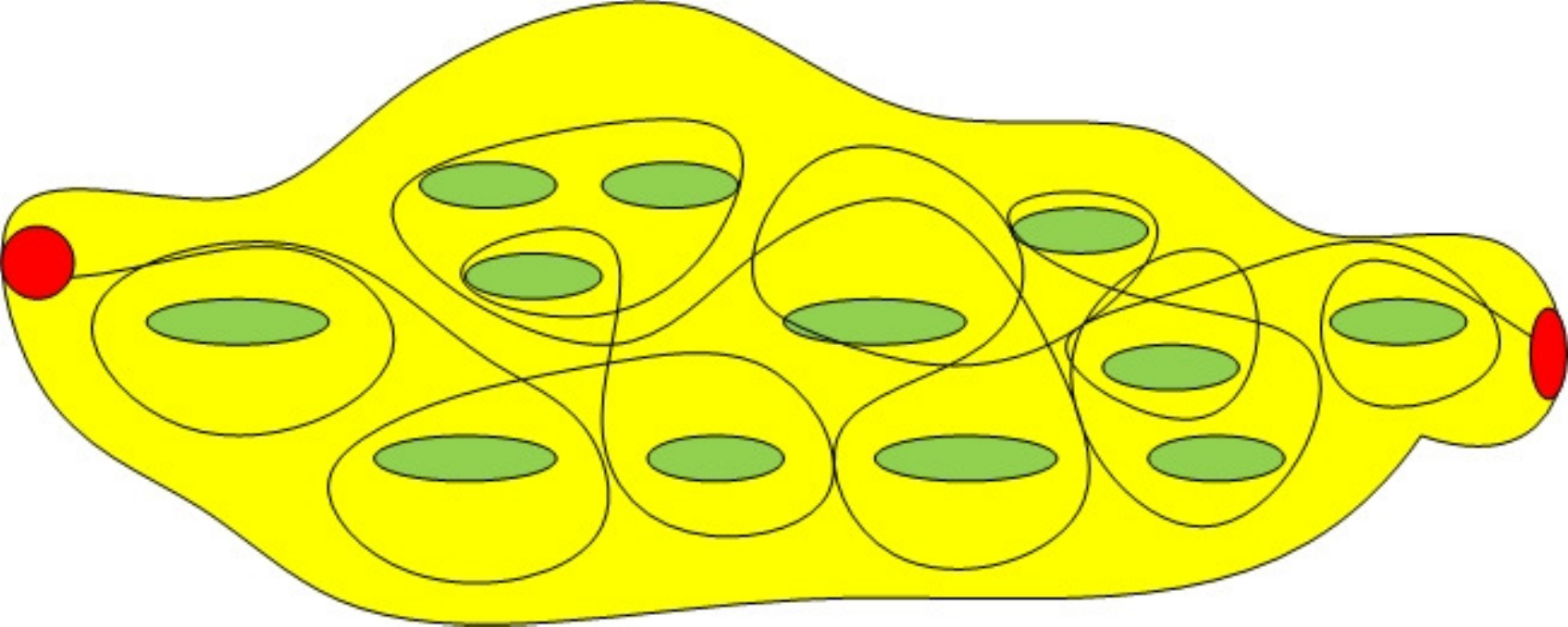,width=8cm}
 \caption{In  the left picture, it is represented the time series of  knots and links  in the 
 secondary protein  as an open string with curvature between 3 types of principal bundle. 
The link operators in skein relation of knotted protein are parametrized by time series data of knotted protein in the  Khovanov cohomology for biology. The open 2 end points of string are kernel elements projected from higher dimensions
  of cohomology moduli boundary conditions of starting and ending gene expression in amino-acid 
sequence of geneon state in DNA. \label{timeseries}. The picture shown on the right is the transformation of the left figure from one dimension in 
modified Khovanov cohomology to third modified Khovanov cohomolgy group of surface with 12 obstructions components shown as genus of surface. The knotted time series data is a trajectory
 of curve around these genus of surface in moduli state space model of time series data. With cohomology theory, the boundary of surface can be calculated and we can 
get a volume  form   with curvature of  secondary protein surface.}
 \end{figure}

In this section, we give a new definition of time series of protein folding (see  Fig. \ref{timeseries}) as an  infinite long exact sequence of
 modified Khovanov cohomology for biology and axioms of quantum field theories introduced by Michael Atiyah \cite{atiyah}. First we explicitly give a new definition to 3 types of connection over the 
principal  bundle of secondary structure with their underlying  3-co-cycles of geneon, transposon and retrotransposon states in genotype.

\begin{Definition}
Let $[s_{i}]$ be a spin state in DNA (for more details on the  definition of all states of spinor field in time series
 of genetic code see \cite{preprint}).
 Let $[L]^{[s_{i}]}_{A_{\mu}}$ be the  protein link  of the 
 underlying spin state $[s_{i}]$ and gauge field $A_{\mu}$ for the behavior of
 underlying genotype.
 The source of line is a category of free Abelian group over  a
short exact sequence of sheaf cohomology of central dogma.
The behavior of gene expression link and the behavior gauge field is induced from 
 interaction adaptive behavior of protein docking with capsid 
protein and host cell receptor
 in feedback recursive loop space of underlying hyperbolic knot of link.
 We define the Khovanov cohomology for biology the vector space of 
states in genetic code discretized by genetic shift and drift operator, $ \phi^{d_{i}}_{\pm} = 0 \rightarrow  \mathbb{Z} \rightarrow  0 ,i=1,2,3$, with co-differential map for co-chain sequence, $d_{i}^{\ast}$.
Let   an equivalent class of link operator over figure 8 hyperbolic knot $\kappa=4_{1}$  in loop quantum  gravity for biology be
\begin{equation}
[L_{\pm}]^{[s_{i}]}_{A_{\mu}} = \mathcal{F}\{ 0   \rightarrow \mathcal{O}_{D}\stackrel{\phi^{d_{1}}_{\pm}}{  \rightarrow }\mathcal{O}_{R} 
\stackrel{\phi^{d_{2}}_{\pm}} {  \rightarrow }\mathcal{O}_{R^{\ast}} 
    \rightarrow   \mathcal{O}_{P}
 \stackrel{\phi^{d_{3}}_{\pm}}{  \rightarrow }\mathcal{O}_{P^{\ast}}
\stackrel{\phi^{d_{4}}_{\pm}} {  \rightarrow }\mathcal{O}_{D^{\ast}} 
\rightarrow  0\}   
\end{equation}

Let $\Phi_{i}^{\pm}([L]^{[s_{i}]}_{A_{\mu}})$ be ghost  and anti-ghost field from 
docking and undocking state of protein of 
gene $i$ at position of time series of knot $ [L_{\pm}]^{[s_{i}]}_{A_{\mu}}$ in protein . 
A time series of knotted protein over Wilson loop of gauge field of genotype $W_{\kappa,\rho}(A_{\mu})$ is
\begin{equation}
\Phi_{i}^{\pm}([L]^{[s_{i}]}_{A_{\mu}})=\sum_{\mu=1}^{n} [L_{\pm}]^{[s_{i}]}_{A_{\mu}}
W_{\kappa,\rho}(A_{\mu}).
\end{equation}

where $A_{\mu}$  is a connection over gauge transformation of
 a right translation group operation over fiber space of genetic code
 according to the type of fiber space of protein. 
A type 1 with a connection $A_{\mu=1}$ is  a $\alpha$ helix connection, type 2, with
 a conneciton $A_{\mu=2}$, is a connection over fiber of $\beta$ sheet in secondary protein,
 type 3, $A_{\mu=3} $  is a connection over fiber of  loop region in secondary protein.
 When a protein is folding to the  secondary structure, 
the knots of protein are open states obtained by using the skein relation as a modified covariant derivative over parallel transport of connection of genetic code $[A_{\mu}]$. 
Let $[L_{\pm,0}]:=[L_{\pm}]_{A\mu}^{[s_{i}]}$ be the link operator of protein over a modified Grothendieck
 cohomology. Let $H_{\cdot}( \cdot)$ be a functor of cobordism of
 secondary protein satisfying  the modified Atiyah-Segal-Hitchin axioms for 
quantum biology. We have a long exact sequence of Grothendieck cohomology
 induced by short exact sequence of central dogma with value in
 principle bundle of secondary protein fiber $P_{\mathcal{K}}$, that is 
\begin{equation}
  H_{\cdot}([L_{\pm}]_{A\mu}^{[s_{i}]};P_{\mathcal{K}}): 0  \rightarrow \cdots \rightarrow  H_{n}([L_{\pm}]_{A\mu}^{[s_{i}]};P_{\mathcal{K}})  \stackrel{\phi^{d_{1}}_{\pm} }{  \rightarrow }
 H_{n}([L_{\pm}]_{A\mu}^{[s_{i}]};P_{\mathcal{K}})\stackrel{\phi^{d_{2}}_{\pm} }{  \rightarrow } 
 H_{n}([L_{\pm}]_{A\mu}^{[s_{i}]};P_{\mathcal{K}})
\stackrel{\phi^{d_{3}}_{\pm} }{  \rightarrow } \nonumber
\end{equation}
\begin{equation}
  \rightarrow H_{n+1}([L_{\pm}]_{A\mu}^{[s_{i}]};P_{\mathcal{K}})
  \stackrel{\phi^{d_{1}}_{\pm}}{  \rightarrow }
 H_{n+1}([L_{\pm}]_{A\mu}^{[s_{i}]};P_{\mathcal{K}})\stackrel{\phi^{d_{2}}_{\pm}}{  \rightarrow } 
 H_{n+1}([L_{\pm}]_{A\mu}^{[s_{i}]};P_{\mathcal{K}})
\stackrel{\phi^{d_{3}}_{\pm} }{  \rightarrow } \cdots \rightarrow 0
\end{equation}
We call  $  H^{n}([L_{\pm}]_{A\mu}^{[s_{i}]};P_{\mathcal{K}})=Hom(H_{n}([L_{\pm}]_{A\mu}^{[s_{i}]},P_{\mathcal{K}}) $ 
a modified Khovanov cohomology for quantum biology.
It is possible to give a new definition of  differential one-form for knotted protein structure in every connection of 
gauge field over genotype by
$\Phi^{\pm}_{i}(A_{\mu})  \mapsto   d\Phi^{\pm}_{i}(A_{\mu})$ defined by the 
 skein relation over link $[L]_{A\mu}^{[s_{i}]} $,
\begin{equation}
 \bigtriangledown_{\Phi^{-}_{i}(A_{\mu})  }^{[L_{+}]^{[s_{i}]}_{A_{\mu}}}\Phi^{+}_{i}(A_{\mu})=\frac{ [L_{+}]^{[s_{i}]}_{A_{\mu}}
W_{\kappa,\rho}(A_{\mu})- [L_{-}]^{[s_{i}]}_{A_{\mu}}
W_{\kappa,\rho}(A_{\mu})}{ [L_{0}]^{[s_{i}]}_{A_{\mu}}
W_{\kappa,\rho}(A_{\mu}) },
\end{equation}
and
\begin{equation}
  \bigtriangledown_{\Phi^{+}_{i}(A_{\mu})  }^{[L_{+}]^{[s_{i}]}_{A_{\mu}}} \Phi^{-}_{i}(A_{\mu})=\frac{ [L_{-}]^{[s_{i}]}_{A_{\mu}}
W_{\kappa,\rho}(A_{\mu})- [L_{+}]^{[s_{i}]}_{A_{\mu}}
W_{\kappa,\rho}(A_{\mu})}{ [L_{0}]^{[s_{i}]}_{A_{\mu}}
W_{\kappa,\rho}(A_{\mu}) },
\end{equation}
A modified Grothendieck cohomology with $(\alpha,\beta)$ 
cocycle is given by
\begin{equation}
\Phi_{i}^{\pm}(A_{\mu=k}) = \alpha_{\pm} + \beta_{\pm} k+\epsilon_{k},
\end{equation}
for every gene of viral glycoprotein.
\end{Definition}

From this definition, we can measure $\alpha$ and $\beta$ for
regression parameters of trend in $J^{\mu}(A_{\mu})$ in order to  calculate the band gap between 
the transition of adaptive curvature in self-dual form of  Seiberg-Witten equation for protein docking
 $F^{+}_{A_{\mu}}=0, D^{+}_{\Phi^{+}(A_{\mu})}\Phi^{-}=0$ for the 
 evolutional field produced by the  feedback loop of
 adaptive changing curvature of secondary protein folding fiber with  3 types of connection, i.e.
 $\beta$ sheet for $A_{\mu=1}$, $\alpha$ helix, $A_{\mu=2}$  and variation loop region, $A_{\mu=3}$ in
 docking system between   viral glycoprotein and host cell receptor.

Let the cell superspace  be 
\begin{equation}
\mathcal{M}_{X} = X_{t}^{N} \coprod Y_{t}^{N} \coprod  X_{t}^{N}/Y_{t}^{N}
\coprod X_{t}^{O}
 \coprod Y_{t}^{O}
\coprod N_{t}
\coprod O_{t}
\coprod Cyte_{t}
\end{equation}
where

\begin{itemize}
\item $X_{t}^{N}$ inner space of nuclear membrane.
\item $Y_{t}^{N}$ outer space of nuclear membrane.
\item $X_{t}^{N}/Y_{t}^{N}$  space of nuclear pore.
\item $X_{t}^{O}$  inner space of double layer organelle membrane.
\item $Y_{t}^{O}$ outer space of double layer organelle membrane.
\item $N_{t}$ space of nucleoplasm.
\item $O_{t}$ space of matrix inside organelle.
\item $Cyte_{t}$ space of cytoplasm.
\end{itemize}

\begin{figure}[!t]
 \centering
 \epsfig{file=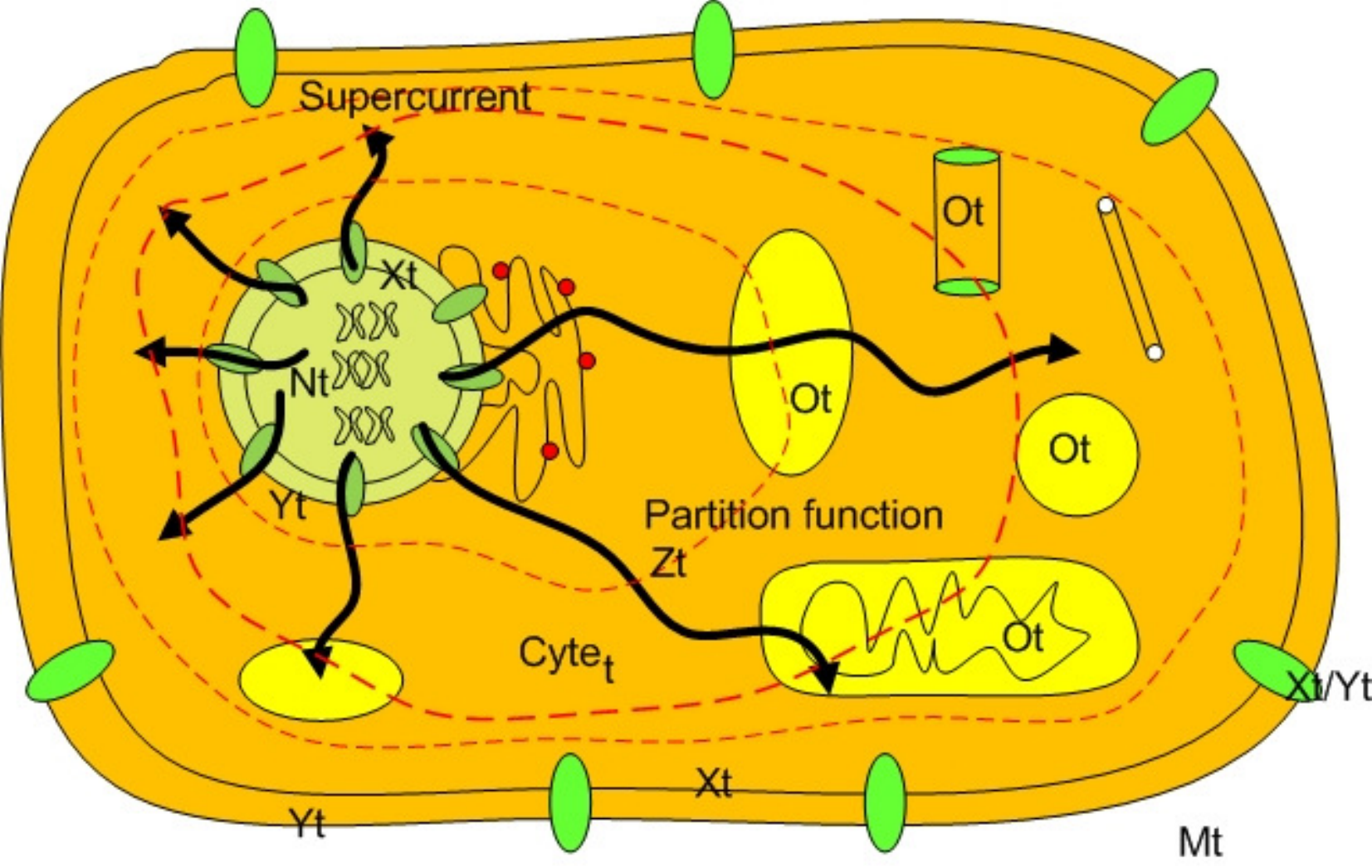,width=8cm}
\epsfig{file=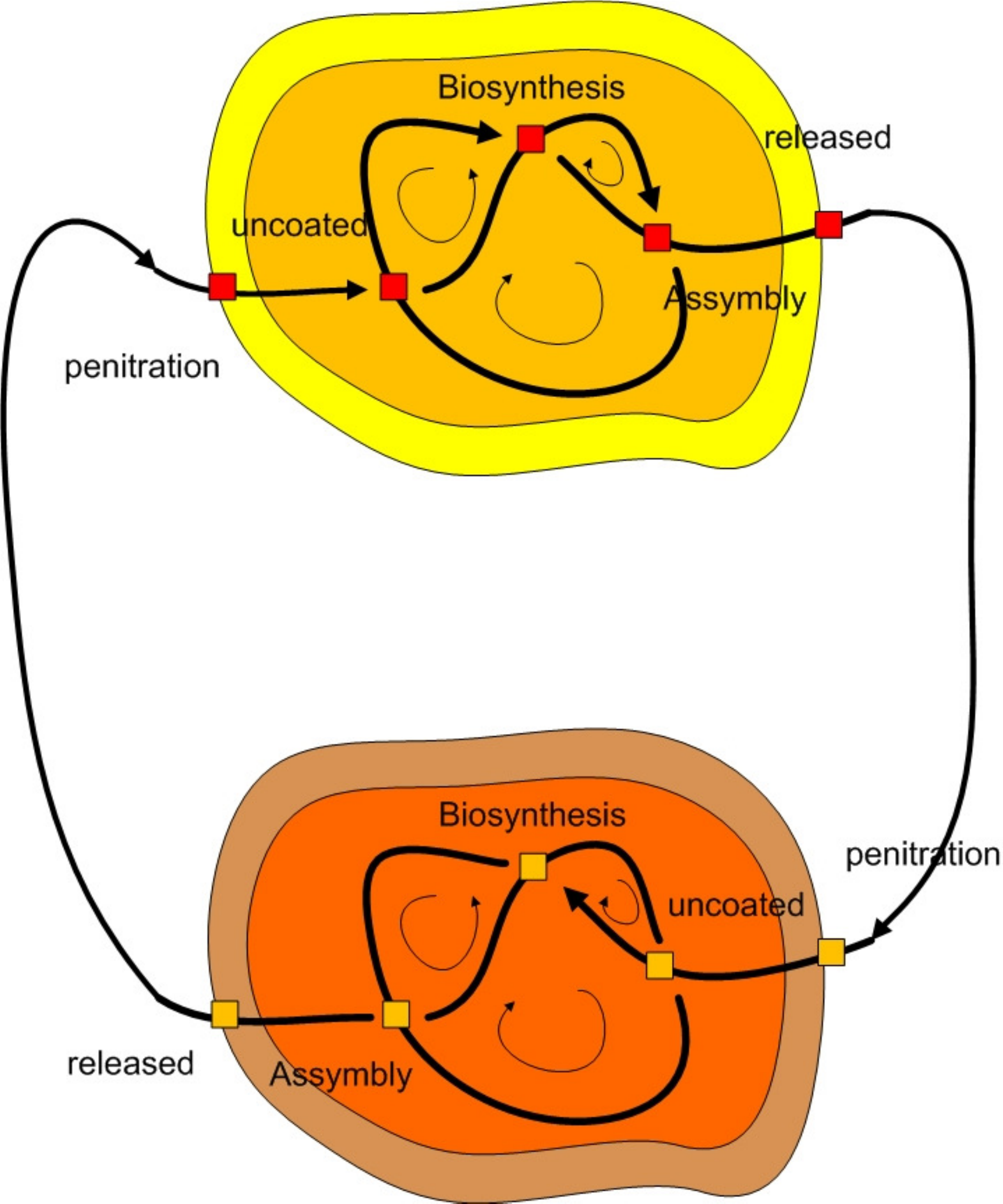,width=8cm}
 \caption{The  picture on the left panel represents the    superspace of host
 cell with supercurrent from its genotype. We can separate the  supermanifold of cell $\mathcal{M}$
 into subspaces with partition function associated with their phase transition in the  form of Wilson loop 
for quantum observables. The Chern-Simons current of DNA can be transferred around the whole  subspace as a phase transition in  analogy with Kelvin knotted model  for cell.
The  picture on the right panel is a viral replication circle  connecting  2 hosts cells
 with retroposon state in $4_{1}$ linked between 2 couplings of areplication circle. In the above cell,
 red dots means junctions between the interaction of parasitism state of viral particle and the host cell.
The exact sequence of replication circle induces infinite Khovanov cohomology sequence in immunosystem.
%The right picture   , we show the diagram of gene expression for icosahedral capsid protein $VP0,VP1,VP2,VP3,VP4$.
% We can notice that in this diagram the expression  for genotype of capsid proteins start at 
%nucleotide number 741.
 \label{cell}}
 \end{figure}

We have 2 types of dimensions, i.e.  the  first is the dimension of space, the second is the dimension of superspace with a given  number of layers. The first is double layers. The second is a single layer of spinor network of genetic code  and spinor network of protein secondary structure folding.
There exists 2 types of double layers for dbrane and anti-dbrane. Nuclear membrane and some organelle 
membrane like ER and mitochrondia membrane see (Fig. \ref{cell} for detail). The last type is a single  dbrane of cell membrane.
The maps between those 5 layers of cytoplasm define the coordinates  of all enzyme as a node in ribbon graph 
of a supersymmetric support  of Dirac network for quantum atomic cell. We give a simple quantum mechanical  equation 
derived from the  interaction of protein and gene between organelle and cell membrane for a single cell.
We have a modified attach and docking operator for  ghost field in protein state $+$ and
 undocking state with sign $-$ in the  feed back loop of protein
 receptor  over the cell superspace $\mathcal{M}_{X}$.
This new Dirac operator is defined as
%%%%%%%%%%%%%%%%%

\begin{eqnarray} 
D^{\pm}_{A_{\mu}}|\Phi_{i}^{\pm}([L]^{[s_{i}]}_{A_{\mu}})>=
 \oint_{M} \Pi_{k}  \int_{Nt}A\wedge \partial A + A\wedge A\wedge A+
\int_{Cyte} A\wedge A\wedge A\wedge A +
 \int_{X_{t}/Y_{ts}}\Pi_{j} W_{i} \Gamma_{i}^{jk}(\omega_{1},
 \omega_{2},\cdots ,\omega_{n} )+E(k)
\nonumber
\end{eqnarray} 
\begin{eqnarray} 
-\xi  \frac{d^{n}}{d^{n}\omega^{n}}|_{C_{n}(V_{ts})}\Pi_{j} W( \Gamma^{i}_{jk}(\omega^{1},
\omega^{2},\cdots ,\omega^{n} ))-E^{\ast}(k)=J_{N}(\kappa;q;A_{\mu})|\Phi_{i}^{\pm}([L]^{[s_{i}]}_{A_{\mu}})>.
  \end{eqnarray}

\section{Seiberg-Witten invariant for knots and links in proteins}
 
 The connection $A_{\mu}$
  has also some extra properties related to the  behavior of docking and undocking states  of protein signals
 of receptor and transmission along intercellar and intracellular as partition function in form of 
Chern-Simon current of life energy encryption in underlying 
 methyl transfer of codon in tRNA.  
   We have an expectation of gene expression as quantum gauge observable. The  wave function 
is the orbit of principal bundle with  group action while the translation process in gene expression is 
 the fixed point of the gauge group. The
 mirror symmetry gauge group act on  transposon and retrotransposon 
along the ribbon graph of knotted DNA folding. They  dock to each other and the  curvature is the strength of
 connection. We measure the change of curvature by using the modified  Dirac docking operator
 for protein receptor $D^{-}_{\mathcal{O}_{D},\mathcal{O}_R,\mathcal{O}_P}: (\Phi^{-}, \Phi^{+})\mapsto
  Adj_{\Phi^{-}} \Phi^{+}=\epsilon_{t}\in \mathbb{R}$,
\begin{equation}
<\Phi^{+}(A_{\mu})>_{k}=
\frac{1}{Z_{k}} \int \Pi_{+}W_{\rho,\beta,D^{-}_{\mathcal{O}_{D},\mathcal{O}_R,\mathcal{O}_P} }(A_{\mu})e^{-iS_{CS}},\nonumber
\end{equation}
\begin{equation}
=J_{k}(N;q=e^{\pm \frac{2\pi i}{N}}),
\end{equation}
where $J_{k}$ is a Jones polynomial.

%%%%

\begin{Definition}
 A Ramanujan-Jones-Laurent knot polynomial in genetic code over protein docking state 
with connection $A_{\mu}$
 for a  modified  Wilson loop $\Phi^{\pm}(A_{\mu}):=W_{\rho,\kappa,D^{\pm}}(A_{\mu})$ with a
knot $\kappa$ of coupling constant  between geneon, transposon and retrotransposon with link $L$ is
\begin{equation}
<\Phi^{\pm}(A_{\mu})>_{k}=J_{k}(q,W_{\rho,\kappa ,D^{\pm}}(A_{\mu});N)
\end{equation}
where $q=e^{\frac{2\pi i }{k+h} }$. Let $<[L^{[s_{i}]}_{A_{\mu}}]>$ (or we write shortly $<L>$) be an 
expectation value of the corresponding modified Wilson loop for the Chern-Simons
 current of level $k$. The Jones polynomial for the link $L$ in $S^{3}$ needs 
to satisfy the skein relation written in the general form of Khovanov cohomology
 of link $<L>$,
\begin{equation}
q_{1} <L_{+}>+q_{2} <L_{0}>+q_{3} <L_{-}>=0
\end{equation}
where the coefficients $q_{1},q_{2},q_{3}$ are given
 as Verlinde-Laurent-Jones-HOMFLY polynomial $P(q_{1},q_{2},q_{3})$, that is

\begin{equation}
q_{1}=-e^{\frac{2\pi i}{h(h+k)}},\hspace{0.5cm} 
q_{2}=-e^{\frac{\pi i(2-h-h^{2})}{h(h+k)}}  -e^{\frac{\pi i(2+h-h^{2})}{h(h+k)}},
\hspace{0.5cm} q_{3}=e^{\frac{2\pi i(1-h^{2})}{(h(h+k))}}.
\end{equation}
The coefficient  $q_{1},q_{2},q_{3} $ are obtained from 
Verlinde fusion rule \cite{fusion} for 2d conformal field theories.
\end{Definition}

%%%
 
From the definition, the modified  Wilson loop is a gauge invariant \cite{preprint2} of spinor field in genetic code.
 Let $\mathcal{G}_{X}$ be a gauge group.
 We have a orbit space $\mathcal{O}_{X_{t}}=\mathcal{A}_{X_{t}}/\mathcal{G}_{X_{t}} $
 with $\Phi\in \mathcal{O}_{X_{t}}.$ 
Let $\beta_{1}^{geneon}$ be the cocycle of geneon state of principal  bundle of spinor field in gene.
Let  $\beta_{2}^{anti-geneon}$ be the cocycle of 
anti-geneon state in active side of DNA. 
Let  $\beta_{3}^{transposon}$ be the cocycle of 
transposon state in inactive side of trash DNA with momentum $k$ in cycle.
 Let  $\beta_{4}^{retrotransposon}$ be the cocycle of retrotransposon state in gene
 with momentum $-k$ in trash DNA. 
These 4 cocycles are loop and knot properties in $\kappa =4_{1}$ which induces hidden 8 states in DNA,
 e.g. the figure 8 hyperbolic knotted DNA in bacteriophage.
Each state can have its own knotted state denoted by $\kappa_{1},\kappa_{2},\kappa_{3},\kappa_{4}$.
 They have representations $\rho_{1},\rho_{2} ,\rho_{3},\rho_{4}$ of $G$ with link together in form
 of $L^{genotype}$. We define the modified Wilson loop with some extra-property of underlying protein docking and undocking states, measured by $D^{+}$ and $D^{-}$,  as 

\begin{equation}
\Phi(A_{\mu})^{\pm}:=W_{L^{genotype},D^{\pm}}(A_{\mu})=W_{\rho_{1},\kappa_{1},D^{\pm}}(A_{\mu})W_{\rho_{2},\kappa_{2},D^{\pm}}(A_{\mu})W_{\rho_{3},\kappa_{3},D^{\pm}}(A_{\mu})W_{\rho_{4},\kappa_{4},D^{\pm}}(A_{\mu}),\hspace{0.5cm} 
\forall A_{\mu} \in \mathcal{A}_{X}
\end{equation}

The partition function  of genotype is  a phase transition 
of gene expression at all level in the cell, primary protein, secondary protein folding and also cell signal gene expression in intercellular transmission between cell membrane.
 It is a  gauge invariant over canonical form in cell. We give   $\Phi$ as an element in the 
 moduli state space of cell under the gauge
 potential of gene expression $\mathcal{M}_{X}=\mathcal{A}_{X}/\mathcal{G}_{X}$.
 Let $k\in \mathbb{R}$  be a coupling constant of transition states between 3 orbitals of genotype: geneon, transposon and retrotransposon.
 Then the Chern-Simons action $S_{CS}$ of this coupling state takes the form of Chern-Simons current for life energy
\begin{equation}
J^{\mu}:= S_{CS}(A_{\mu})= \frac{k}{4\pi}\int_{X} tr(A_{\mu}\wedge dA_{\mu}  
+\frac{2}{3} A_{\mu}\wedge A_{\mu} \wedge A_{\mu} )
\end{equation}
where $J^{\mu}$ is a Chern-Simons current, in biology analogy, with free energy of different 
of spinor fields between starting  and end points of parallel transport of geneon along the link $L$ in lattice gauge field $A_{\mu}$: this is the 
 so called holomony of ribbon graph in path integral. 
 
Let us define a partition function of these current in all small units of inside cell, i.e. organelle,
 nucleous, cell membrane and etc. by  
\begin{equation}
Z_{k}(\Phi^{\pm})=\int_{\mathcal{M}_{X}} e^{-i S_{CS}(A_{\mu})} \Phi^{\pm}(A_{\mu}) D\mathcal{A} 
\end{equation}
where $\Phi^{\pm} :(\mathcal{A},s)\rightarrow \{1,-1\}$ is a quantum observable of genotype
 in ghost field of docking state with no variation in gene and anti-ghost field of undocking 
state with induced feedback loop of  a genetic variation $\epsilon_{k}$.
 For moduli space $F_{CS}:\mathcal{M}_{X}\rightarrow \mathbb{R}/\mathbb{Z}$ 
with coupling constant $k$
The expectation value of genotype in genetic code $ <\Phi^{\pm}>_{k}$
 of the observable $\Phi^{\pm}$ is given by
\begin{equation}
 <\Phi^{\pm}>_{k}=\frac{Z_{k}(\Phi^{\pm})}{Z_{k}(S^{3})}=
\frac{ \int_{\mathcal{M}_{X}} e^{-iS_{CS}}  \Phi^{\pm}(A_{\mu}) D\mathcal{A}  }
{    \int_{\mathcal{M}_{X}} e^{-iS_{CS}}D\mathcal{A}    }.
\end{equation}

\begin{figure}[!t]
 \centering
\epsfig{file=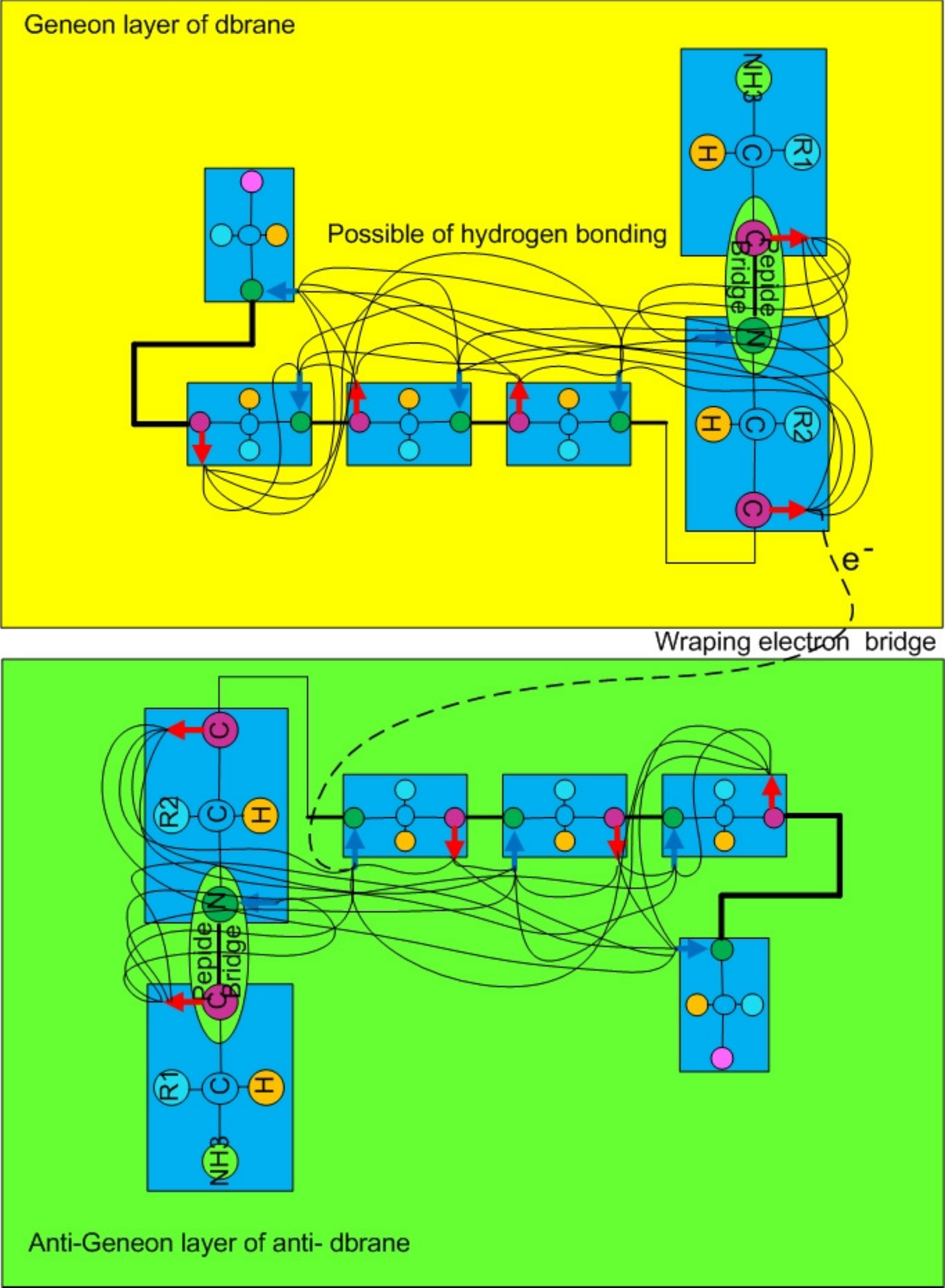,width=8cm}
\epsfig{file=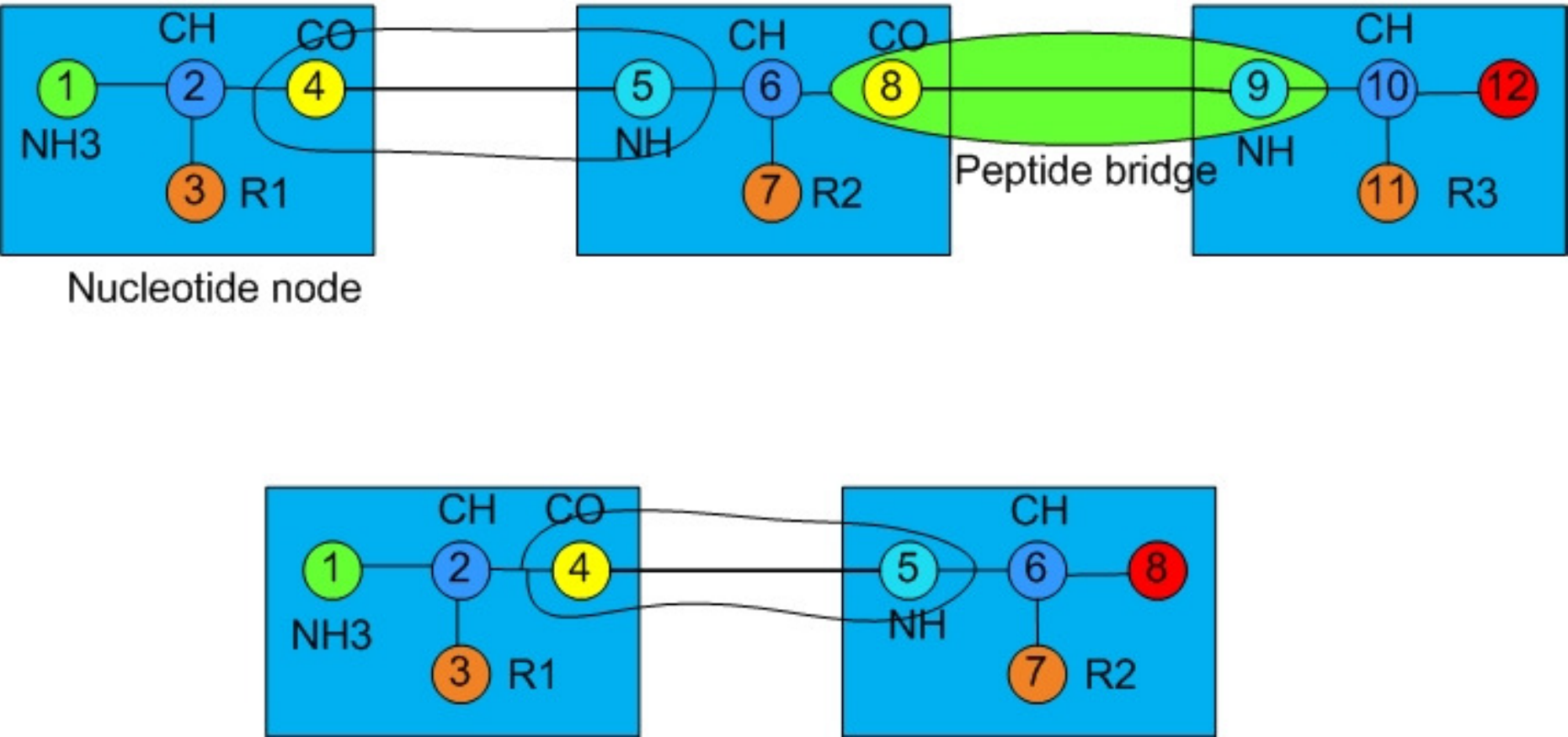,width=8cm}
 \caption{In the  picture on the left , we represent a   ribbon graph model of  nucleotide in genotype. There are many possibilities of knotted form from the quantization of hydrogen bond in peptide chain. It can be visualized as a ribbon graph for protein structure with loop space in time series data. 
In the picture on the right, we show the peptide bridge in protein. The hydrogen bonding in ribbon graph twists the  plane of secondary protein into the principal bundle with connection. In our model, when we connect 2 amino-acids, the ribbon graph  has 8 nodes reflected with  8 states 
with peptide bridge.  The node of ribbon graph is an amino-acid molecule with the edge of ribbon graph as hydrogen bonding.\label{ribbon} }
 \end{figure}

If $X=S^{3}$ and $G=SU(2)$ ,Witten obtains the explicit
form of partition function of the level $k$
\begin{equation}
Z_{k}(S^{3})=\sqrt{ \frac{2}{k+2} } \sin (\frac{\pi}{k+2}).
\end{equation}
This expression is the so called normalized Witten invariant $W_{G,\kappa}(X)$ of 2 frames in which the state of  geneon is defined by the Jones polynomial with the knotted property of the skein relation in fusion rule, 
closed, oriented 3-manifold $X$ with link $\kappa$ where 
$W_{G,\kappa}(X)=\frac{\Xi_{G,\kappa}(X) }{\Xi_{G,\kappa}(S^{3})}$. More details of derivation of 
this formulas can be found in \cite{witten}. In a previous work \cite{preprint}, we used this representation for 64 states in codon explicitly as partition function of 20 amino-acids over the genetic code in 4 dimensional manifold of living cell.

 Here we apply the same  theoretical approach  to compute the time series of knotted protein with 
Ramanujan-Witten-Jones-Laurent polynomials  in the time series data of secondary structure of protein.
Let $\Phi^{\pm}(t_{0})$ be the ghost field of starting point of the fiber of secondary protein and $\Phi^{\pm}(t)$ be the ghost field at position  $t$ in the sequence of amino-acids in the peptide chain.
Let $q_{N}(P_{\mathcal{K}}; \kappa;\Phi^{\pm}(t))=[e^{i\alpha(t-t_{0})}/e^{i\beta(t-t_{0})}/ ,  1]\simeq e^{i(\alpha-\beta)(t-t_{0})}  $ .
Every ghost field and anti-ghost field of protein is parameterized by the partition function with order parameter $\beta^{geneon}$ in protein 
sequence with associate quantum biological observable as the link operators $<L_{0}>,<L_{+}>,<L_{-}>$ satisfying the  skein relation.
In a ring of Laurent polynomial, we can associate  2 types of loop space with left and right supersymmetry in reversed direction.
The change of direction is done by the Wilson loop operator. We have modified the skein relation to recover the  relation for parameterized time 
series of knots and links in proteins, that is 
\begin{equation}
q_{N}^{1}(t) <L_{+}>-q_{N}^{-1}(t) <L_{-}>=q_{N} <L_{0}>(t-1),t=1,2,3,\cdots ,n.
\end{equation}
This relation start from the  first sequence to the end of protein structure. The measurement of the polynomial can be done by using the algorithm 
of Chern-Simons current for biology over genetic code $[A_{\mu}]$.

The Seiberg-Witten equation for biology can be used  for the protein transport  across the cell membrane. 
We have the following
\begin{Definition}
Let $\mathcal{M}_{X}$ be the moduli state space of cell membrane with 2 phospholipid layers.
Let $P^{\ast}$ be the induced structure of protein in passive hidden state. We have a complex connection for principle bundle of protein receptor embedded in the cell membrane with 2 states, active state $P$ with connection  $ \mathcal{A}_{\mu}$ with  the structure of complex principal  bundle of protein in hidden state $P_{\mathbb{C},\mathcal{K}}$,
\begin{equation}
\mathcal{A}_{\mu}=A_{\mu}(P_{\mathcal{K}})+ i A_{\mu}(P^{\ast}_{\mathcal{K}}) , 
\end{equation}
where 
\begin{equation}
A_{\mu}(P^{\ast}_{\mathcal{K}})\in \Omega^{1}(\mathcal{M}_{X})\otimes Ad(P_{\mathcal{K}}),
\end{equation}
is an inactive state of channel. We have the Chern-Simons current  $J^{\mu}(A_{\mu})$  of flow of protein through the channel in cell membrane with cocycle of state in geneon descripted by the Wilson loop of protein $W_{\kappa,\beta}(A_{\mu})$ with
\begin{equation}
h=-Re(e^{i\beta } J^{\mu}(\mathcal{A_{\mu}})).
\end{equation}
The metric for time scale in time series data of protein transmission is parametrized by
\begin{equation}
ds^{2}=\int_{\mathcal{M}_{X}} Tr \delta  \mathcal{A}_{\mu} \wedge \ast \delta \mathcal{A}_{\mu}.
\end{equation}
with the modified Seiberg-Witten equation for the cell membrane,
\begin{equation}
\frac{d\Phi^{\pm}(A_{\mu})}{ds}=\bigtriangledown h(\Phi^{\pm}(A_{\mu}) ) .
\end{equation}
In this situation, the system of receptor protein $\Phi^{+}(A_{\mu})$ in cell membrane is in equilibrium state of docking between the
 curvature of active protein from viral particle $P$ with curvature $F^{[A_{\mu}]}=dA_{\mu}+A_{\mu}\wedge A_{\mu}$
 and curvature of passive protein $\Phi^{-}(A_{\mu})$ of host cell receptor in cell membrane  
 $P^{\ast}$ with curvature $\ast F^{[A_{\mu}]}$.
Actually, the Seiberg-Witten equation for biology is a system of 2 equations,
\begin{equation}
F^{[A_{\mu}]}=0,
\end{equation}
and
\begin{equation}
D^{+}_{\Phi^{+}(A_{\mu})}|\Phi^{-}(A_{\mu}) >=0  ,\hspace{0.5cm} D^{-}_{\Phi^{-}(A_{\mu})}|\Phi^{+}(A_{\mu}) >=0,
\end{equation}
if we add extra-dimensions of time scale to the system of
 $\mathcal{M}_{X}\times \mathbb{I}$ homotopy path of  cell membrane with 2 phospholipid layers.
Let $[A_{\mu}]$ be an equivalent class of genetic code of viral glycoprotein in which  attaches the 
host receptor docking protein. We can define an evolution signal of  $[A_{\mu}]$ over time scale in which the
analogy with genetic shift and genetic drift is
\begin{equation}
\frac{\partial [A_{\mu}]  }{\partial t}= - \frac{\delta J^{\mu}([A_{\mu}])}{\delta [A_{\mu}]} =- \ast F^{[A_{\mu}]}.
\end{equation}
The solution for this equation is 
solved for the Hitchin system of quantum field in biology by  the pair
 $(\Phi^{\pm}(A_{\mu}),A_{\mu}(P_{\mathcal{K}}))$.
\end{Definition}

 \section{Computation  in synthetic  time series of knots and links in proteins }
\begin{figure}[!t]
 \centering
\epsfig{file=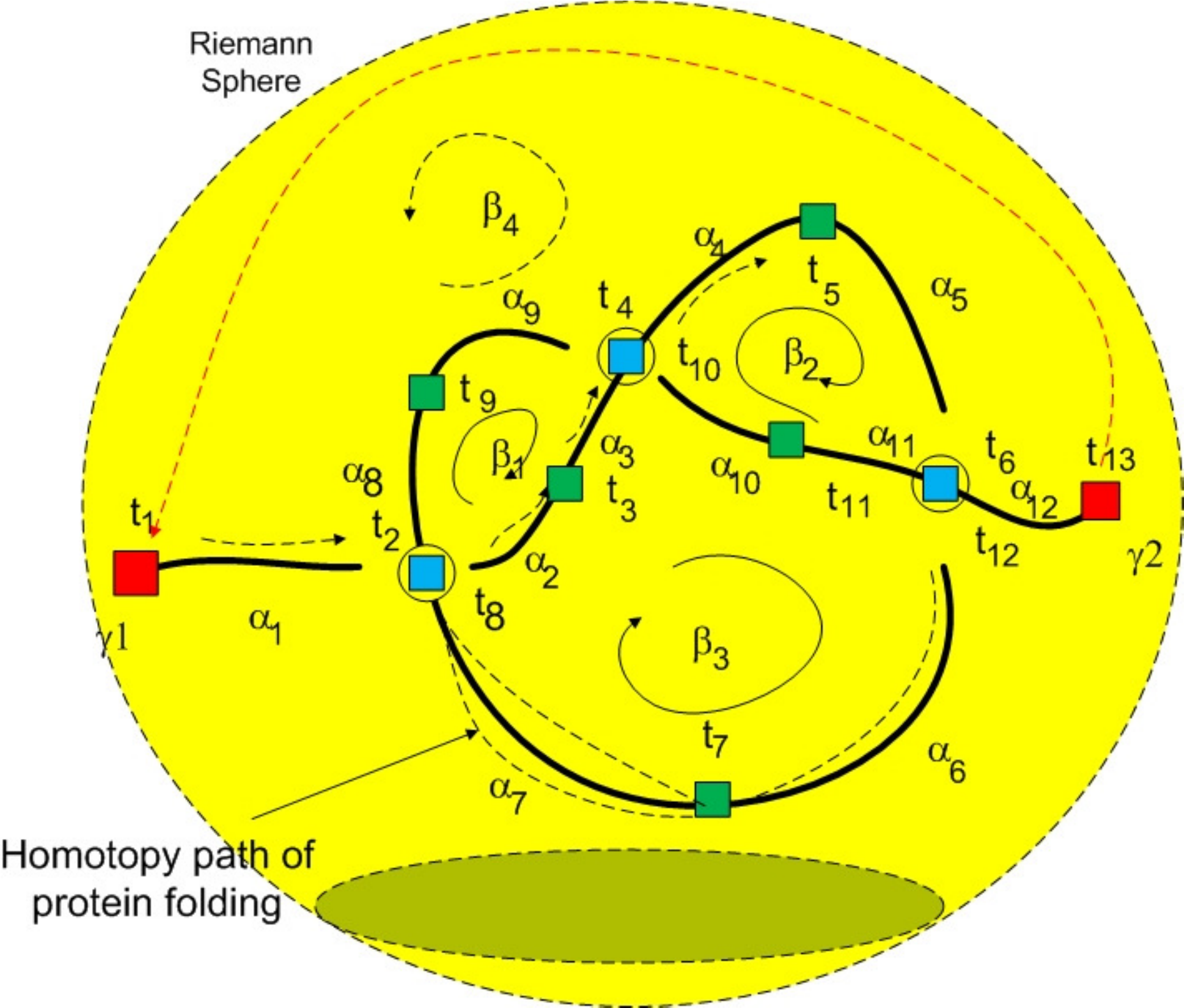,width=8cm}
\epsfig{file=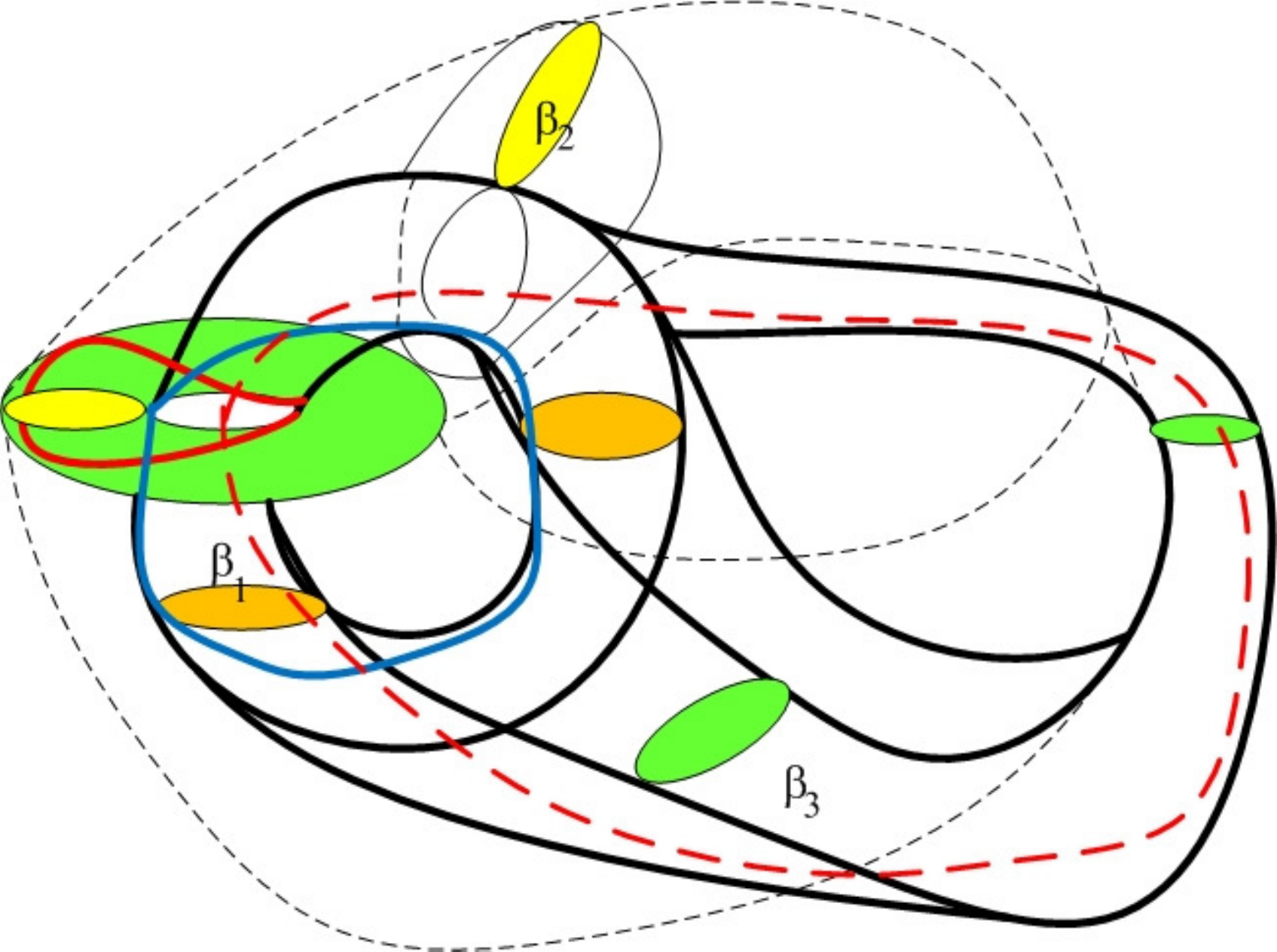,width=8cm}
 \caption{ The picture  on the left is an example of synthetic  time series data of knots and links in proteins  embedded in a  Riemann sphere of 3 dimensions. We use this  example with 13 synthesis amino-acids connected to open knots and links of proteins  in extended closed knotted with figure 8 $4_{1}$ hyperbolic knot. The picture represents  our new algorithm to show how to compute the modified Khovanov cohomology for biology.
 The right picture is a transformation of left picture into the loop space of time series data.
 The center of 4-torus are $\beta$ co-cycle of obstruction components in time series of knotted protein.
 We can connect all obstruction components with a 4-loop of knots and links  labelled  with red dots and blue line.
It is a pictorial representation of  modified Khovanov cohomolgy for biology as the loop space of time series data.  \label{knot_protein}}
\end{figure}
Let $L$ be a superspace of link operator in time series of knots and links in proteins   
induced by the adaptive behavior of evolutional factor from mutation of crossing in DNA while  
the recombination process of  replication cycle is going on. 
Let topology of cell be a hyperbolic knot $\kappa =4_{1}$.
Let $P$ be unknots and links in proteins  with principal fiber $P_{\mathcal{K}}$. The length of amino-acid
 is assigned to $n$ with homotopy path parameterized by the  time series as short exact sequence.
It is  $ t_{1} \stackrel{\alpha_{1}} {\rightarrow}t_{2} \stackrel{\alpha_{2}} {\rightarrow} t_{3}
 \stackrel{\alpha_{3}} {\rightarrow} 
t_{4} \stackrel{\alpha_{4}} {\rightarrow} \cdots  \stackrel{\alpha_{12}} {\rightarrow}  t_{t_{13}} $.
Let $[s_{1}],[s_{2}], \cdots [s_{8}] $ be 8 states of spinor field in time series of genetic code.
We use the fact that figure 8 knots generate an   hyperbolic octahedral volume wihere $[s_{i}],i=1,2,\cdots 8$ are 
in the corner of lattice of the octahedron for codon.
In Axiom 7th of quantum field biology, the path $\alpha_{i},i=1,2,\cdots 12$ in figure 8 knot 
 belong to the zeros cohomology group $H^{0}(End(P_{\mathcal{K}})\otimes L)$.  

This path induces a homotopy path in $H^{1}(End(P_{\mathcal{K}}))$ of the axiom by the homotopy equivalent under the 
influence factor of evolution due to the  tensor product in superspace $<L_{0}>,<L_{+}>,<L_{-}>\in L$.
It is worth noticing  that we have the homotopy path in loop over $4_{1}$ with 4 equivalent classes of loop space in time series data
 of knotted protein.
\begin{equation}
  [\beta_{1}],[\beta_{2}],[\beta_{3}],[\beta_{4}]  \in H^{1}( End(P_{\mathcal{K}}))
\end{equation}
 where the path in loop is a generator of knotted group for secondary protein structure with
\begin{equation}
\beta_{1}: 0\rightarrow \alpha_{2}\rightarrow \alpha_{3}\rightarrow \alpha_{9}^{-1}\rightarrow \alpha_{8}^{-1}\rightarrow 0 
\end{equation}
\begin{equation}
\beta_{2}: 0\rightarrow \alpha_{4}\rightarrow \alpha_{5}\rightarrow \alpha_{11}^{-1}\rightarrow \alpha_{10}^{-1}\rightarrow 0 
\end{equation}
\begin{equation}
\beta_{3}: 0\rightarrow \alpha_{2}\rightarrow \alpha_{3}\rightarrow \alpha_{10}\rightarrow
 \alpha_{11}  \rightarrow \alpha_{6}\rightarrow \alpha_{7}       \rightarrow 0 
\end{equation}
\begin{equation}
\beta_{4}: 0\rightarrow \alpha_{1}\rightarrow \alpha_{8}^{-1}\rightarrow \alpha_{9}^{-1}
\rightarrow
 \alpha_{4}^{-1}  \rightarrow \alpha_{5}^{-1}\rightarrow \alpha_{12}       \rightarrow 0
\end{equation}

There exists  2 trivial elements in open state for 2 ends terminal of protein secondary structure folding.
 It is  an unknotted state  of protein generates from the kernel map in higher dimensional knotted state of protein loop superspace.
\begin{equation}
   [\gamma (t_{1})],[\gamma (t_{13})]  \in H^{2}( End(P_{\mathcal{K}})\,.
\end{equation}

From this example,  we have a time series of link operator parametrized by the underlying ghost  and anti-ghost fields of knots and links in proteins 
 with modified Wilson loop over the exact sequence of link operator $<L>(t)$

\begin{equation}
<\Phi^{\pm}(A_{\mu})>_{k}(t): 0\rightarrow  [L_{0}]^{[s_{i}(t_{1})]_{A_{\mu}}}W_{4_{1},\beta_{4}}(A_{\mu=3};t_{1})
\rightarrow [L_{-}]^{[s_{i}(t)]_{A_{\mu}}}W_{4_{1},\beta_{1},\beta_{4}}(A_{\mu=3};t_{2})  
\rightarrow      \nonumber
\end{equation}

\begin{equation}
 \rightarrow   [L_{0}]^{[s_{i}(t_{3})]_{A_{\mu}}}W_{4_{1},\beta_{1}}(A_{\mu=3};t_{3})
 \rightarrow    [L_{+}]^{[s_{i}(t_{4})]_{A_{\mu}}}W_{4_{1},\beta_{1},\beta_{2}}(A_{\mu=3};t_{4})  
\rightarrow  [L_{0}]^{[s_{i}(t_{5})]_{A_{\mu}}}W_{4_{1},\beta_{2}}(A_{\mu=3};t_{5})
    \rightarrow \nonumber
\end{equation}

\begin{equation}
 \rightarrow   [L_{+}]^{[s_{i}(t_{6})]_{A_{\mu}}}W_{4_{1},\beta_{2}, \beta_{3}}(A_{\mu=3};t_{6})
 \rightarrow    [L_{0}]^{[s_{i}(t_{7})]_{A_{\mu}}}W_{4_{1},\beta_{3}}(A_{\mu=3};t_{7})  
 \rightarrow    [L_{-}]^{[s_{i}(t_{8})]_{A_{\mu}}}W_{4_{1},\beta_{3},\beta_{1}}(A_{\mu=3};t_{8})         
 \rightarrow \nonumber
\end{equation}

\begin{equation}
 \rightarrow   [L_{0}]^{[s_{i}(t_{9})]_{A_{\mu}}}W_{4_{1},\beta_{1}}(A_{\mu=3};t_{9})
 \rightarrow    [L_{-}]^{[s_{i}(t_{10})]_{A_{\mu}}}W_{4_{1},\beta_{1},\beta_{2}}(A_{\mu=3};t_{10})  
\rightarrow  [L_{0}]^{[s_{i}(t_{11})]_{A_{\mu}}}W_{4_{1},\beta_{2}}(A_{\mu=3};t_{11})
    \rightarrow \nonumber
\end{equation}

\begin{equation}
 \rightarrow   [L_{+}]^{[s_{i}(t_{12})]_{A_{\mu}}}W_{4_{1},\beta_{2},\beta_{4}}(A_{\mu=3};t_{12})
 \rightarrow    [L_{+}]^{[s_{i}(t_{4})]_{A_{\mu}}}W_{4_{1},\beta_{1},\beta_{2}}(A_{\mu=3};t_{13})  
    \rightarrow 0.
\end{equation}
This  short sequence can be written  as a linear combination  of bases,
\begin{equation}
<\Phi_{i}^{\pm}([L]^{[s_{i}]}_{A_{\mu}})>_{k}(t)=\sum_{j=1}^{12}\sum_{i=1}^{4}\lambda_{ij} [L_{\pm}]^{[s_{i}]}_{A_{\mu}}
\Pi_{\mu }W_{\kappa,\beta_{i}}(A_{\mu})(t_{j}).
\end{equation}

\begin{equation}
  H^{\cdot}([L_{\pm}]_{A\mu}^{[s_{i}]};P_{\mathcal{K}}): 0 
 \rightarrow \cdots \rightarrow  H^{3}([L_{\pm}]_{A\mu}^{[s_{i}]};P_{\mathcal{K}}) 
 \stackrel{\phi^{d_{1}^{\ast}}_{\pm} }{  \rightarrow }
 H^{3}([L_{\pm}]_{A\mu}^{[s_{i}]};P_{\mathcal{K}})\stackrel{\phi^{d_{2}^{\ast}}_{\pm} }{  \rightarrow } 
 H^{3}([L_{\pm}]_{A\mu}^{[s_{i}]};P_{\mathcal{K}})
\stackrel{\phi^{d_{3}}_{\pm} }{  \rightarrow } \nonumber
\end{equation}
\begin{equation}
  \rightarrow H^{2}([L_{\pm}]_{A\mu}^{[s_{i}]};P_{\mathcal{K}})
  \stackrel{\phi^{d_{1}}_{\pm}}{  \rightarrow }
 H^{2}([L_{\pm}]_{A\mu}^{[s_{i}]};P_{\mathcal{K}})\stackrel{\phi^{d_{2}}_{\pm}}{  \rightarrow } 
 H^{2}([L_{\pm}]_{A\mu}^{[s_{i}]};P_{\mathcal{K}})
\stackrel{\phi^{d_{3}}_{\pm} }{  \rightarrow } \cdots \rightarrow 0
\end{equation}

The short exact sequence of time series of knotted protein induces an infinite sequence of 
modified Khovanov cohomology  with the 3d-cohomology group associated with the volume form
 of  the Chern-Simons current in the protein  docking and undocking states.

 \section{Computation of Chern-Simons current in viral capsid glycoproteins}
 
\begin{figure}[!t]
\centering
 \epsfig{file=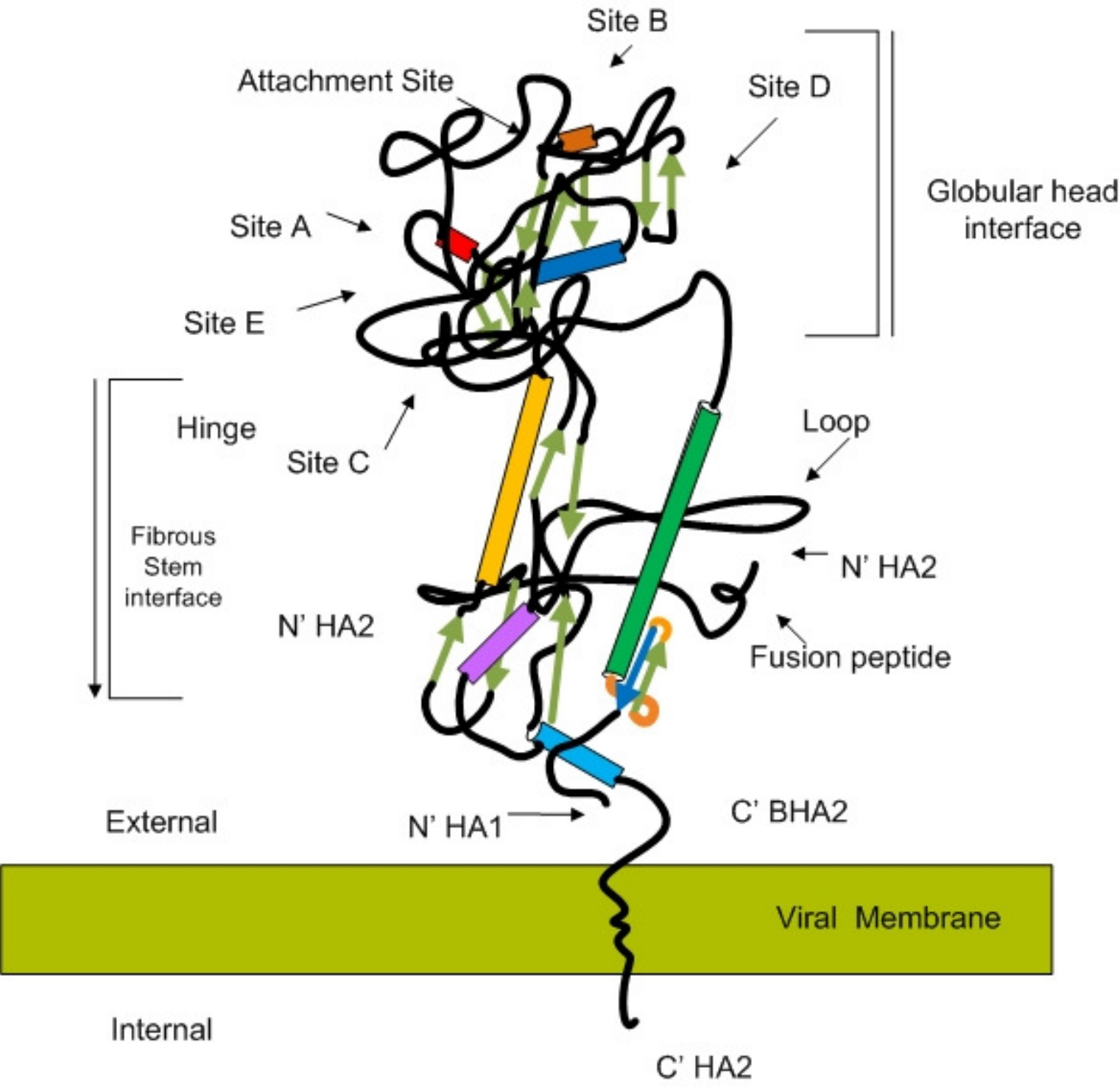,width=8cm}
\epsfig{file=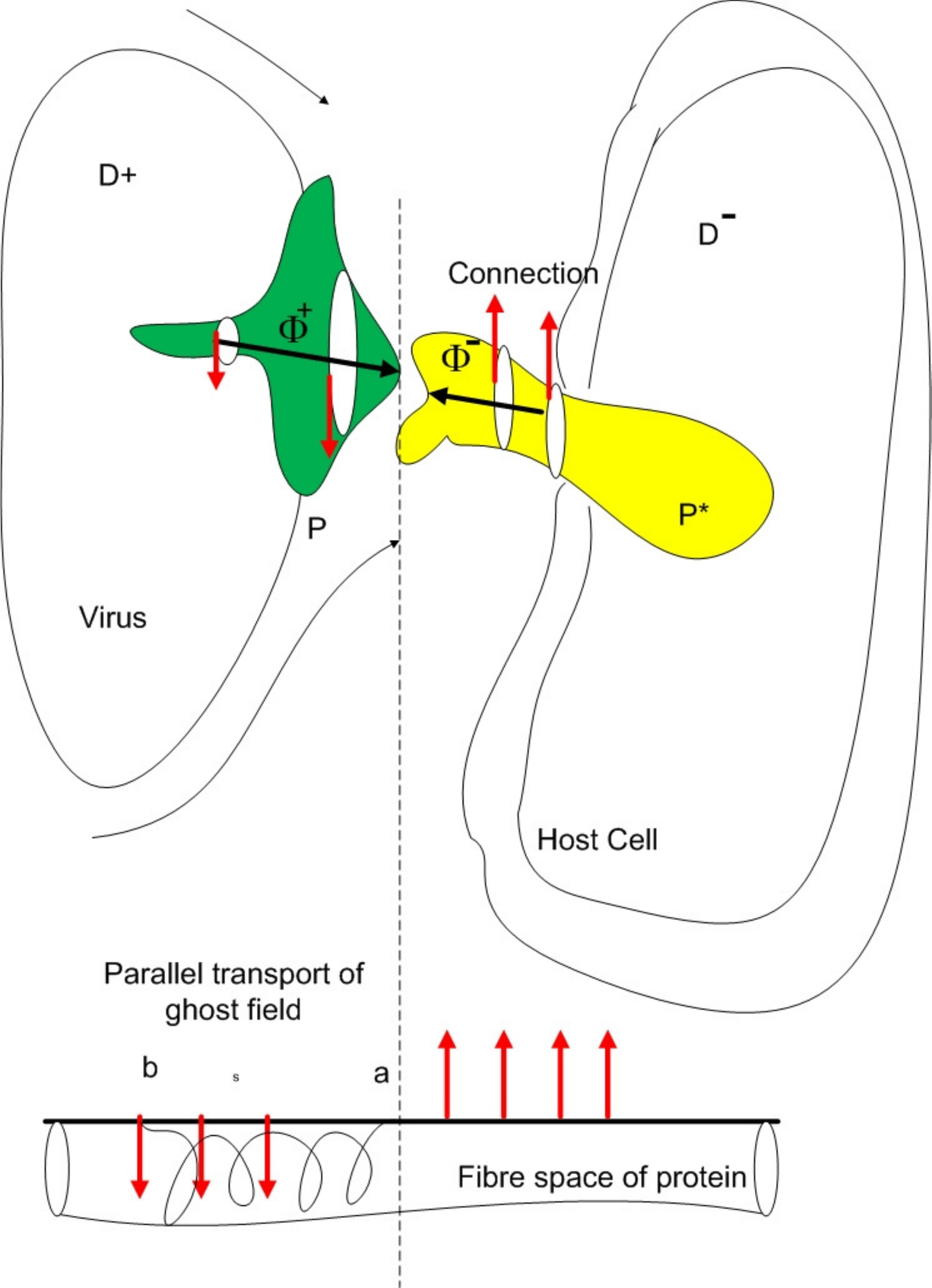,width=8cm}
\caption{ The picture  on the left shows an infuenza viral glycoprotein
 model from x-ray crytallography. The glycoprotein 
 is drawing from experiments.
  An icosahedral group $E_{8}\times E_{8}$ model of ghost field and antighost field for this component is shown in Table \ref{glycoprotein}. 
The axiom of quantum field for biology is satisfied.
 We use this glycoprotein  to  calculate, with a similar algorithm,  other artificial  icosahedral 
   viral capsid proteins, i.e.
 VP1,VP2,VP3,  which are particles in hidden 14 dimensional model of super-statistics in BV-cohomology theory.
 A glycoprotein of virus with 14 sides with recognized name from x-ray crystallography is reported. 
In the right picture, we show the docking system of $D^{+}_{\Phi^{+}(x_{t};A_{\mu})}|\Phi^{-}(y_{t};A_{\mu})$.
 The viral glycoprotein is denoted by the ghost field $\Phi^{+}(x_{t};A_{\mu})$ with the attach docking operator derived from the  behavior of gene expression in 
virus, denoted as $D^{+}$ operator. We move $\Phi_{+}$ to dock with static place of $\Phi^{-}$ in host cell membrane. This situation is 
specified with the notation $D^{+}_{\Phi^{+}}(-)$.
 We specify also the parallel transport of gauge field of docking protein.
In this model,  we have defined the  docking operator by using the Grothendieck coordinates with continuous co-adjoint functor with their 
limit   $\lim_{b\rightarrow a}\frac{\Gamma_{a}^{b\mu}(\gamma)\Phi^{+}_{b}-\Phi_{a}^{-}}{t_{b}-t_{a}}=0,$
 as modified covariant derivative of proteins docking system in principal fiber bundle.
 This equation of docking is a parts of  the modified
 Seiberg-Witten equation for biology.}
\label{fig8}
\end{figure}

In this section, we use the  notation in Table \ref{glycoprotein} for 
integrate and differentiate  over cohomology of loop space in homotopy path along the 
hidden 14 dimensional model. The ghost fields of 14 glycoprotein sides (see Fig. \ref{fig8}
 for details) are adopted  in  super-integrals to represent the  viral attach cell.
 The   14 dimensional loop space of  cell is derived from docking and undocking states of ghost  and anti-ghost fields that model  the protein-protein interactions. 
 This is just a  simple example of how to use the Grothendieck topology to coordinate the protein docking equation in a unified theory $E_{8}\times E_{8}$ model
 for living organism. We use  synthetic ghost fields of protein secondary structure from influenza viral glycoprotein and antibody light  and heavy chains of human cell as example for 
computing the Khovanov cohomology for time series of knots and links in proteins .

Let $\mathcal{A}$ be a superspace for artificial viral particle in $S^{-k,k=14}$ attached with artificial cell structure of mathematical structure of antibody in $S^{k,k=3}$ model.

\begin{table}[!t]
\renewcommand{\arraystretch}{1.3}
\caption{  The table below shows  artificial viral glycoprotein for synthetic signal of time series in knots and links in proteins  according to 
the definition of first 14-ghost fields 
in $E_{8}\times E_{8}$ model of artificial capsid VP1, VP2, VP3 viral supermanifold structure of secondary protein.
\label{glycoprotein}}
\centering
 \begin{tabular}{|c|c|c|c|} \hline \hline
 Ghost field $\Phi_{i}$  &   site name &   state space variable&type of fiber    \\ \hline \hline
  $\Phi_{1}^{+}$& $ C' HA2  $& $x_{t}$ &$A_{\mu=3},$ loop    \\ \hline

$\Phi_{2}^{+}$& $N' HA1  $& $x_{t}$  &$A_{\mu=3},$ loop    \\ \hline

$\Phi_{3}^{+}$& $CB HA2  $& $x_{t}$ &$A_{\mu=3}, $ loop     \\ \hline

$\Phi_{4}^{+}$& $site C  $& $x_{t}$ &$A_{\mu=3},$ loop     \\ \hline

$\Phi_{5}^{+}$& $Hinge  $& $x_{t}$ &$A_{\mu=3},$ loop     \\ \hline

$\Phi_{6}^{+}$& $Site E  $& $x_{t}$ &$A_{\mu=3},$ loop     \\ \hline

$\Phi_{7}^{+}$& Attachment Site  & $x_{t}$   &$A_{\mu=3},$ loop   \\ \hline

$\Phi_{8}^{+}$& $site B  $& $x_{t}$  &$A_{\mu=3},$ loop    \\ \hline

$\Phi_{9}^{+}$& $Site D  $& $x_{t}$ &$A_{\mu=1},\beta$ sheet     \\ \hline

$\Phi_{10}^{+}$& $Loop   $& $x_{t}$   &$A_{\mu=3},$  loop   \\ \hline

$\Phi_{11}^{+}$& $N' HA2  $& $x_{t}$ &$A_{\mu=2},\alpha $ helix     \\ \hline

$\Phi_{12}^{+}$& $C'BHA2$(Bromelain cleavage)  & $x_{t}$ &$A_{\mu=3},$ loop     \\ \hline

$\Phi_{13}^{+}$& $Fusion peptide  $& $x_{t}$  &$A_{\mu=3},$ loop    \\ \hline

$\Phi_{14}^{+}$& $C'HA1  $& $x_{t}$ &$A_{\mu=3},$ loop     \\ \hline

$\Phi_{1}^{-}$& Antigent Binding Site& $y_{t}$  &$A^{\mu=3},$ loop    \\ \hline

$\Phi_{2}^{-}$& $ Light Chain  $& $y_{t}$ &$A^{\mu=1},\beta$ sheet     \\ \hline

$\Phi_{3}^{-}$& Heavy Chain& $y_{t}$ &$A^{\mu=1}, \beta$  sheet     \\ \hline
  \hline\hline
	 \end{tabular}
\end{table}

In this example, we have
 14 bases (see Fig. \ref{fig8}) 
for ghost fields 11 bases for anti ghost fields 
(for detail of definition of ordering of basis, see table \ref{glycoprotein})
\begin{equation}
<\Phi_{1}^{+}(A_{\mu})> = \frac{1}{Z_{k=1}}\int_{ H^{1}([L_{\pm}]_{A\mu}^{[s_{i}]};
P_{\mathcal{K}})      }\Pi_{\mu} W_{\beta_{1},\kappa}( A_{\mu})e^{-j2\pi\beta_{1}(C' HA2 )}e^{-iS_{SC}} d\beta_{1} ,
D^{-}_{\Phi^{-}(A_{\mu})}|\Phi_{1}^{+}(A_{\mu})> =0,\nonumber
\end{equation}

\begin{equation}
<\Phi_{2}^{+}(A_{\mu})> = \frac{1}{Z_{k=2}}\int_{ H^{1}([L_{\pm}]_{A\mu}^{[s_{i}]};P_{\mathcal{K}})      }
\Pi_{\mu} W_{\beta_{2},\kappa}( A_{\mu})e^{-j2\pi\beta_{2}(N' HA1 )}e^{-iS_{SC}} d\beta_{2},D^{-}_{\Phi^{-}(A_{\mu})}|\Phi_{2}^{+}(A_{\mu})> =0 \nonumber
\end{equation}

\begin{equation}
<\Phi_{3}^{+}(A_{\mu})> = \frac{1}{Z_{k=3}}\int_{ H^{1}([L_{\pm}]_{A\mu}^{[s_{i}]};P_{\mathcal{K}}) }
\Pi_{\mu} W_{\beta_{3},\kappa}( A_{\mu})e^{-j2\pi\beta_{3}(CB HA2 )}e^{-iS_{SC}} d\beta_{3},D^{-}_{\Phi^{-}(A_{\mu})}|\Phi_{3}^{+}(A_{\mu})> =0 \nonumber
  \end{equation}

\begin{equation}
<\Phi_{4}^{+}(A_{\mu})> = \frac{1}{Z_{k=4}}\int_{ H^{1}([L_{\pm}]_{A\mu}^{[s_{i}]};P_{\mathcal{K}}) }
\Pi_{\mu} W_{\beta_{4},\kappa}( A_{\mu})e^{-j2\pi\beta_{4}(site C )}e^{-iS_{SC}} d\beta_{4},D^{-}_{\Phi^{-}(A_{\mu})}|\Phi_{4}^{+}(A_{\mu})> =0 \nonumber
 \end{equation}

\begin{equation}
<\Phi_{5}^{+}(A_{\mu})> = \frac{1}{Z_{k=5}}\int_{ H^{1}([L_{\pm}]_{A\mu}^{[s_{i}]};P_{\mathcal{K}}) }
\Pi_{\mu} W_{\beta_{5},\kappa}( A_{\mu})e^{-j2\pi\beta_{5}(Hinge )}e^{-iS_{SC}} d\beta_{5},D^{-}_{\Phi^{-}(A_{\mu})}|\Phi_{5}^{+}(A_{\mu})> =0 \nonumber
\end{equation}

\begin{equation}
<\Phi_{6}^{+}(A_{\mu})> = \frac{1}{Z_{k=6}}\int_{ H^{1}([L_{\pm}]_{A\mu}^{[s_{i}]};P_{\mathcal{K}}) }
\Pi_{\mu} W_{\beta_{6},\kappa}( A_{\mu})e^{-j2\pi\beta_{6}(Site E )}e^{-iS_{SC}} d\beta_{6},D^{-}_{\Phi^{-}(A_{\mu})}|\Phi_{6}^{+}(A_{\mu})> =0 \nonumber
 \end{equation}

\begin{equation}
<\Phi_{7}^{+}(A_{\mu})> = \frac{1}{Z_{k=7}}\int_{ H^{1}([L_{\pm}]_{A\mu}^{[s_{i}]};P_{\mathcal{K}}) }
\Pi_{\mu} W_{\beta_{7},\kappa}( A_{\mu})e^{-j2\pi\beta_{7}(Attach Site )}e^{-iS_{SC}} d\beta_{7},D^{-}_{\Phi^{-}(A_{\mu})}|\Phi_{7}^{+}(A_{\mu})> =0 \nonumber
 \end{equation}

\begin{equation}
<\Phi_{8}^{+}(A_{\mu})> = \frac{1}{Z_{k=8}}\int_{ H^{1}([L_{\pm}]_{A\mu}^{[s_{i}]};P_{\mathcal{K}}) }
\Pi_{\mu} W_{\beta_{8},\kappa}( A_{\mu})e^{-j2\pi\beta_{8}(Site B )}e^{-iS_{SC}} d\beta_{8},D^{-}_{\Phi^{-}(A_{\mu})}|\Phi_{8}^{+}(A_{\mu})> =0 \nonumber
 \end{equation}

\begin{equation}
<\Phi_{9}^{+}(A_{\mu})> = \frac{1}{Z_{k=9}}\int_{ H^{1}([L_{\pm}]_{A\mu}^{[s_{i}]};P_{\mathcal{K}}) }
\Pi_{\mu} W_{\beta_{9},\kappa}( A_{\mu})e^{-j2\pi\beta_{9}(Site D )}e^{-iS_{SC}} d\beta_{9},D^{-}_{\Phi^{-}(A_{\mu})}|\Phi_{9}^{+}(A_{\mu})> =0 \nonumber
 \end{equation}

\begin{equation}
<\Phi_{10}^{+}(A_{\mu})> = \frac{1}{Z_{k=10}}\int_{ H^{1}([L_{\pm}]_{A\mu}^{[s_{i}]};P_{\mathcal{K}}) }
\Pi_{\mu} W_{\beta_{10},\kappa}( A_{\mu})e^{-j2\pi\beta_{10}(Loop )}e^{-iS_{SC}} d\beta_{10},D^{-}_{\Phi^{-}(A_{\mu})}|\Phi_{10}^{+}(A_{\mu})> =0 \nonumber
\end{equation}

\begin{equation}
<\Phi_{11}^{+}(A_{\mu})> = \frac{1}{Z_{k=11}}\int_{ H^{1}([L_{\pm}]_{A\mu}^{[s_{i}]};P_{\mathcal{K}}) }
\Pi_{\mu} W_{\beta_{11},\kappa}( A_{\mu})e^{-j2\pi\beta_{11}(N' HA2  )}e^{-iS_{SC}} d\beta_{11},D^{-}_{\Phi^{-}(A_{\mu})}|\Phi_{11}^{+}(A_{\mu})> =0 \nonumber
 \end{equation}

\begin{equation}
<\Phi_{12}^{+}(A_{\mu})> = \frac{1}{Z_{k=12}}\int_{ H^{1}([L_{\pm}]_{A\mu}^{[s_{i}]};P_{\mathcal{K}}) }
\Pi_{\mu} W_{\beta_{12},\kappa}( A_{\mu})e^{-j2\pi\beta_{12}(C'BHA2  )}e^{-iS_{SC}} d\beta_{12},D^{-}_{\Phi^{-}(A_{\mu})}|\Phi_{12}^{+}(A_{\mu})> =0 \nonumber
  \end{equation}

\begin{equation}
<\Phi_{13}^{+}(A_{\mu})> = \frac{1}{Z_{k=13}}\int_{ H^{1}([L_{\pm}]_{A\mu}^{[s_{i}]};P_{\mathcal{K}}) }
\Pi_{\mu} W_{\beta_{13},\kappa}( A_{\mu})e^{-j2\pi\beta_{13}(Fusion peptide )}e^{-iS_{SC}} d\beta_{13},D^{-}_{\Phi^{-}(A_{\mu})}|\Phi_{13}^{+}(A_{\mu})> =0 \nonumber
 \end{equation}

\begin{equation}
<\Phi_{14}^{+}(A_{\mu})> = \frac{1}{Z_{k=14}}\int_{ H^{1}([L_{\pm}]_{A\mu}^{[s_{i}]};P_{\mathcal{K}}) }
\Pi_{\mu} W_{\beta_{14},\kappa}( A_{\mu})e^{-j2\pi\beta_{14}(C'HA1 )}e^{-iS_{SC}} d\beta_{14},D^{-}_{\Phi^{-}(A_{\mu})}|\Phi_{14}^{+}(A_{\mu})> =0 \nonumber
\end{equation}
Considering the expectation ghost  field  $ H^{3}(Y_{t})/H^{14}(X_{t})\simeq  H^{-11}(Y_{t}/X_{t}) \ni \epsilon_{t}^{\ast}$ with superderivative with respect to anti-ghost field in cell  defined by
\begin{equation}
<\Phi_{1}^{-}(A_{\mu})>=\frac{d}{dS^{1}} g^{11}\Phi_{1}^{+}(A_{\mu})=\frac{g^{11}de^{2\pi i\alpha_{2}(Binding Site)}}{d\alpha_{1}}
,D^{+}_{\Phi^{+}(A_{\mu})}|\Phi_{1}^{-}(A_{\mu})> =0 \nonumber
\end{equation}

\begin{equation}
    <\Phi_{2}^{-}(A_{\mu})>    =\frac{d}{dS^{1}} g^{22}\Phi_{2}^{+}(A_{\mu})=\frac{g^{22}de^{2\pi i\alpha_{2}(Light Chain)}}
{d\alpha_{2}},D^{+}_{\Phi^{+}(A_{\mu})}|\Phi_{2}^{-}(A_{\mu})> =0 \nonumber
\end{equation}

\begin{equation}
 <\Phi_{3}^{-}(A_{\mu})>=\frac{d}{dS^{1}}g^{33} \Phi_{3}^{+}(A_{\mu})=\frac{g^{33}de^{2\pi i\alpha_{3}(Heavy  Chain)}}{d\alpha_{3}},
D^{+}_{\Phi^{+}(A_{\mu})}|\Phi_{3}^{-}(A_{\mu})> =0 
\end{equation}
 we have
\begin{equation}
\frac{\alpha_{t}([y_{t}])}{\beta_{t}([x_{t}])}\simeq [{\epsilon_{t}}^{\ast}] 
\in H^{-11}(\mathcal{O}_{Y_{t}/X_{t}};P_{\mathcal{K}}). 
\end{equation}
If we write the Dirac operator for docking system  denoted by $D=: D^{-}_{\Phi^{-}(A_{\mu})}$
 and $d:=D^{+}_{\Phi^{+}(A_{\mu})}$,
when docking state is in equilibrium, all curvatures do not change and equal to each other from duality sites.  So we have

\begin{equation}
 Ad_{\Phi_{-} (A_{\mu})   } \Phi_{+} (A_{\mu}):=  \bigtriangledown_{\Phi^{-}(A_{\mu})} \Phi^{+}(A_{\mu}):=\{ ddd\Phi_{1}^{-}(y_{t})\Phi_{2}^{-}(y_{t})\Phi_{3}^{-}(y_{t}),
\underbrace{DDDDDDDDDDDDDD}_{14}\Phi_{1}^{+}(x_{t})\cdots \Phi_{14}^{+}(x_{t})\}
\end{equation}
If docking is in equilibrium and viral can penetrate to host cell, that means we have just one part of mirror symmetry 
of Dirac operator to be zero according to our new definition. Therefore, we have
 \begin{equation}
 D\Phi_{1}^{+}D\Phi_{2}^{+}D\Phi_{3}^{+}D\Phi_{4}^{+}D\Phi_{5}^{+}\cdots D\Phi_{14}^{+}=0.
\end{equation}
If we defined our adjoint operator with extra properties of algebraic operator by

 \begin{equation}
 DDDD\cdots DDDD\Phi_{1}^{+}\Phi_{2}^{+}\Phi_{3}^{+}\Phi_{4}^{+}
   \cdots \Phi_{14}^{+}=D\Phi_{1}^{+} DDDD\cdots DDD\Phi_{2}^{+}\cdots\Phi_{13}^{+}  \Phi_{14}^{+}  
  \Phi_{14}\nonumber
\end{equation}
\begin{equation}
=\cdots =D\Phi_{1}^{+} D\Phi_{2}^{+} D\Phi_{3}^{+} \cdots D\Phi_{12}^{+} D \Phi_{13}^{+}  D\Phi_{14}^{+} ,
\end{equation}
therefore we have a solution for each gene in each ghost field given by the system of equations
\begin{equation}
D\Phi_{1}^{+}=0, D\Phi_{2}^{+}=0 ,D\Phi_{3}^{+}=0, \cdots, D\Phi_{12}^{+}=0, D \Phi_{13}^{+}=0 , D\Phi_{14}^{+}=0.
\end{equation}
We can factorize the gene in time series of knotted glycoprotein with obstruction components for each gene. It can be solved by using  the modified Wilson loop for gene expression.

In the model of protein complex surface, the   differential 2 forms induced by the equilibrium $d^{2}=0$ over 8 states and hidden states  generate the codon with 64 bases in exact sequence. Then we use the triplet state over the layer of protein $P$ to generate a Chern-Simons 3 form over the canonical form of exact sequence of triplet state $P$ with its dual $P^{\ast}.$ The computation is as follows.
For the  equilibrium docking,  we have $H^{-k}( \mathcal{O}_{A},s)=0$ , for all $k>0$.  This means that  we cannot notice ghost  and anti-ghost fields if the system of docking is in equilibrium. There will exists hidden negative dimension area of trash DNA active only when the state of undocking in protein system appears. The immunosystem will induce feedback loops 
of retrotransposon and transponson by cohomology in negative dimension in feedback path to cohomology in positive dimension to  find a new equilibrium point of docking until it  holds then the
 negative path is zeros again recursively. 
The main expression of gene can be written as a  super-regression $[\alpha] \sim [\beta]\in H^{2}(\mathcal{O}_{\mathcal{A}};\mathcal{O}_{X/Y})$, so we induce an evolutional field in trash area as a hidden field to control the second round of recursive form of induced long infinite exact sequence along superspace of time series data. The  super-regression is  
given  by co-states that couple with  2 feedback loop of geneon and retrotransposon state in genotypes, that is 
\begin{equation}
\varphi_{Y}^{retrotransposon}=[\alpha]  +[\beta] \varphi_{X}^{geneon}+[\epsilon_{t}]_{X/Y} 
\end{equation}

The copy process  of retrotransposon loop space is a characteristic class of cohomology in negative dimensions induced by a  master equation with adjoint functor as cohomology functor .
 The extended Chern-Simons current for trash DNA is possible by cosaxter number $h$ in the 
Laurent series of poles of state $[s_{i}]^{\ast},i=1,2,\cdots, 8$  with icosahedral group $E_{8}$ we get $h=30.$ Therefore 
we can extend this approach to represent genetic code as Laurent polynomials in the variable $q$
 with integer coefficients, that is for trash area with knot $K$ over sheaf cohomology of DNA.
 
%\begin{table}[!t]
% \renewcommand{\arraystretch}{1.3}
% \caption{  The table below  show amount of nucleotide in 5 Rhinovirus with icosahedral nucleocapsid structural glycoprotein $VP1,VP2
%,VP3,VP4$. The capsid protein VP4 is come from $1A$  gene.The capsid protein VP2 is come from $1B$  gene.The capsid protein VP3 is come from $1C$  gene.The capsid protein VP1 is come from $1D$  gene.   } 
%\label{table_capsid}
%\centering
% \begin{tabular}{||c|c|c|c|c|c|}   \hline\hline
%Protein & Poliovirus& Coxsackievirus & Hepatitis A virus &  Rhinovirus &Foot-and-mouth  virus \\ \hline\hline
%VP1 &302&284 &274 & 290 &212 \\
%VP2 &271 & 261 & 222&  262 & 218\\
%VP3 &238 & 238 & 246 & 236 & 221 \\
%VP4 & 69 & 69 & 23 & 69 & 81 \\
%2A  &149 & 147 & 71 & 145 & 16 \\
%2B & 97 & 99 & 215 & 97& 154\\
%2C & 329 & 329 & 335 & 330& 317\\
%3A & 87 & 89 & 74& 85 & 154 \\
%3B &22 & 22 & 23 & 23 & 23/24 \\
%3C & 182 & 183 & 219 & 182 & 214 \\
%3D &461 & 462 & 489 & 460 & 470\\  \hline\hline
%\end{tabular}
% \end{table}

\begin{table}[!t]
 \renewcommand{\arraystretch}{1.3}
 \caption{  The table  shows  the Chern-Simons current for  genetic code of the  first 20 peptides. We calculate  the average codon of predefined Chern-Simons current for amino-acids considering  real values  of parameter  $k$.} 
\label{table_current}
\centering
  \begin{tabular}{||c|c|c|c|c|c|c|}   \hline\hline
no.&Amino-acid & Abbreviation & Code & Average Chern-Simons Current &state space(k) &Average state for one pixel\\ \hline \hline
1&Alanine   & Ala & A &0.0320&29,30,31,32& 30.5\\
2&Cysteine   & Cys & C & 0.0120&49,50&49.5    \\
3&Aspartic acid   & Asp  & D & 0.0136&41,42  &41.5 \\
4&Glutamatic acid   &Glu & E & 0.0128 &47&47  \\
5&Phenylalanine   & Phe & F &  0.6036 &1,2&1.5 \\
6& Glycine   & Gly & G & 0.0086  &61,62,63,64  &62.5 \\
7&Histidine   & His & H &0.0179  &37,38  &37.5\\
8&Isoleucine   & IIe & I & 0.1066 &9,10,11&10\\
9&Lysine   & Lys & K & 0.0145 &43,44&43.5\\
10&Leucine   & Leu & L & 0.2305 &5,6,7,8&6.5\\
11&Methionine   & Met& M &0.0841 &12&12 \\
12&Asparagine   & Asn & N &0.0154  &41,42&41.5\\
13&Proline   & Pro & P & 0.0367 &21,22,23,24&22.5\\
14&Glutamine   & Gln & Q &0.0166 &39,40&39.5 \\
15&Arginine   & Arg & R &0.0100  &53,54,59,60&56.5\\
16&Serine   & Ser & S &0.0352  &17,18,19,20,57,58&31.5\\
17&Threonine   & Thr & T &0.0292 &25,26,27,28&26.5 \\
18&Valine   & Val & V & 0.2658 &13,14,15,16&14.5\\
19&Tryptophan   & Trp & W & 0.0112&52&52 \\ 
20& Tyrosine   & Tyr & Y &0.0210 &33,34&33.5\\ \hline\hline
\end{tabular}
\end{table}

A hyperbolic manifold is a manifold with a metric of constant negative curvature. A system
 of viral replication cycle in host cell can be visualized as a hyperbolic manifold with 
2 states space variables $(x_{t},y_{t})$ over ghost field $\Phi^{+}(A_{\mu};x_{t})$ 
with co-cycle $\beta$ of viral particle and 
anti-ghost field
  $\Phi^{-}(A_{\mu};y_{t})$ over co-cycle $\alpha$ of host cell over protein docking system on the
surface of host cell. We give an simple hyperbolic equation of viral attach to
host cell with some explanation of transition state between geneon and anti-geneon in docking system 
as evolutional gauge field of genetic code given as the connection $[A_{\mu}]$.

Let trajectory of path in replication cycle of viral components penetrate to host cell and fuse RNA 
with DNA of host cell until go out from host cell.  It can be  parameterized by figure 8 knot $\kappa$
 with hyperbolic manifold in host cell $S^{3}-\kappa$ with finite hyperbolic volume $Vol(\kappa)$.
 These volumes are invariant in host cell and can be, specifically,  thought as invariant volumes of host cell 
receptor knotted protein if we visualized the viral glycoprotein as a trajectory in
 time series of knotted docking protein along $\kappa$.

\begin{Definition}
Let $\mathcal{M}_{X_{t}/Y_{t}}$ be a moduli state space of parasitism state between host cell with viral replication cycle.
It is a configuration space of coupling transition state between geneon and anti-geneon.
Let $S$ be spin state span by 8 bases of spinor field $[s_{i}]$ and gauge field $A_{\mu}$ for behavior of
 underlying genotype in time series data of genetic code.
A coupling parasitism system is the statistical system $(\mathcal{M}_{X_{t}/Y_{t}},S)$ with a 
transition between $[s_{i}]$ to $[s_{j}^{\ast}]$ in evolution dual state
 $[\epsilon_{t}]=<[s_{j}^{\ast}(t)]|[s_{i}(t)]>\in  \mathcal{F}(\mathcal{M}_{X_{t}/Y_{t}},S)$.
 The energy $\mathcal{E}_{k}$ of the system is measured by coupling constant $k\in K$ as
 the transition level
 in partition function of internal state of parasitism replication cycle, that is
\begin{equation}
\mathcal{E}_{k}: \mathcal{F}(\mathcal{M}_{X_{t}/Y_{t}},S)\rightarrow \mathbb{R}
\end{equation}
The partition function of the system is defined as
\begin{equation}
Z_{k}=\sum \mathcal{E}_{k} ([\epsilon_{t}]) \varphi^{geneon}(t)
\end{equation}
where $\varphi^{geneon}(t) :\mathcal{F}(\mathcal{M}_{X_{t}/Y_{t}},S)\rightarrow \mathbb{R} $
is a wave function of geneon state of the transition between ground states $[s_{i}]$
 to excited state $[s_{j}^{\ast}]$ in   genotypes of docking system.
\end{Definition}

\newtheorem{Theorem}{Theorem}
\begin{Theorem}
The state of docking between viral capsid protein and host receptor protein is in 
equilibrium  when  the Seiberg-Witten equation for cell membrane of host cell holds, that is 
\begin{equation}
F^{[A_{\mu}]}=0,D^{-}_{\Phi^{-}}|\Phi^{+}_{k}(A_{\mu};x_{t})>=[s_{k}] |\Phi^{+}_{k}(A_{\mu};x_{t})> =0 ,
 \hspace{0.5cm } D^{+}_{\Phi^{+}}|\Phi^{-}_{k}(A_{\mu};y_{t})>=[s_{k}^{\ast}]|\Phi^{-}_{k}(A_{\mu};y_{t})>=0
\end{equation}
The solution for Hitchin pair of  this equation can be described in hyperbolic equation over 2 variables in moduli state space model
 of parasitism $(x_{t},y_{t})$ with parameter $(\alpha, \beta)$ as a coupling co-cycle.
The transition function is defined by the holomony of   $\Phi_{in}^{\pm}(A_{\mu})$ 
and $\Phi_{out}^{\pm}$ between quantum 
 states of in docking system over the cell membrane
\begin{equation}
Hol_{\mathcal{M}_{X/Y}}(<\delta \Phi^{\pm}(A_{\mu})>_{k}):=\Pi_{\mu}W_{\beta,P_{X},\kappa }(A_{\mu})W_{
\gamma,P_{Y},\kappa}(A_{\mu})=0=<\Phi_{out}(A_{\mu})>-<\Phi_{in}(A_{\mu})>=<\delta \Phi>_{k}\,.
\end{equation}
\end{Theorem} 
{\em Proof:  Let $[A_{\mu}]=\Gamma^{\mu}_{ij}$ be the equivalent class of gauge field gene $[s_{i}]$
 with transition hidden state to $[s_{j}]^{\ast}$. 
From $F^{[A_{\mu}]}=0$,  we have $\partial_{\mu}A_{\nu} - \partial_{\mu}A_{\nu} +A_{\mu}
 \wedge A_{\nu}=0$ or, shortly,  $dA=-A\wedge A$.
Let $x_{t}$ be a state variable of geneon in viral
 capsid protein genotype and $y_{t}$ be state variable of geneon in host receptor protein.
We choose the connection $A_{\mu}$ to represent a gauge field where the  source of field is a spinor field 
$[s_{i}]=\varphi^{geneon}= [e^{i\beta_{i}}]$ located at the center of obstruction component in $S^{1}$. 
The  state spinor variable $x_{t}-[s_{i}]$ is a radius of circle $S^{1}$.
We choose a representation of Wilson loop with connection in principal fiber $P_{X}$ adopting a  M\"obius map from the circle to 
the fiber $F_{[s_{i}]}\in P_{\mathcal{X},[A_{\mu}]}$ defined by
\begin{equation}
W_{\beta , P_{X},\kappa}(A_{\mu})=Tr_{\rho} P e^{\oint A_{\mu}} = 
Tr_{\rho} P e^{ \oint \frac{1}{x_{t}-[s_{i}]}dx_{t}}=e^{ \oint \frac{1}{x_{t}-[s_{i}]}dx_{t}}\,.
\end{equation}
For $\alpha $ co-cycle, we choose, as above,  the connection $A_{\nu}$ to represent a gauge field where the source of field is a spinor field 
$[s_{j}]^{\ast}=\varphi^{anti-geneon}= [e^{-i\alpha_{i}}]$ located at the center
 of $S^{-1}$ and state spinor variable $y_{t}-[s_{j}]^{\ast}$ is a 
tangent to the same unit circle with opposite direction of orientation $S^{1}$. So we have 

\begin{equation}
W_{\alpha,P^{\ast}_{Y}}(A_{\mu},\kappa)=Tr_{\rho} P e^{\oint A_{\nu}} = Tr_{\rho}  P e^{ 
\oint \frac{1}{y_{t}-[s_{j}]^{\ast}}dy_{t}}=e^{ 
\oint \frac{1}{y_{t}-[s_{j}]^{\ast}}dy_{t}}
\end{equation}
and
\begin{equation}
Hol_{\mathcal{M}_{X/Y}}(<\delta\Phi^{\pm}(A_{\mu})>_{k}):=
W_{\beta , P_{X},\kappa}(A_{\mu})W_{\alpha,P^{\ast}_{Y}}(A_{\mu},\kappa)=
e^{ \oint \frac{1}{x_{t}-[s_{i}]}dx_{t}}e^{ 
\oint \frac{1}{y_{t}-[s_{j}]^{\ast}}dy_{t}}\nonumber
\end{equation}
\begin{equation}
=e^{ ln|x_{t}-[s_{i}]|+\epsilon_{x_{t}}} e^{ ln|y_{t}-[s_{j}]^{\ast}|+\epsilon_{y_{t}}}=0.
\end{equation}
Hence, we have
\begin{equation}
|x_{t}-[s_{i}]||y_{t}-[s_{j}]^{\ast}|=0,
\end{equation}
so $x_{t}=\pm [s_{i}],y_{t}=\pm[s_{j}]^{\ast}.$ 
We call 2 hidden states with minus sign from these solutions of docking equation,
 $-[s_{i}]$ a transponson state and $-[s_{j}]^{\ast}$ a retrotransposon state. 

From this solution, if we assign $\pm [s_{i}], \pm[s_{j}]^{\ast} $ a  constant  real value 
by using the projection from spin space to real value and the  transition energy function

\begin{equation}
\varphi^{geneon}(t) :\mathcal{F}(\mathcal{M}_{X_{t}/Y_{t}},S)\rightarrow \mathbb{R},[s_{i}]\mapsto  \varphi^{geneon}([s_{i}];t)
\end{equation}
\begin{equation}
\varphi^{anti-geneon}(t) :\mathcal{F}(\mathcal{M}_{X_{t}/Y_{t}},S)\rightarrow \mathbb{R},[s_{j}]^{\ast}\mapsto  \varphi^{anti-geneon}([s_{i}];t)
\end{equation}
\begin{equation}
\varphi^{transposon}(t) :\mathcal{F}(\mathcal{M}_{X_{t}/Y_{t}},S)\rightarrow \mathbb{R},[-s_{i}]\mapsto  \varphi^{transposon}(-[s_{i}];t)
\end{equation}
\begin{equation}
\varphi^{retrotransposon}(t) :\mathcal{F}(\mathcal{M}_{X_{t}/Y_{t}},S)\rightarrow \mathbb{R},-[s_{j}]^{\ast}\mapsto 
 \varphi^{retrotransposon}(-[s_{j}]^{\ast};t)
\end{equation}
then we can draw solution into $xy$ plane in $\mathbb{R}^{2}$.
The solution will cut $xy$-axis in 4 points.
We use these 4 points as vertices  of 2 hyperboles with the same center point at $(h,k)$. The  
hyperbolar graph  is a simple example of hyperbolic structure of co-state between geneon and 
anti-geneon, that is 
\begin{equation}
\frac{(x_{t}-h)^{2}}{ \varphi^{geneon}([s_{i}];t)}  -  
\frac{(y_{t}-k)^{2}}{  \varphi^{anti-geneon}(   [s_{j}]^{\ast};t )}=1, 
\end{equation}
and hyperbolic structure of co-state between transposon and retrotransposon, that is 
 \begin{equation}
\frac{(y^{\ast}_{t}-h)^{2}}{ \varphi^{transposon}(-[s_{i}];t)} -
  \frac{(x^{\ast}_{t}-k)^{2}}{ \varphi^{retrotransposon}(-[s_{j}]^{\ast};t)}=1. 
\end{equation}
We can write the hyperbolic equation of solution in a general form as
\begin{equation}
Ax_{t}^{2}- By_{t}^{2}+Cx_{t}+Dy_{t}+E=0.
\end{equation}
}
These 5 parameters can be estimated by an  empirical analysis with the Chern-Simons current in genetic code.
The partition function can be used to measured transition states between these 4 states with some energy band gap between
ground state and excited state in the hyperbolic curve derived from the estimation of genetic code in genotype.

If we  $d^{\ast}:=D^{-}_{\Phi^{+}} , D^{\ast}=D^{+}_{\Phi^{-}}$, we  have a mirror symmetry of docking operators for
 transposon and retrotransposon as repeated obstruction solution since the docking system is not in equilibrium,
There exist 2 types of evolutional field as eigenvalues of life energy induced from adaptation of curvature changing  while docking.
The first is the evolutional field induced from the obstruction component of  genetic variation adaptation in genotype of receptor protein in  host cell. The docking operator  
changes direction of spin from adjoint (reversed direction of antiparallel of spin)$-$ to
 co-adjoint direction (parallel direction of spin) $+$ from host cell $y_{t}$ to dock again with viral 
particle $x_{t}$ with ghost field in protein $\Phi^{+}(x_{t})$. This system is  undocking state so virus cannot
  penetrate the cell anymore since host cells can adapt them  and change curvature  in transposon transition state
 with eigenvalue $ \varphi_{transpon}$. In quantum biology,  we use Hermitian notation for excitation states  
$D\mapsto D^{\ast}$.
This eigenvalue brings  the life ground state of ghost field to an  excited state:
\begin{equation}
D^{\ast} |\Phi^{+}(x_{t})>= D^{+}_{\Phi^{-}(y_{t})}|\Phi^{+}(x_{t})> =\varphi_{transpon}|\Phi^{+}(x_{t})>\,. 
\end{equation}
 In this situation, we induce an evolution feedback loop with reversed evolutional field down to 
retrotransposon state. The virus needs to survive and changes its  spin state switching  its direction 
from $+$ to $-$ sign of docking Dirac operator. From the memory of  ghost field 
of underlying  protein state,   DNA methylation or RNA methylation can exist  in 
viral particles to switch some part of their inactive gene to retrotransposon states. We have transition from 
$d\mapsto d^{\ast}$ defined by
\begin{equation}
d^{\ast} |\Phi^{-}(y_{t})>= D^{-}_{\Phi^{+}(x_{t})}|\Phi^{-}(y_{t})> =\varphi_{retrotranspon}
|\Phi^{-}(y_{t})> 
\end{equation}
Let $\varphi_{retrotranspon}=\varphi_{transposon}^{\ast}$ and $ \epsilon_{t}^{2}
=\varphi_{transposon}\varphi_{retrotranspon}$
be an evolution feedback gauge field for docking-undocking system.
We define the curvature of docking  and undocking system as a function of evolution
\begin{equation}
F^{\bigtriangledown}_{\mu\nu}=[\bigtriangledown_{\mu},\bigtriangledown_{\nu} ]=adj\{d^{\ast},D^{\ast}\}=
< \Phi^{-}_{\mu}d^{\ast}|D^{\ast}\Phi^{+}_{\nu}>= \epsilon_{t}^{2}<\Phi^{-}_{\mu}|\Phi^{+}_{\nu}>\,.
\end{equation}
When docking system is in equilibrium, the  host cell will die since viral can penetrate in it  with 
curvature changing  to zero. If host cells want to survive, they need to have an evolutional field to 
change their curvature from zero to eigenvalue of retrotransposon: in this case,  the virus  cannot dock to 
host cell in this feedback loop. In trash DNA,  there are a lot of repeated patterns of retrotransponson inactive 
part.   They can  involve  eigenvalues of  feedback loop of docking system to control immunosystem  
for the defense of virus particle in host cell. This retrotransposon state cannot be reset  
with DNA methylation from mother and father cell. Since the active part is in reversed direction 
of ghost field in genetic code,  they will increase size more and and more from father to son and to 
descendents of all living organisms induced from the feedback loop of docking system in
 evolutional field.
 
\section{Results of Data Analysis }
\begin{figure}[!t]
 \centering
 \epsfig{file=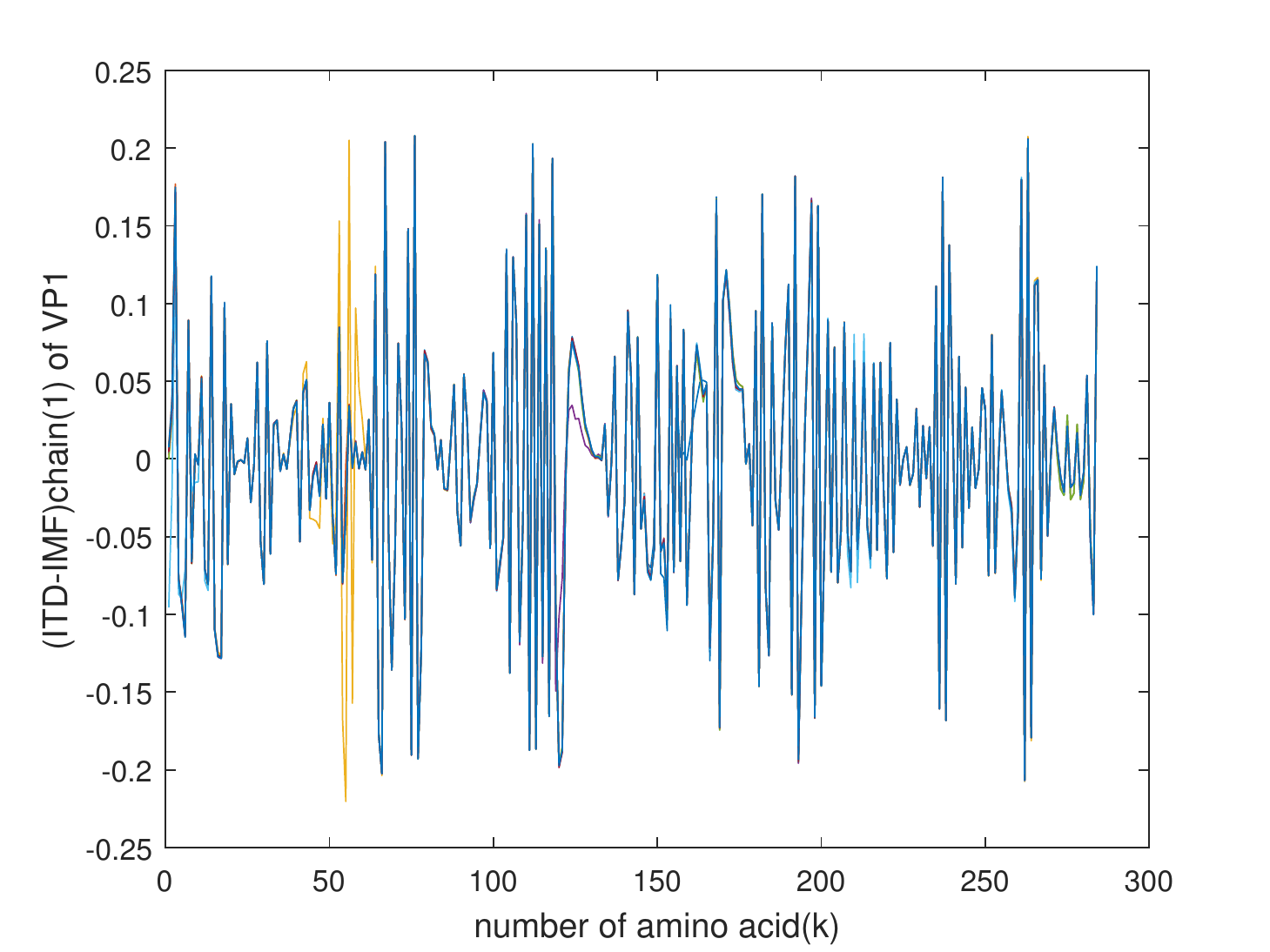,width=8cm}
\epsfig{file=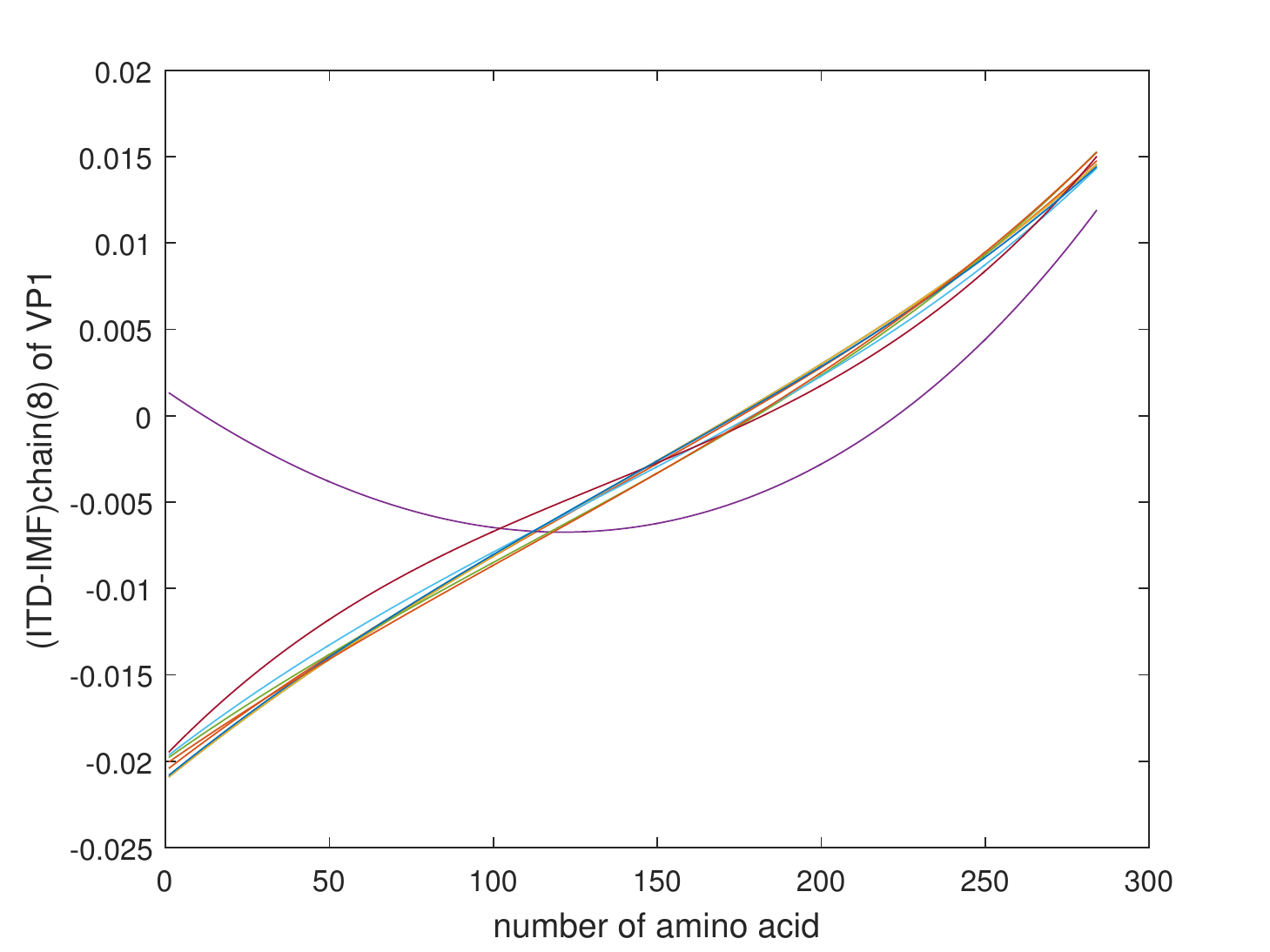,width=8cm}
\caption{On the left, we  plot the time series of $(ITD-IMF)chain(1)$ of gene of $VP1$ in Coxsackievirus  
with 8 samples. On the right, we  plot the time series of $(ITD-IMF)chain(8)$ of gene of $VP1$ in Coxsackievirus  with 8 samples.
\label{result_vp1} }
 \end{figure}

\begin{figure}[!t]
 \centering
  \epsfig{file=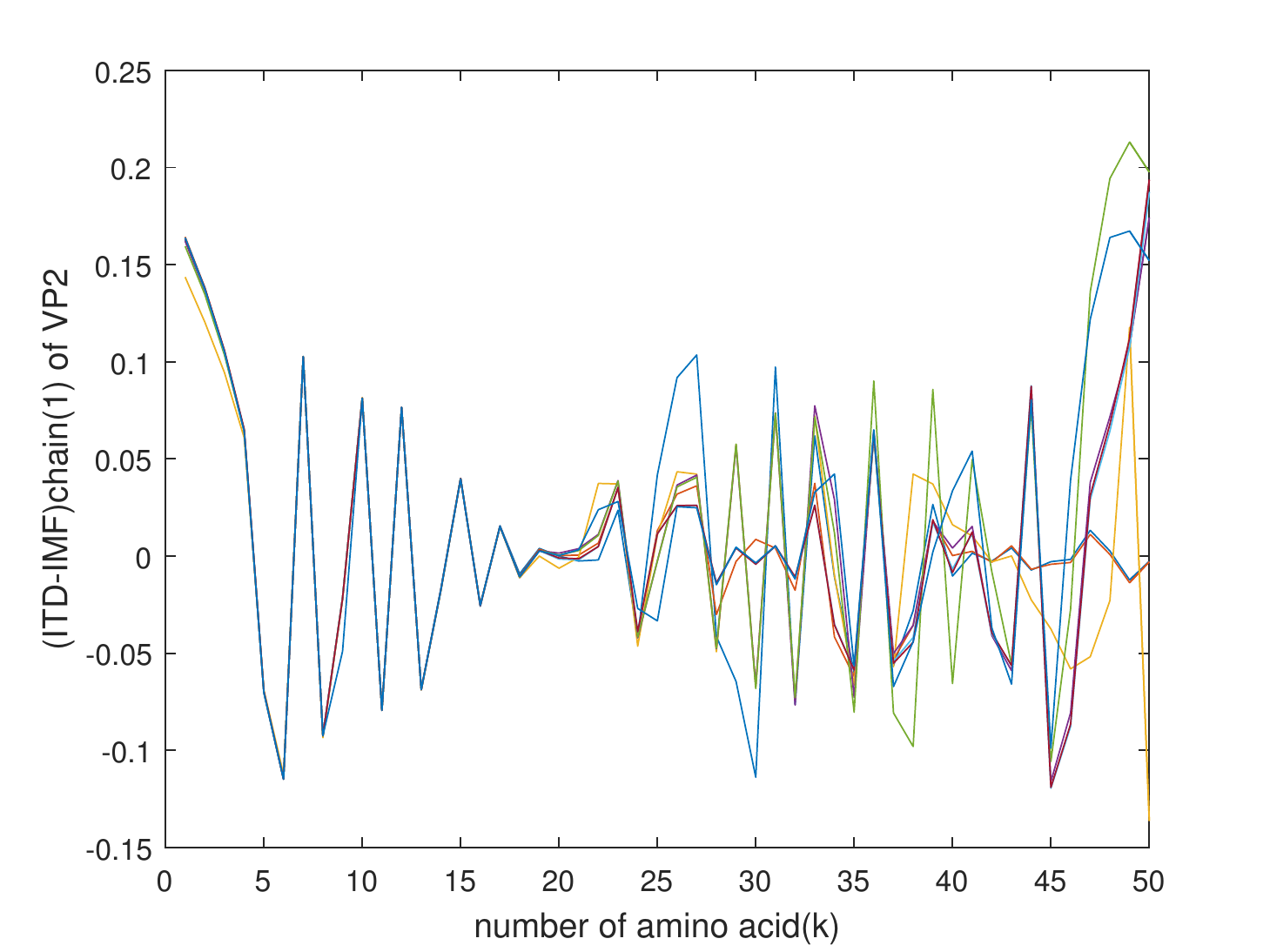,width=8cm}
 \epsfig{file=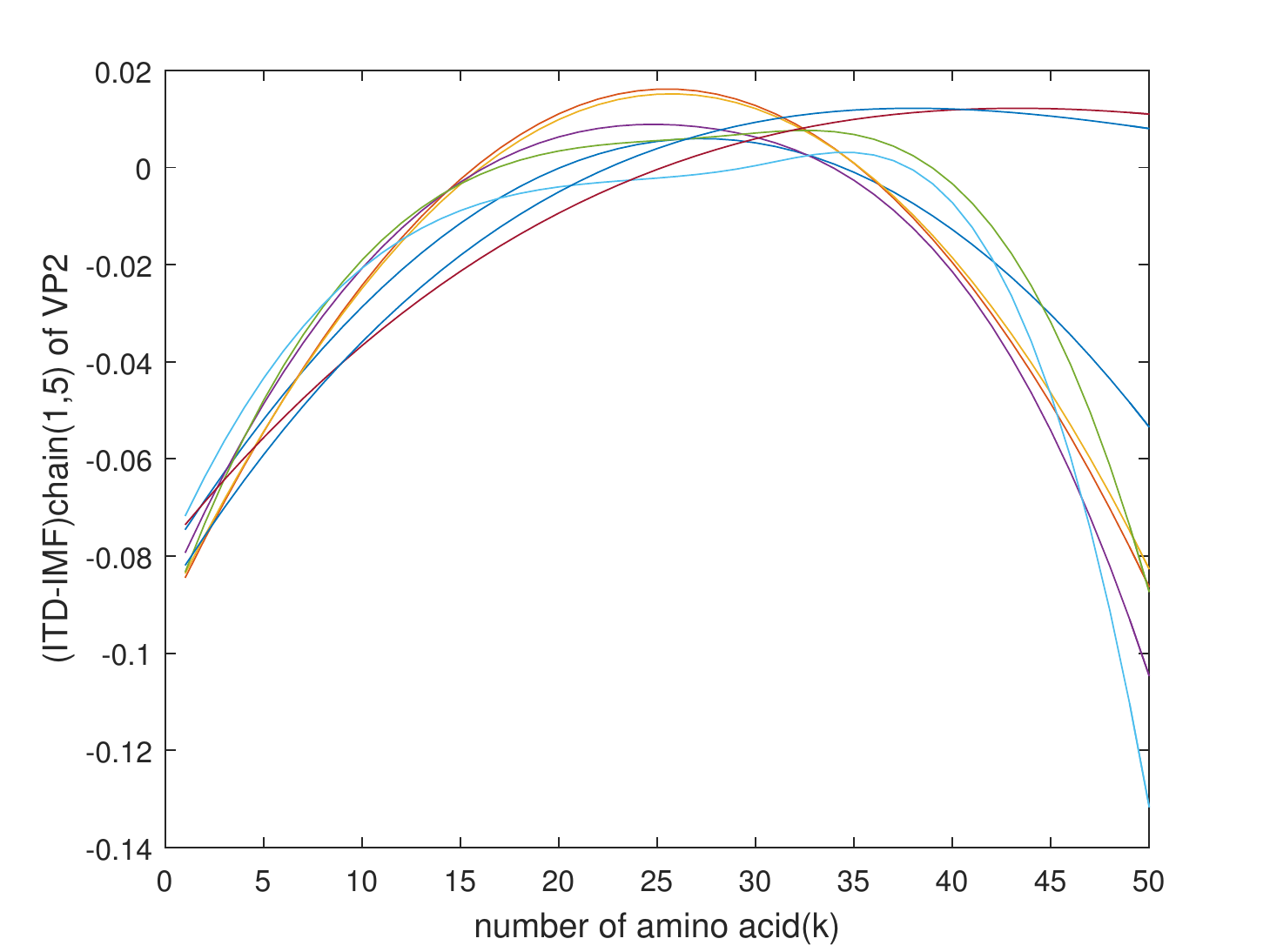,width=8cm}
\caption{On the left, we  plot the time series of   $(ITD-IMF)chain(1)$ of gene of $VP2$ in Coxsackievirus  with 8 samples. On the right, we  plot time series of $(ITD-IMF)chain(1,4),(ITD-IMF)chain(1,5)$ of gene of $VP2$ in Echovirus  with 8 samples, successively.
\label{result_vp2} }
 \end{figure}

\begin{figure}[!t]
 \centering
 
 \epsfig{file=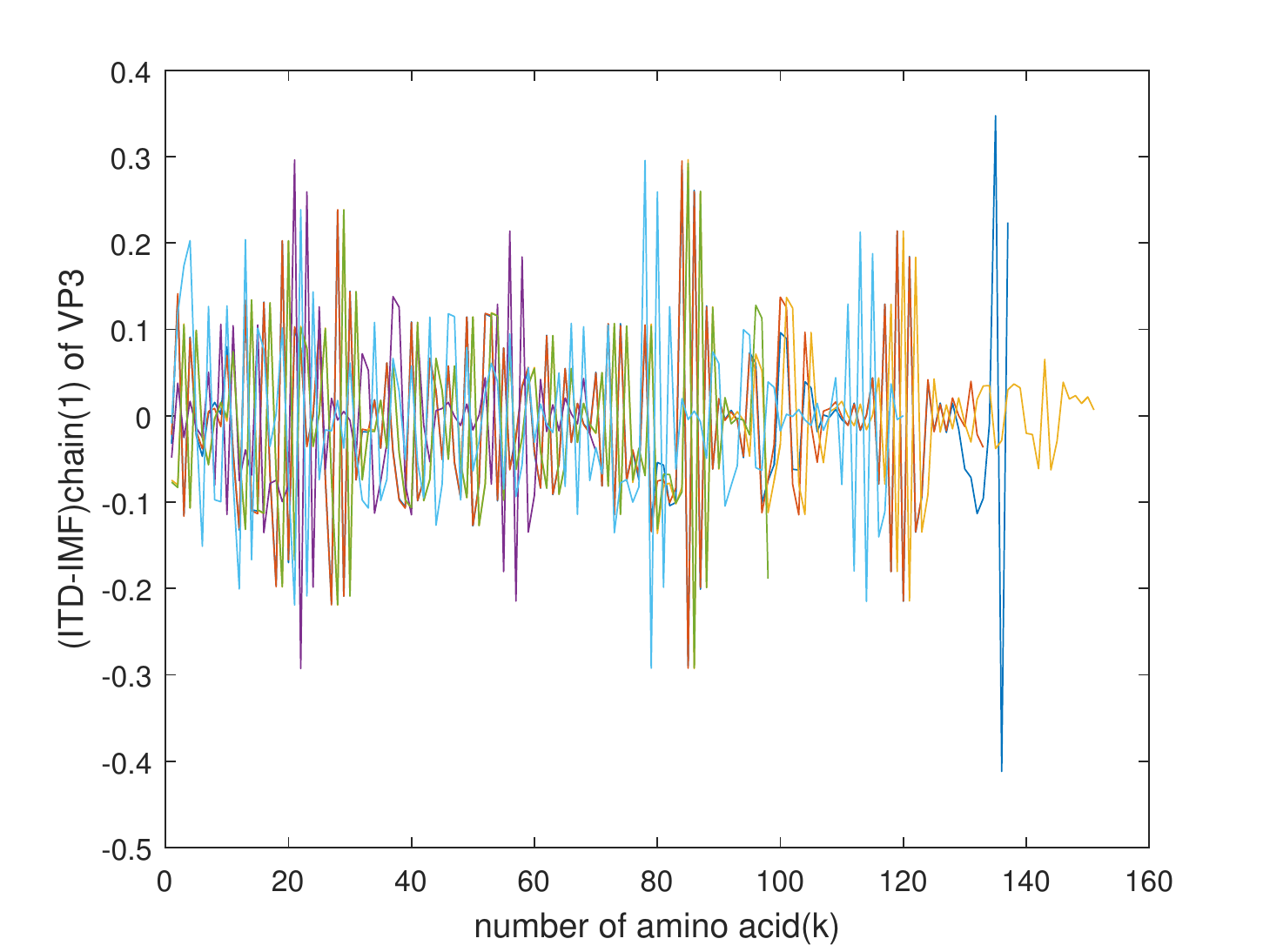,width=8cm}
 \epsfig{file=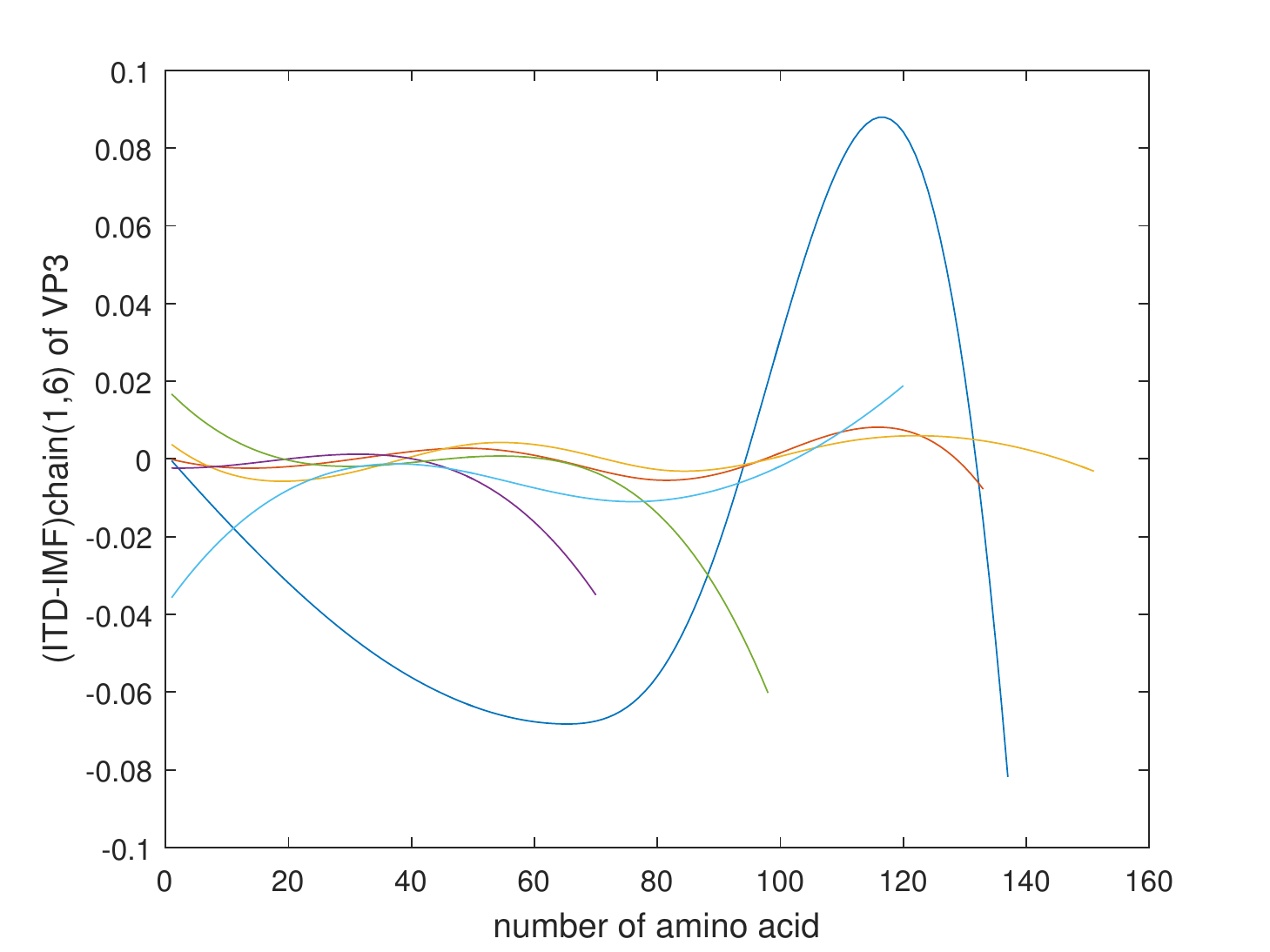,width=8cm}
 \caption{On the left, we  plot time series of gene of $(ITD-IMf)chain(1)$ of $VP3$ in Echovirus by using the average 
value of Chern-Simons current in genotype. On the right, we plot $(ITD-IMF)chain(6)$ of $VP3$
 with different sizes of genome. \label{result_vp3}}
 \end{figure}

\begin{figure}[!t]
 \centering
\epsfig{file= 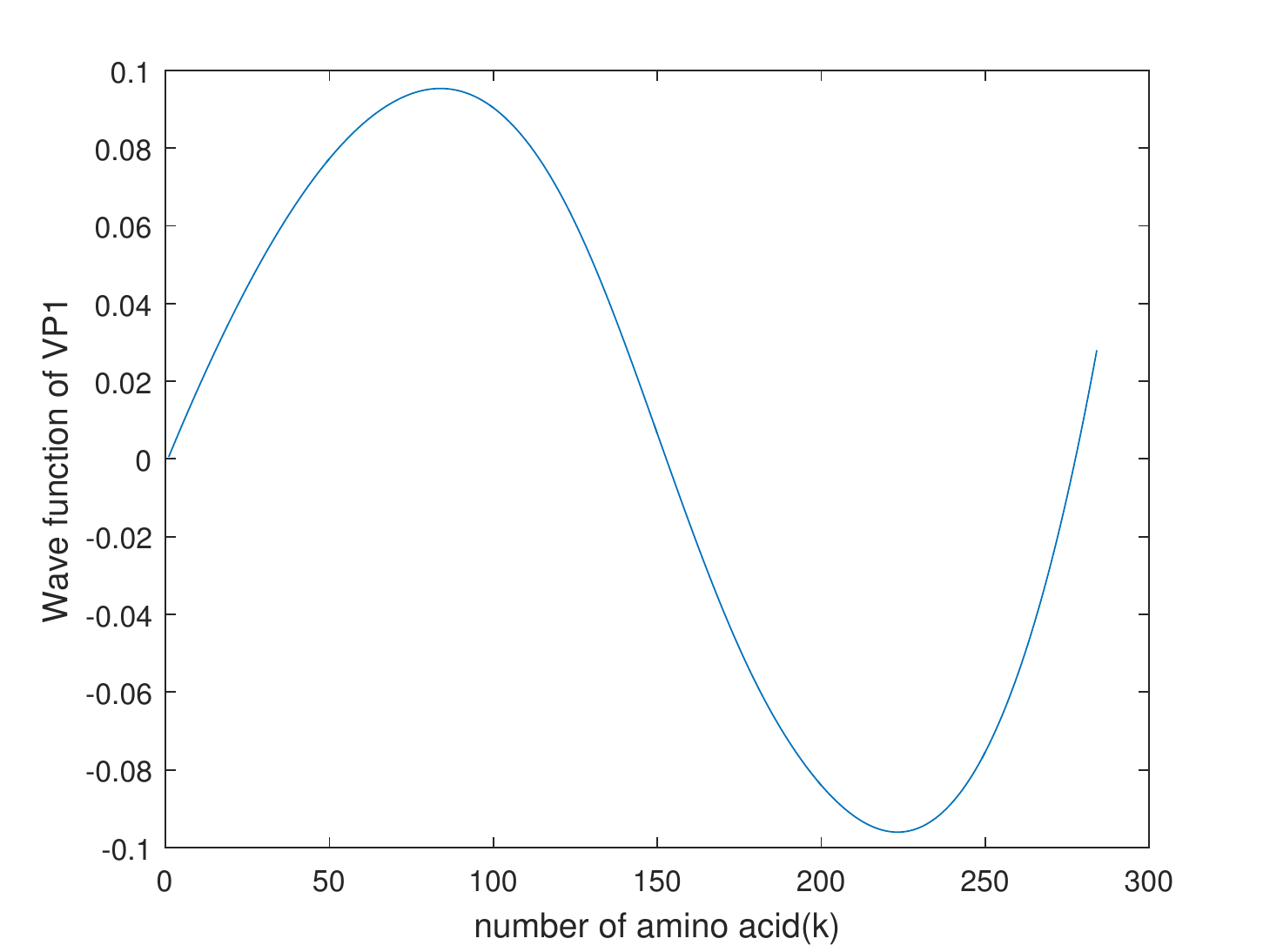,width=8cm}
\epsfig{file= 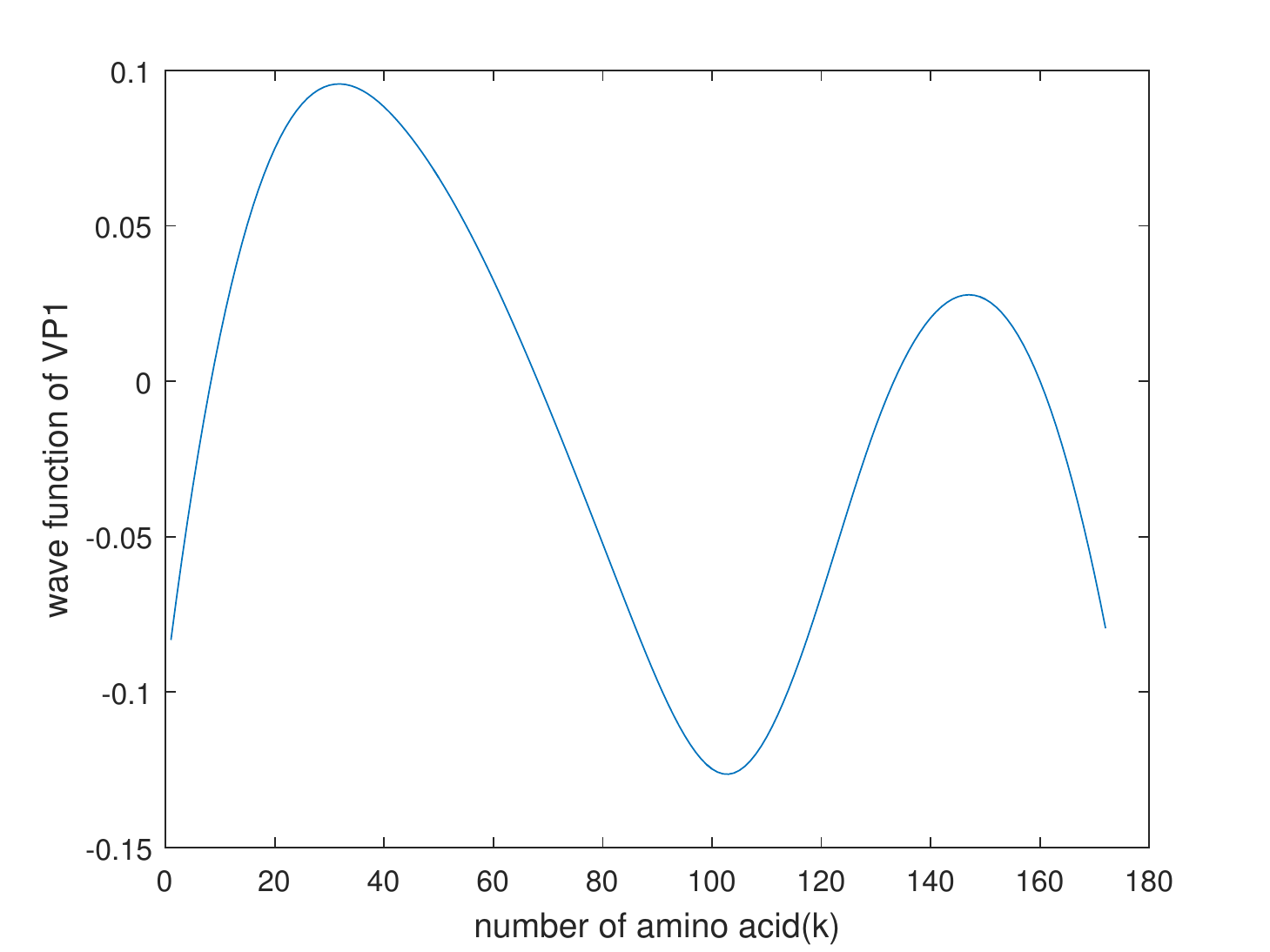,width=8cm}

 \epsfig{file= 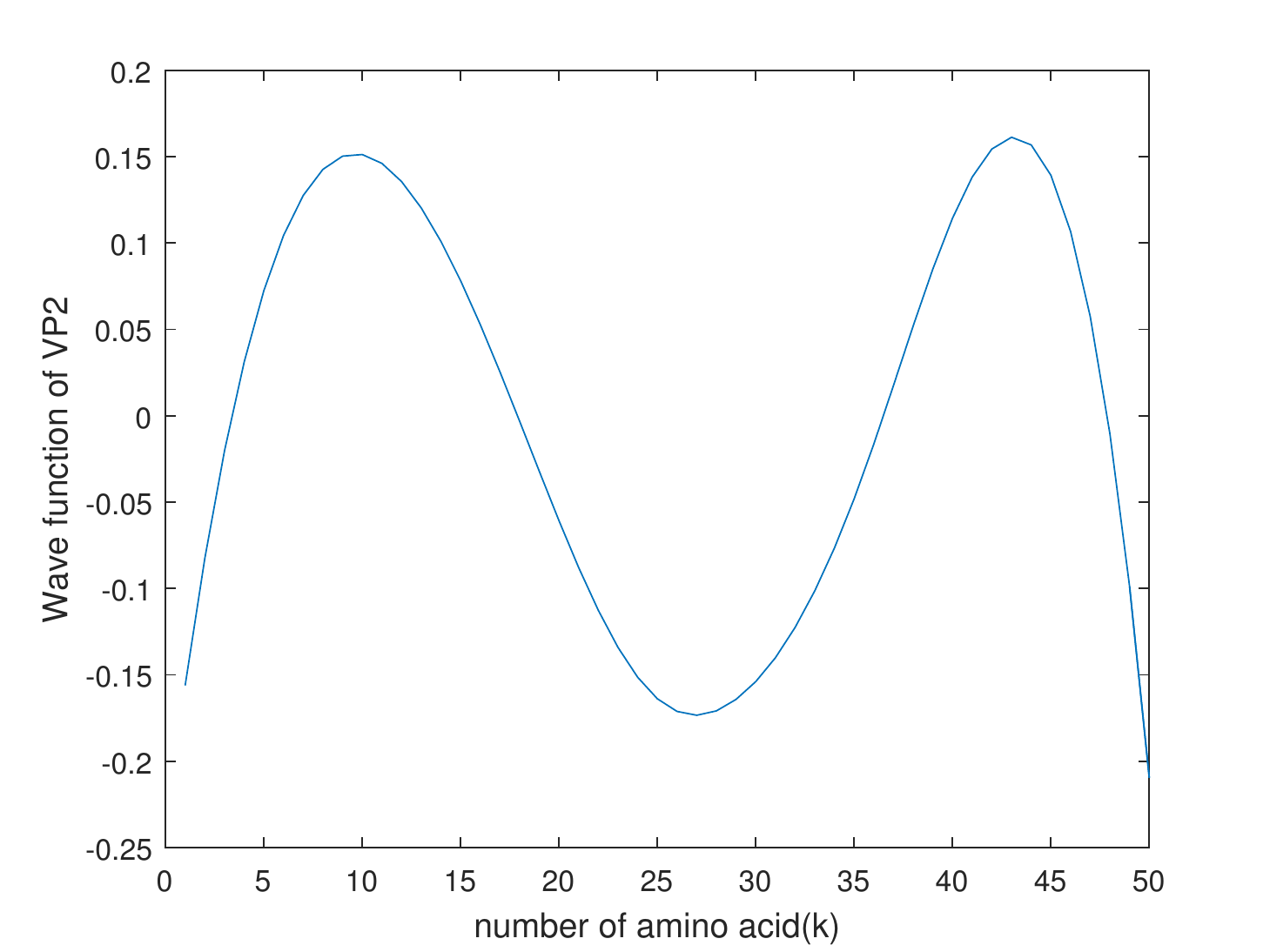,width=8cm}
 \epsfig{file=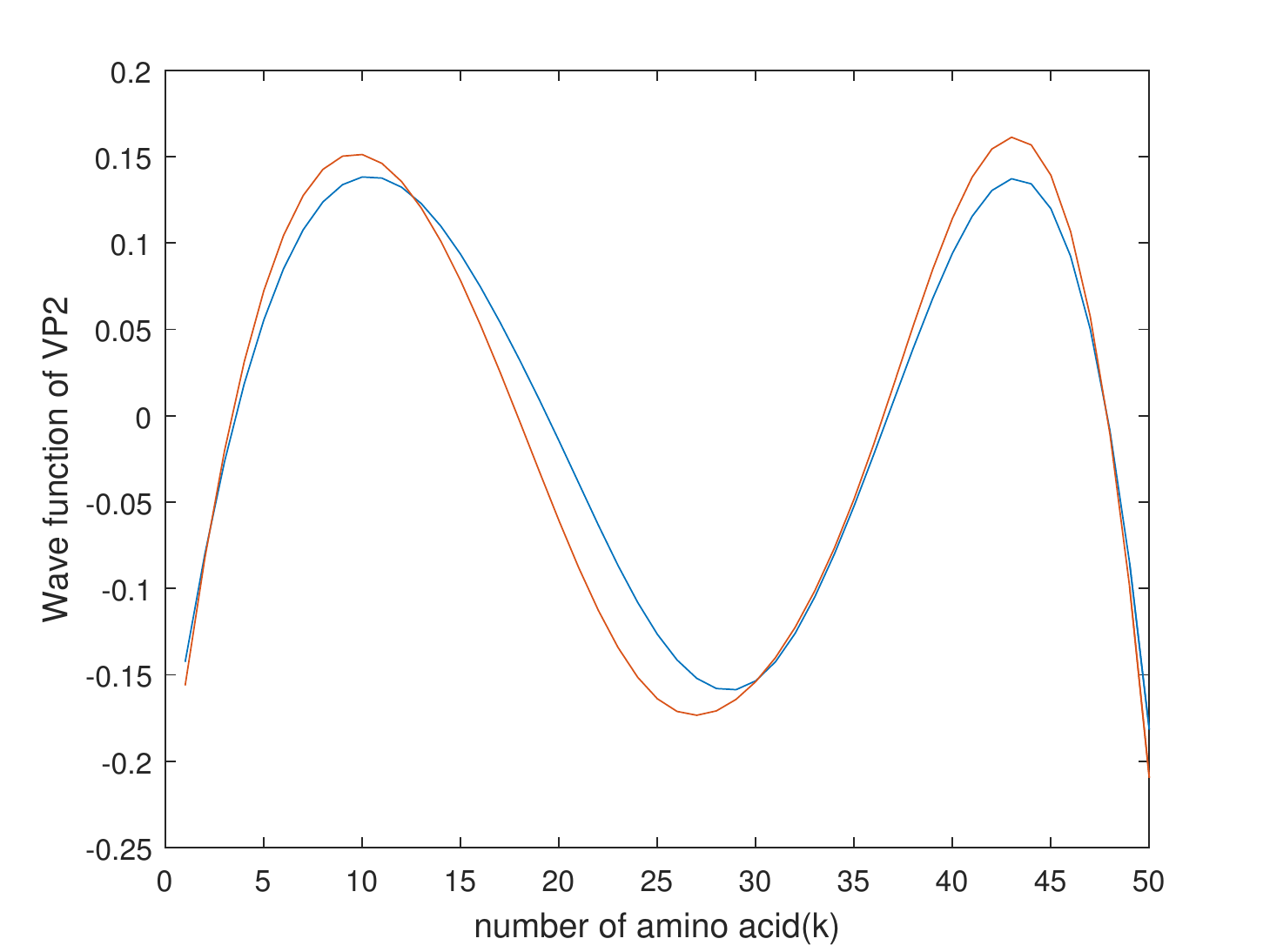,width=8cm}
\caption{In the first row on the left, we plot a  wave function of  geneon in genotype of 
$VP1$ in Coxsackievirus from the  linear combination of 8 species adopting the 
 $(ITD-IMF)chain(1,n)$. In the first row on the right,  we plot
 a wave function of  geneon in genotype of 
$VP1$ in  Echovirus from the  linear combination of 8 species adopting the 
 $(ITD-IMF)chain(1,n)$. In the second row, we plot a wave function of  geneon in genotype of 
$VP1$ in Coxsackievirus from linear combination of 8 species adopting the 
 $(ITD-IMF)chain(1,n)$. In the the second  row on the right, we plot,  in red,  the   wave function of  geneon in genotype of 
$VP1$ in  Echovirus from linear combination of 8 species adopting the 
 $(ITD-IMF)chain(1,n)$  $VP2$ from 8 samples of Echovirus. The blue line is the geneon of $VP2$ in
  Coxsackievirus.\label{wavefunction} }
 \end{figure}

 We  use genotype of viral capside proteins $VP1,VP2,VP3$ and genotype of their host cell receptor protein genotype. The first download sample is a  time series of genetic code  of unknotted 
$VP1, VP2, VP3$ genotypes in 2 species of Rhinovirus, the  so called Echovirus and Coxsackievirus. Second analysis involves their receptor protein, $Ig-CAM$  in host cell genotype.
 We want to dectect the characteristic of geneon wave function as a Holo-Hilbert spectrum \cite{holo} from the trend of super-regression of $(ITD-IMF)chain(1,n)$ with 3 layers of frequency mode modulation, $FM1, FM2, FM3$, 
 from  super-regression of Chern-Simons current in geneotype of unknots and links in proteins.
 We use Rhinovirus and, specifically, perform the  data analysis with 2 species,  Poliovirus and
Escovirus,  with 8 samples
 of different genotype  which contain very high genetic variation and Coxsackievirus.
We compare the amount of nucleotides in 8  samples from   Coxsackievirus and Echovirus 
nucleocapsid structural glycoprotein $VP1,VP2,VP3$. They  are download from genBank with different lengths of genetic code in their geneotypes.
The length of sequence is very short compared with genotype in human gene.
 The capsid protein $VP4$  comes from $1A$  gene and is inside the viral surface protein, therefore we do not
 perform an empirical analysis for this case in this work. The capsid protein $VP2$  comes from $1B$ 
 gene.The capsid protein $VP3$  comes from $1C$  gene.The capsid protein $VP1$  comes from $1D$  gene.  
 To demonstrate the efficiency of   this tool, we perform  data analysis with $(IMF-ITD)chain(1,n)$ 
of Chern-Simons current in geneotype of viral capsid protein genotype with predefined value of one pixel
 to generate the image for pattern matching and to classify the species of organisms. We use 
values from Table \ref{table_current} to  compute the Chern-Simons current 
for 20 amino-acids,  computed from the algorithm of Jones polynomials derived  from the Witten invariant 
 tensor correlation between gene tensor network of Chern-Simons current in genotype of  viral 
capsid protein and host receptor gene. The goal is the visualization of  pattern recognition.  
We use a pixel representation of  Chern-Simons current in genotype for every amino-acid (see Table \ref{table_current} for detail) adopting a  spinor network model.
For genotype,  we use the RGB image of Chern-Simons current for each node of amino-acid.
The capsid proteins of icosahedral virus in  interaction with host receptor 
proteins are only $VP1$, $VP2$.  
The time series of genetic code of $VP2,VP3,VP1$ are induced from gene $1B,1C,1D$
 respectively.
The result of computation of $(ITD-IMF)chain(1)$ and trend of $VP1$ is shown in Fig. \ref{result_vp1}.
In the plot, we notice that  time series of $(ITD-IMF)chain(1)$ of gene of $VP1$ in Coxsackievirus  
with 8 samples does not change so much compared to results in $VP2$ and $VP3$. The trend of super-resgression of $VP1$ is almost similar with straight line with upper slope where  we can  estimate the parameter of regression. The last sample is only with parabolic  shape. The result of computation of $(ITD-IMF)chain(1)$ and trend of VP2 is shown in Fig. \ref{result_vp2}. From this result,  we find  that  8 sample time series of  Chern-Simons current with $(ITD-IMF)chain(1)$ of
  gene of $VP2$  in Coxsackievirus shows more genetic variations. The $VP2$ is on docking site to host receptor with high evolutional field more than $VP1$ site.
The result of computation of $(ITD-IMF)chain(1)$ and trend of $VP1$ is shown in Fig. \ref{result_vp3}.
From the results, we sum up trend from 8 samples into the geneon wave function with transition state. 
The result of geneon wave function  is shown in Fig. \ref{wavefunction}.
 
\begin{figure}[!t]
 \centering
 \epsfig{file=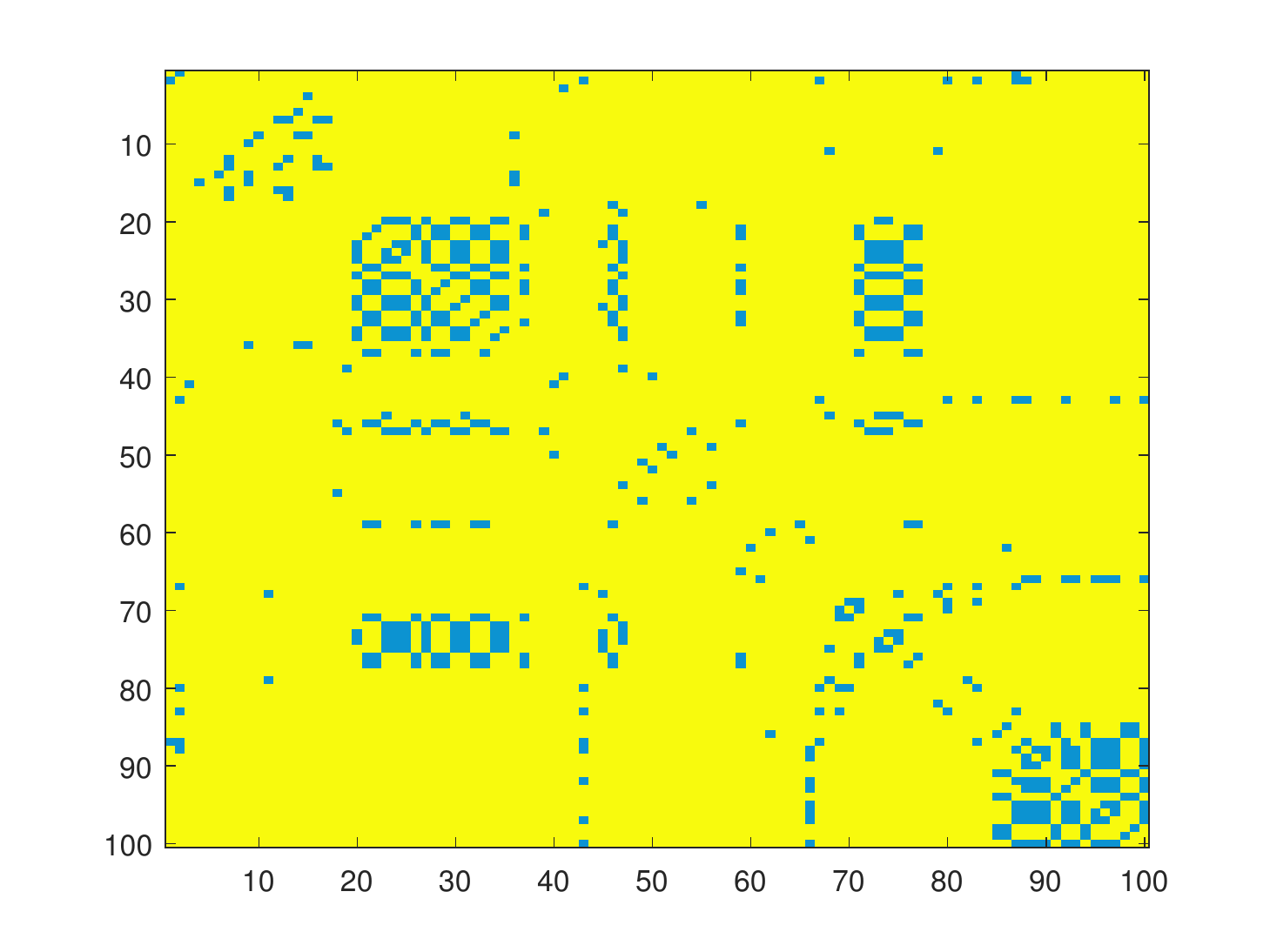,width=5cm}
\epsfig{file=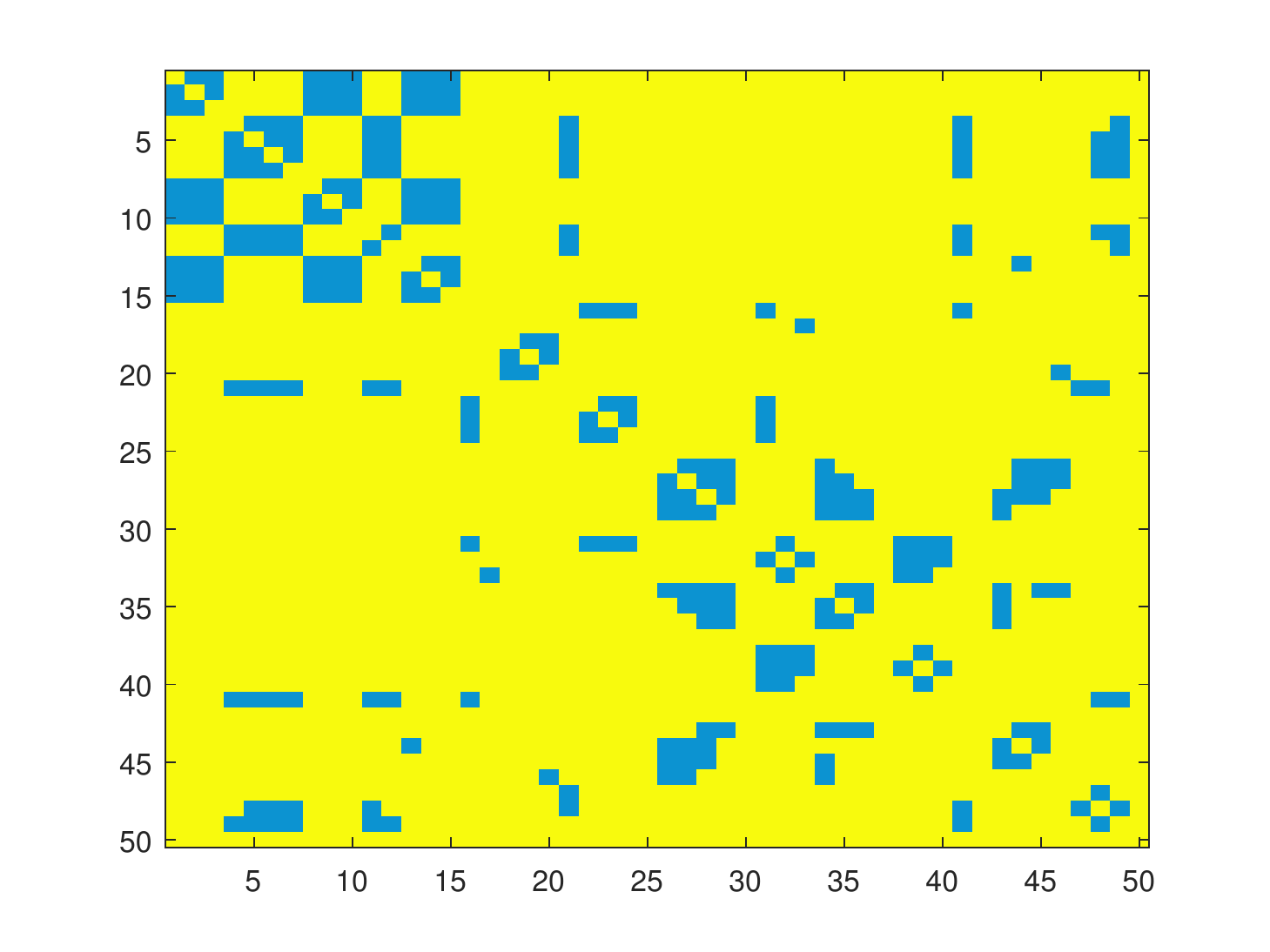,width=5cm}
\epsfig{file=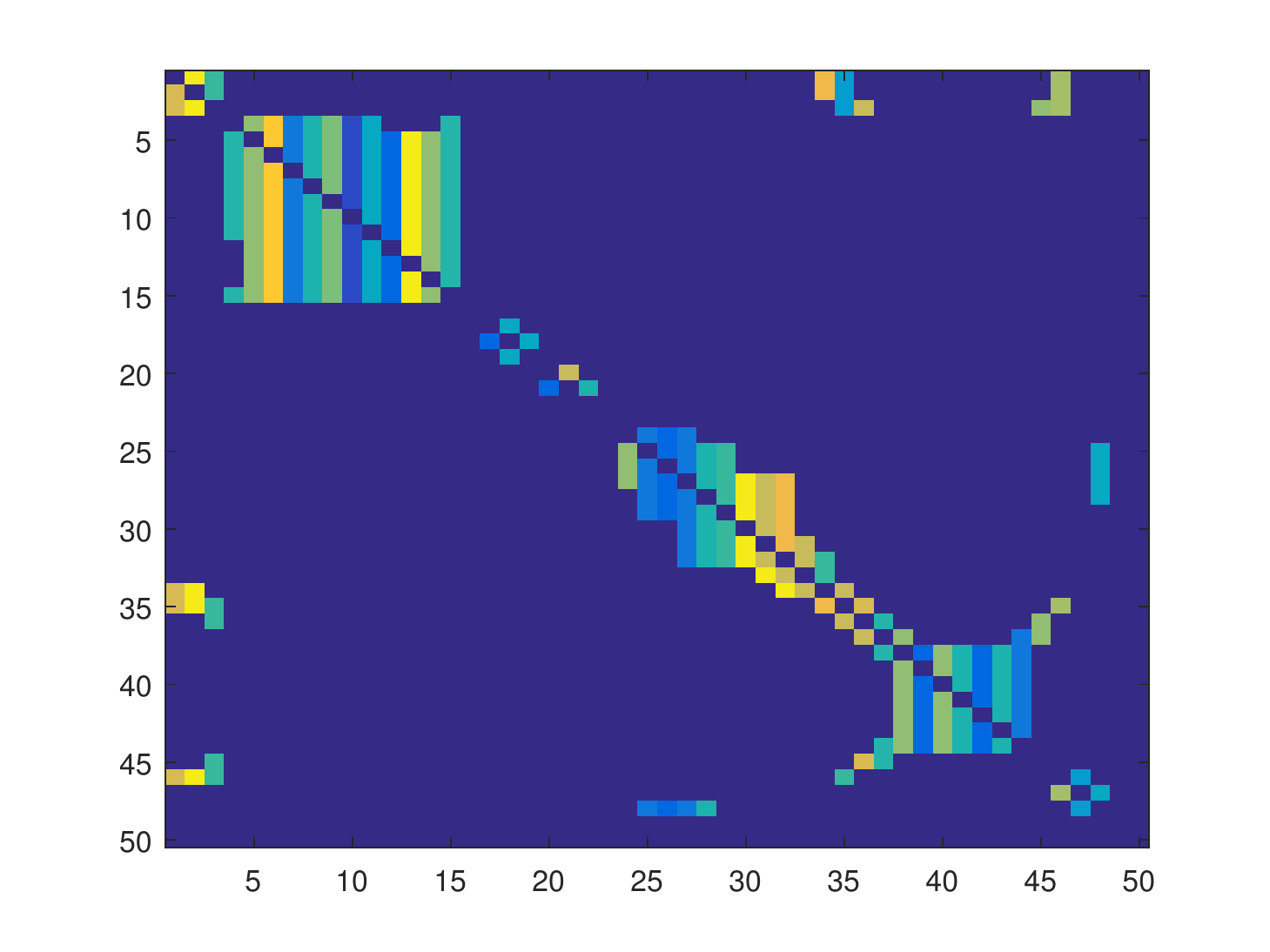,width=5cm}
\caption{The picture represents the   image of geneon in genotype of 
 $VP1,VP2,VP3$  from 8 samples of Echovirus with tensor correlation network. The picture on the left side is 
an artificial image for pattern matching generated from tensor correlation network  between different
 layers of $(ITD-IMF)chain(1,n)$ of $VP1$ between genetic variation among 8 species of Echovirus. The detail of mapping values of one pixel between measurement of
 Chern-Simons current and artificial pixel value is shown in Table \ref{table_current}.
The picture in the middle  is 
an artificial image for pattern matching generated from tensor correlation network  between different
 layers of $(ITD-IMF)chain(1,n)$ of $VP2$ between genetic variation among 8 species of Echovirus. The result of tensor network of $VP1$ and $VP2$ are shown in Fig. \ref{tensor_vp2}. The result of visualization of $VP3$ tensor network is shown in Fig. \ref{tensor_vp3}.
The picture on the right side   is  an artificial image for pattern matching generated from tensor correlation network between different layers of $(ITD-IMF)chain(1,n)$ of $VP3$ between genetic variation among 8 species of Echovirus. 
The dot pattern in image shows tensor correlation between different layers of correlation of different modes in empirical decomposition algorithm. The pattern of dots is fixed for a given  species. We can choose the  
background color for representing the  image by adding a  constant value to every pixel for increasing the  contrast of the 
 image  quality. The yellow background comes from adding  each pixels with 20. On the right picture, we make
 a different background for presenting the image with adding each pixels with 40. The resulting value 
of the background image is the blue color.
\label{image_geneon}}
 \end{figure}
  
\begin{figure}[!t]
 \centering
  \epsfig{file=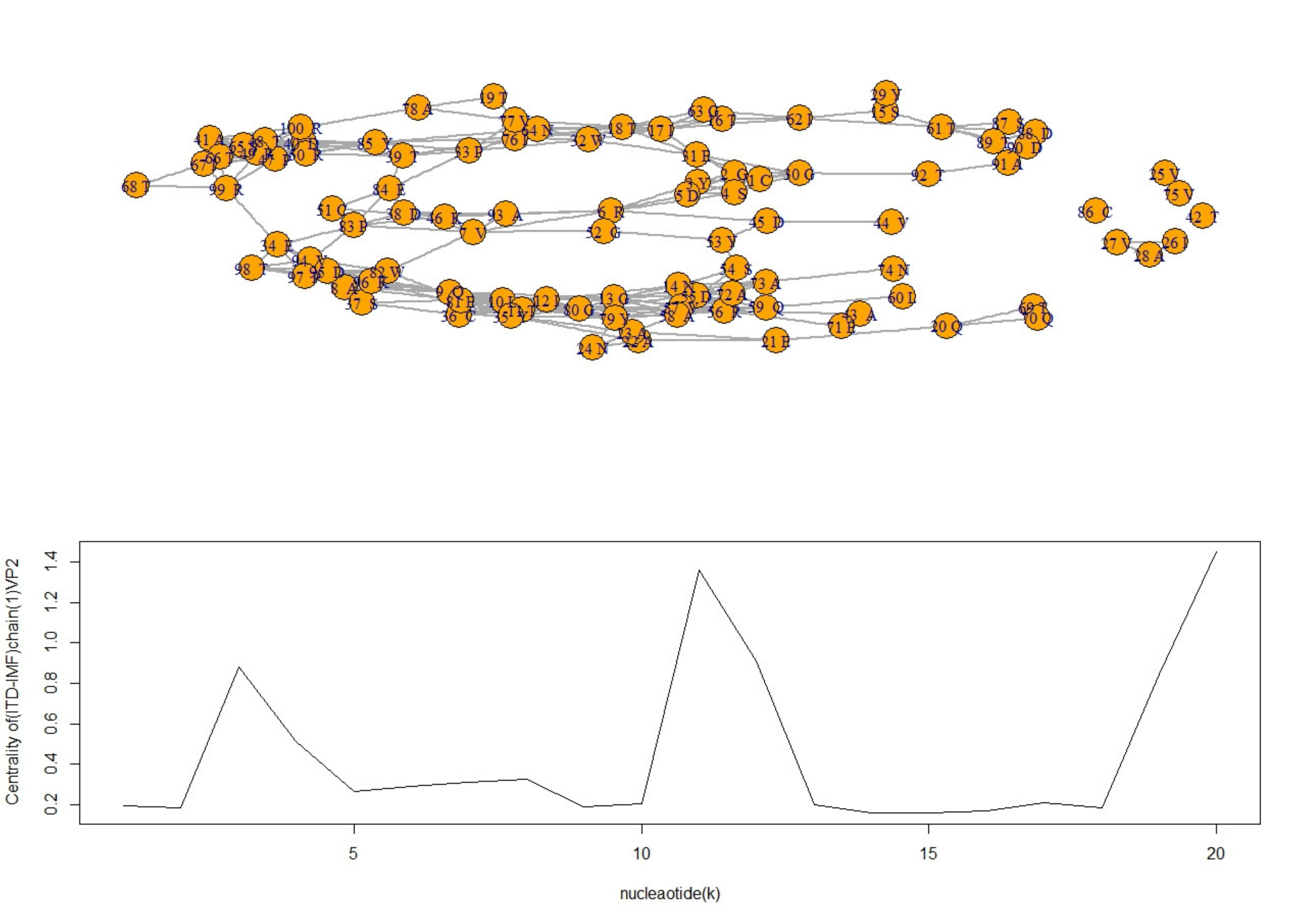,width=8cm}
\epsfig{file=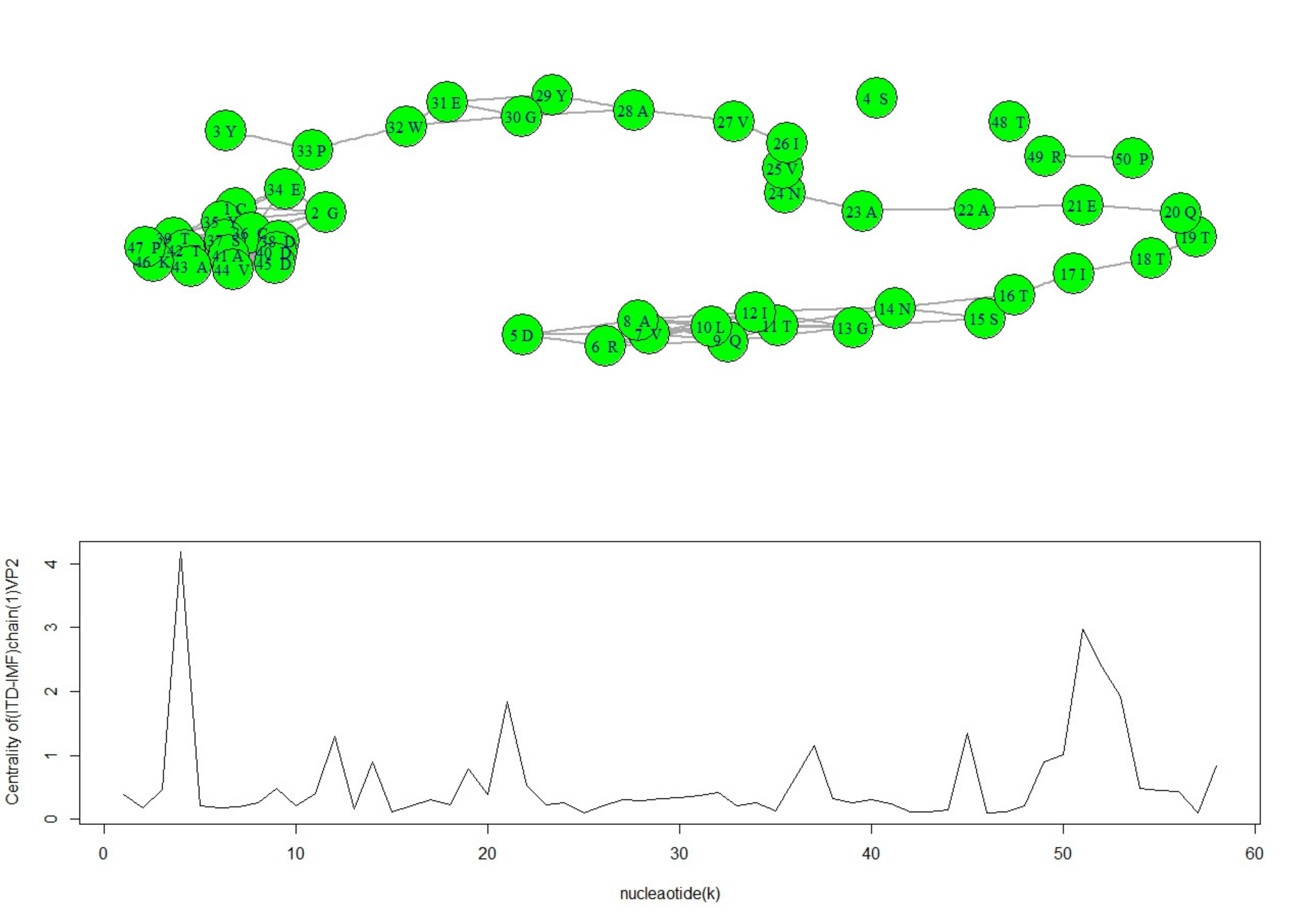,width=8cm}
\caption{On the left, we visualize  the  gene tensor network of $VP1$ in Echovirus with the centrality calculation over genetic variation among 8 samples of species in  Echovirus. The maximum peak of
 centrality measurement shows the  side of protein folding for highest genetic variation.\label{tensor_vp2} 
 On the right, we visualize the    gene tensor network of $VP2$ in Echovirus  plotting the centrality.}
\end{figure}

\begin{figure}[!t]
 \centering
 \epsfig{file=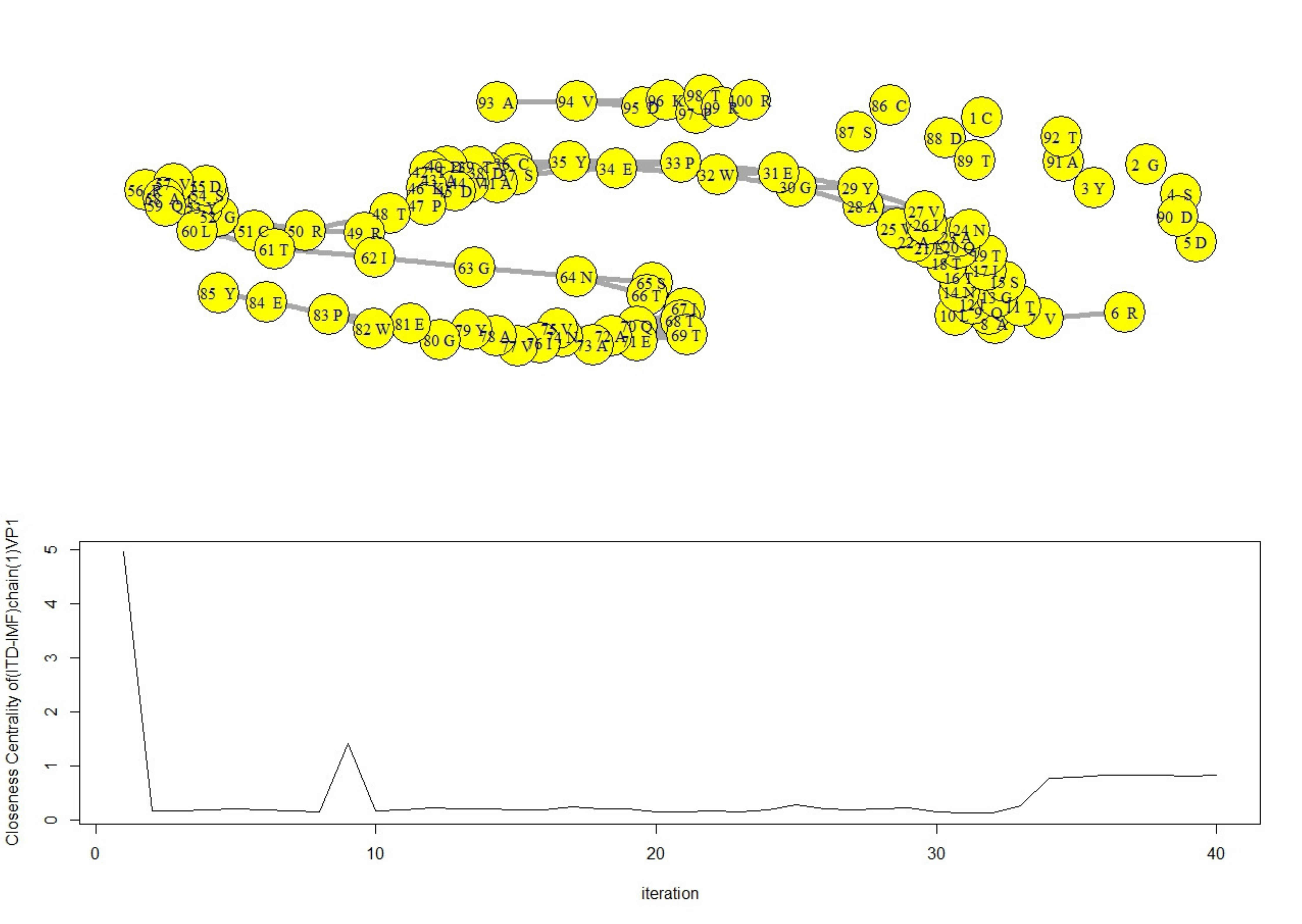,width=14cm}
 \caption{We visualize the    gene tensor network of $VP3$ in Echovirus with the centrality calculation over genetic variation among 8 samples of species in  Echovirus. We notice from the plot that the maximum peak of centrality measurement  appears in the  first area and a small peak at nucleotide number 8-9 of full neoclotides sequence. These 2 peaks show the  area of   highest genetic variation in protein folding structure in $VP3$.\label{tensor_vp3} }
 \end{figure}

\begin{figure}[!t]
 \centering
\epsfig{file=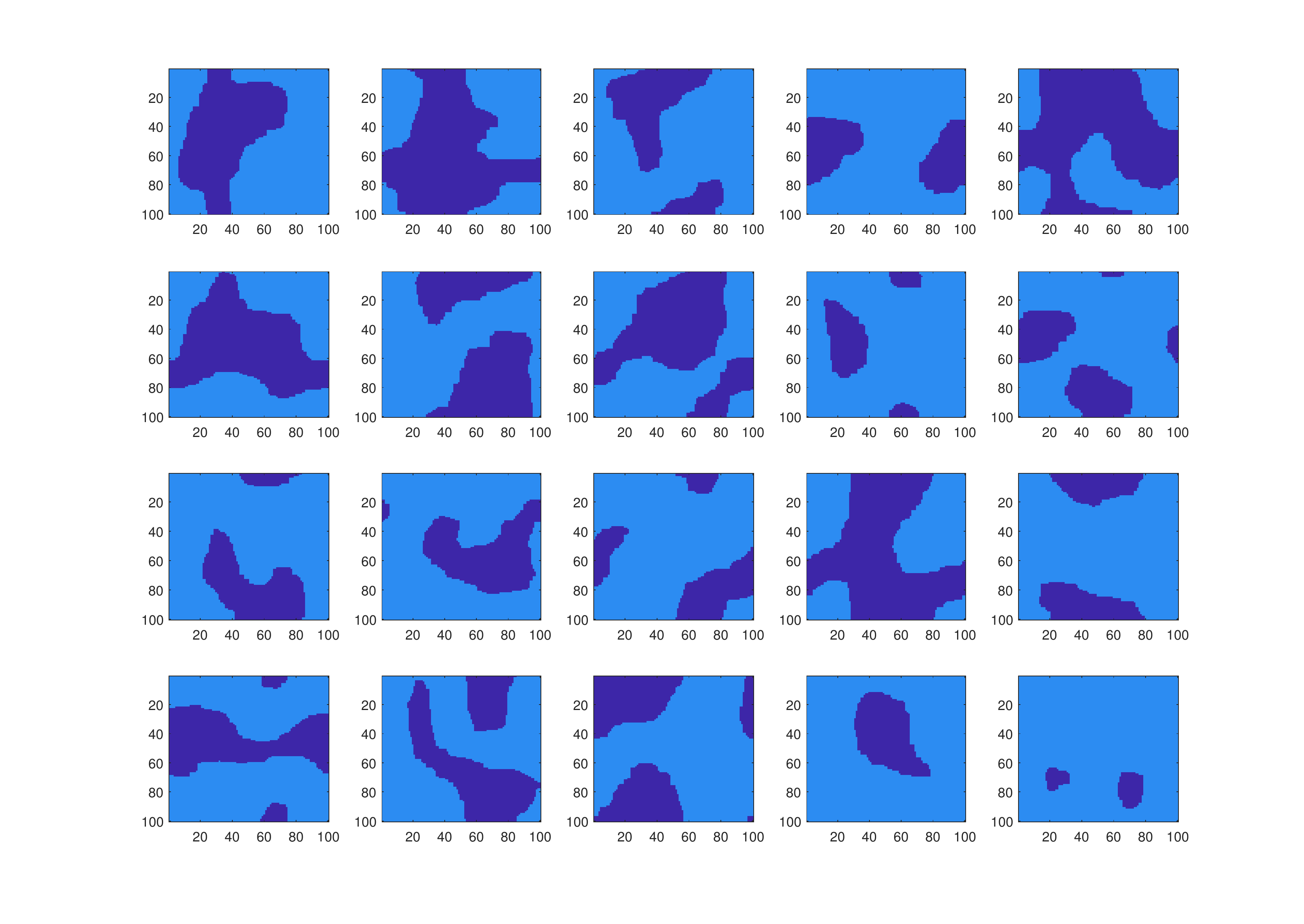,width=5cm}
\epsfig{file=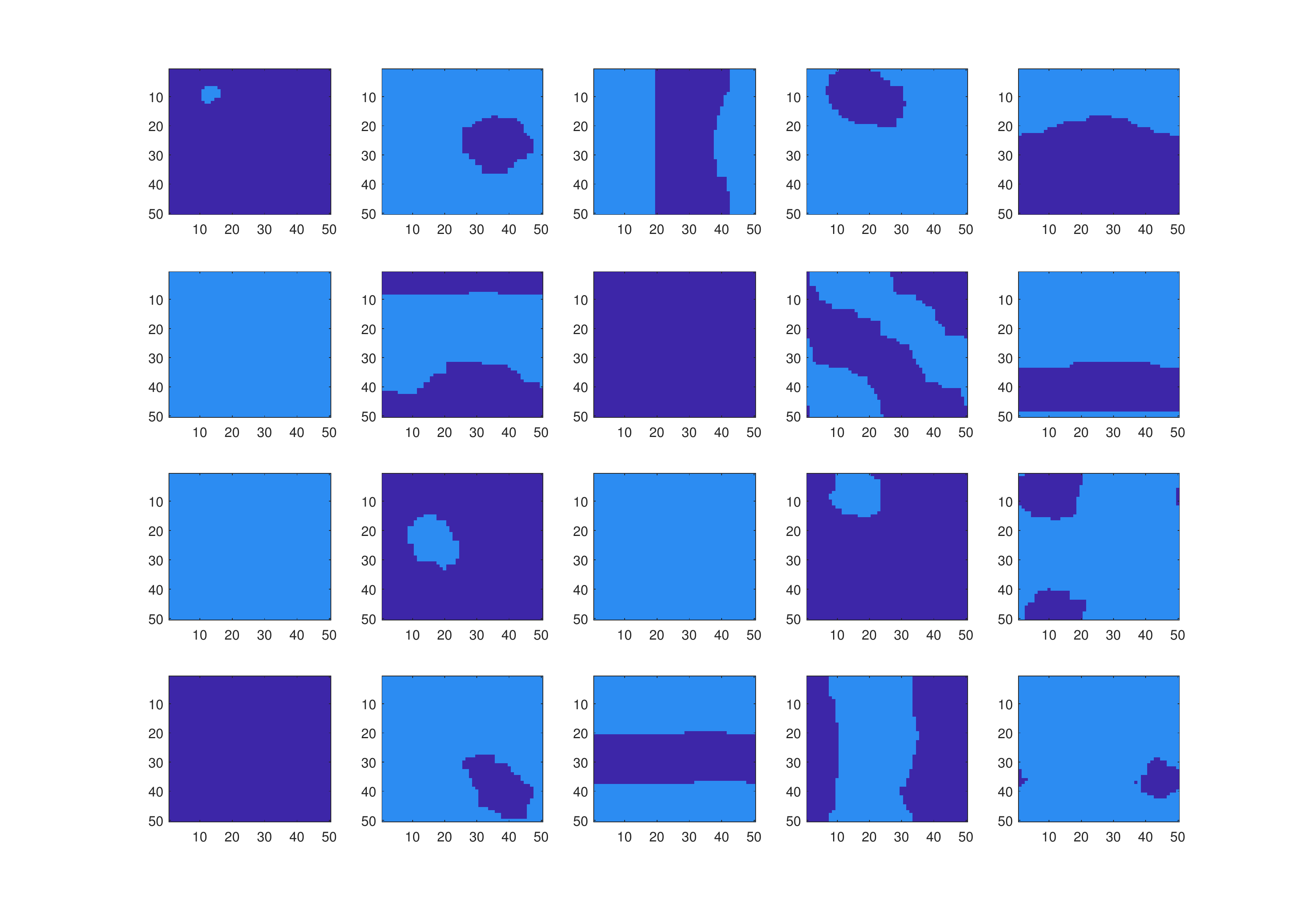,width=5cm}
\epsfig{file=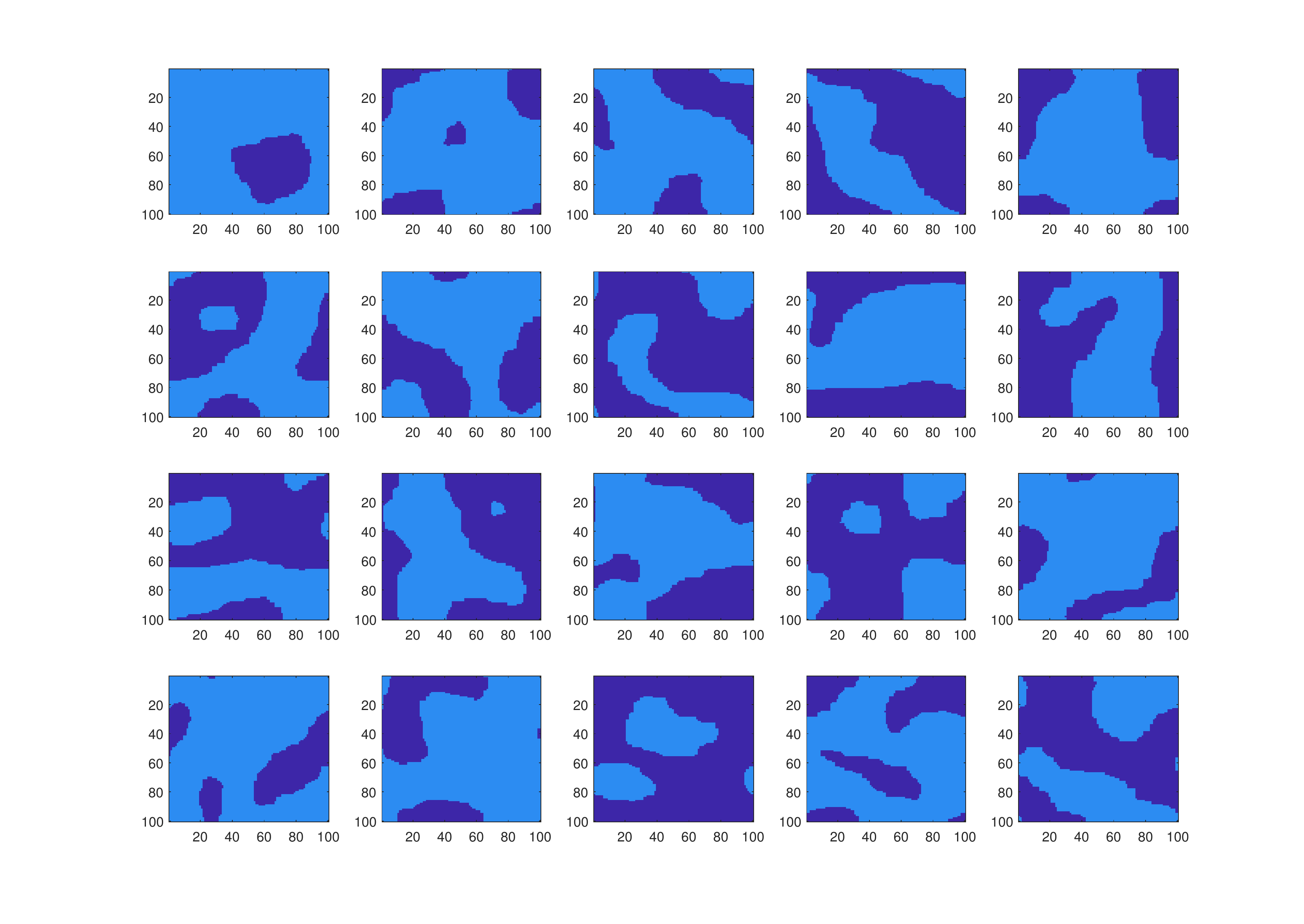,width=5cm}
 \caption{The image  represents network of networks between next 1-20 layers of Ising algorithm calculation of tensor network of 
correlation between geneon of  $VP1$ on the left, $VP2$ in the middle, $VP3$  on the 
right from 8 samples of 
Coxsackievirus. 
  \label{image4}  }
 \end{figure}

\begin{figure}[!t]
 \centering
\epsfig{file=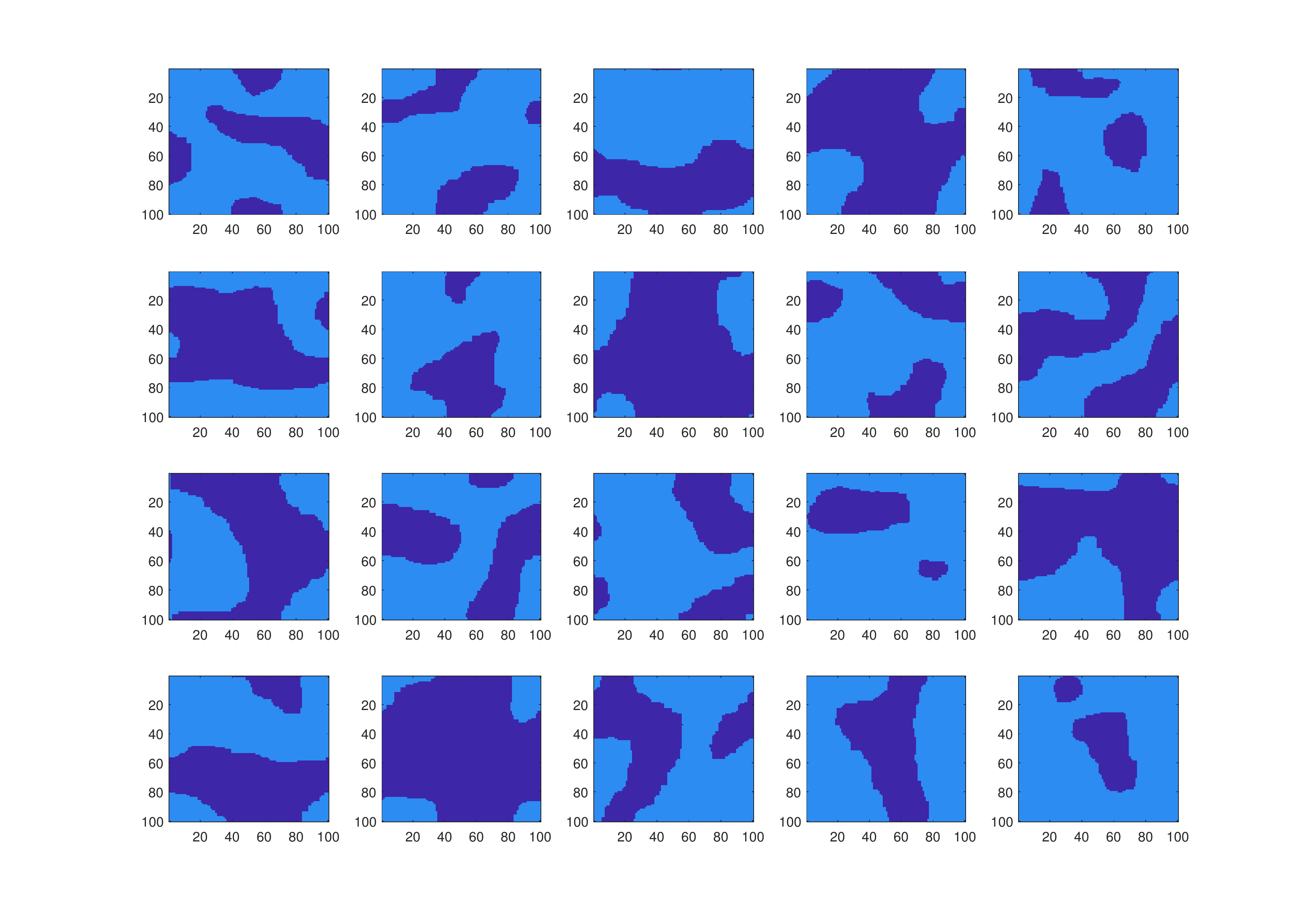,width=5cm}
 \epsfig{file=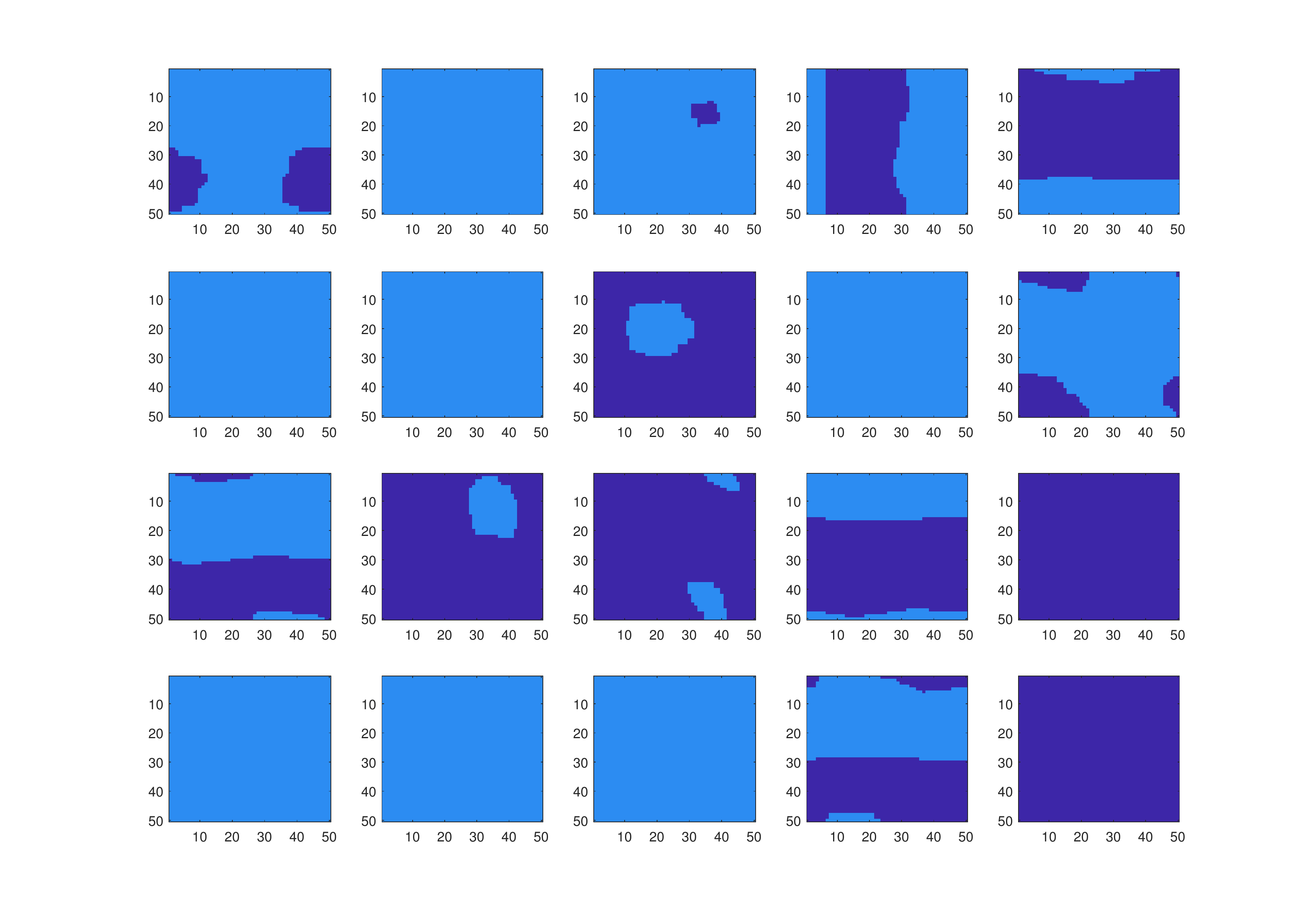,width=5cm}
\epsfig{file=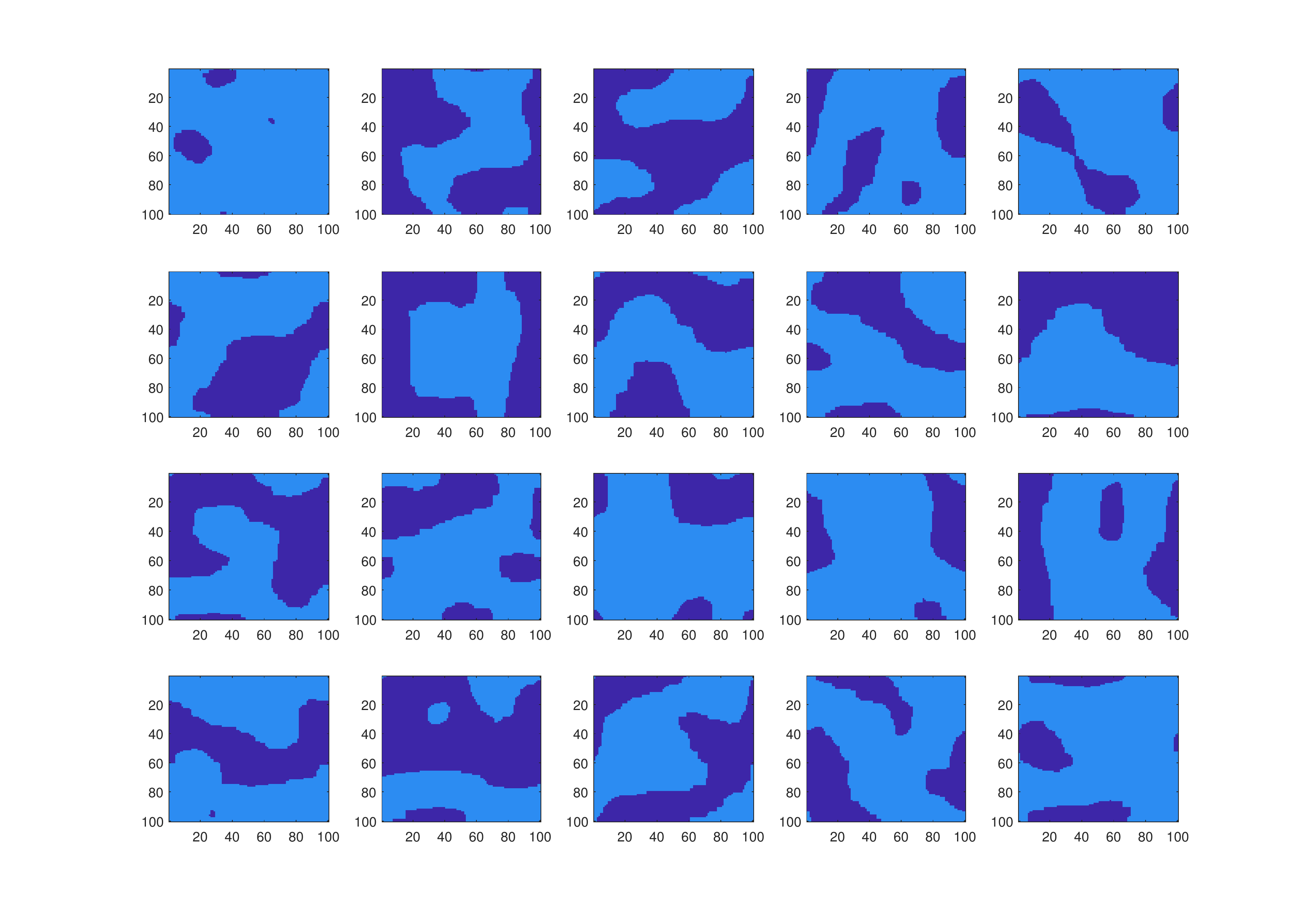,width=5cm}
\caption{ The image represents network of networks between next 21-40 layers of Ising algorithm calculation of tensor network of 
correlation between geneon of  $VP1$ on the left, $VP2$ in the middle, $VP3$  on the 
right from 8 samples of 
Coxsackievirus.  \label{image7}   }
 \end{figure}
  Based on these results, we generate an image for 
species classification of Echovirus and  Coxsackievirus  with tensor network algorithm. The result of image of geneon is shown in Fig. \ref{image_geneon}.
We visualized the tensor network and calculate the closeness centrality for $VP1,VP2,VP3$
 in Echovirus. The only result of $VP2$ in Coxsackievirus.
 is  shown in fig.\ref{tensor_vp2}. The highest peak of closeness centrality is the site of genetic variation in genotype from this analysis.
In order to measure the connection $A_{\mu}$ in geneotype of $VP1,VP2,VP3$ of Coxsackievirus, we used the  Ising algorithm over the  image of tensor network produced from geneon state of Coxsackievirus. The detail of algorithm can be found in \cite{preprint2}.
The result of Ising algorithm over image of geneon in Coxsackievirus  is shown in Fig. \ref{image4} and
Fig. \ref{image7}.    
 
\begin{figure}[!t]
 \centering
\epsfig{file=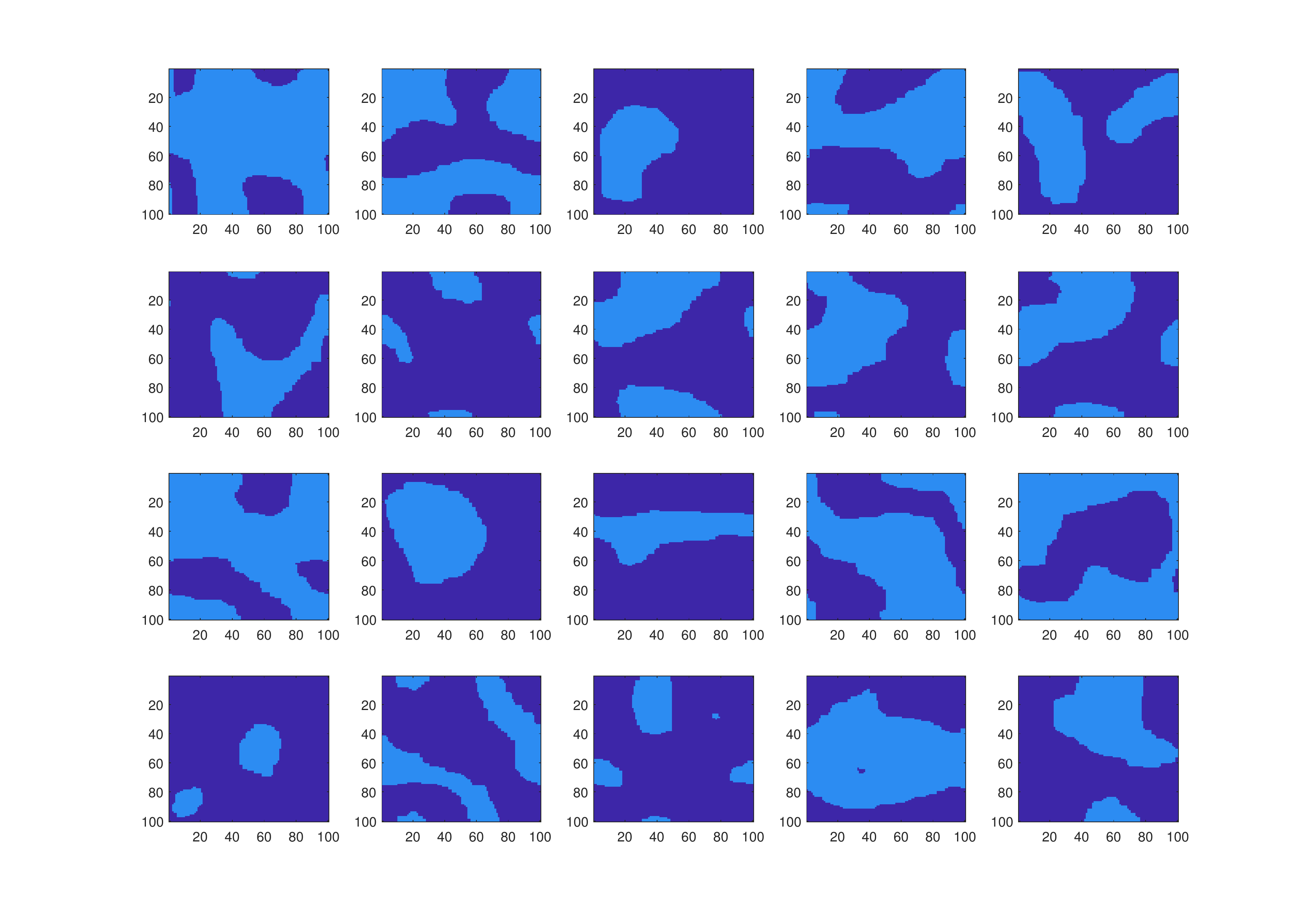,width=4cm}
 \epsfig{file=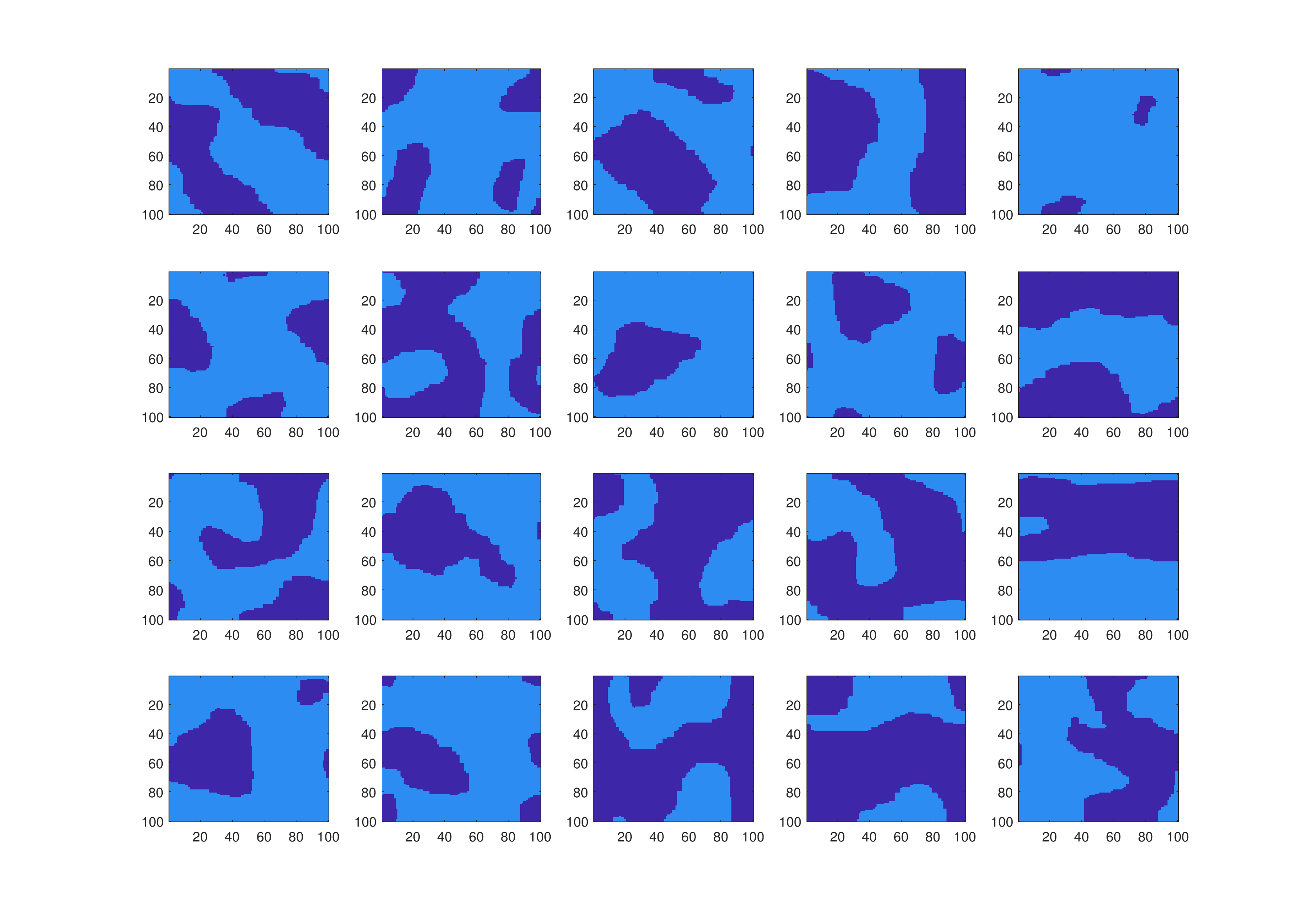,width=4cm}
\epsfig{file=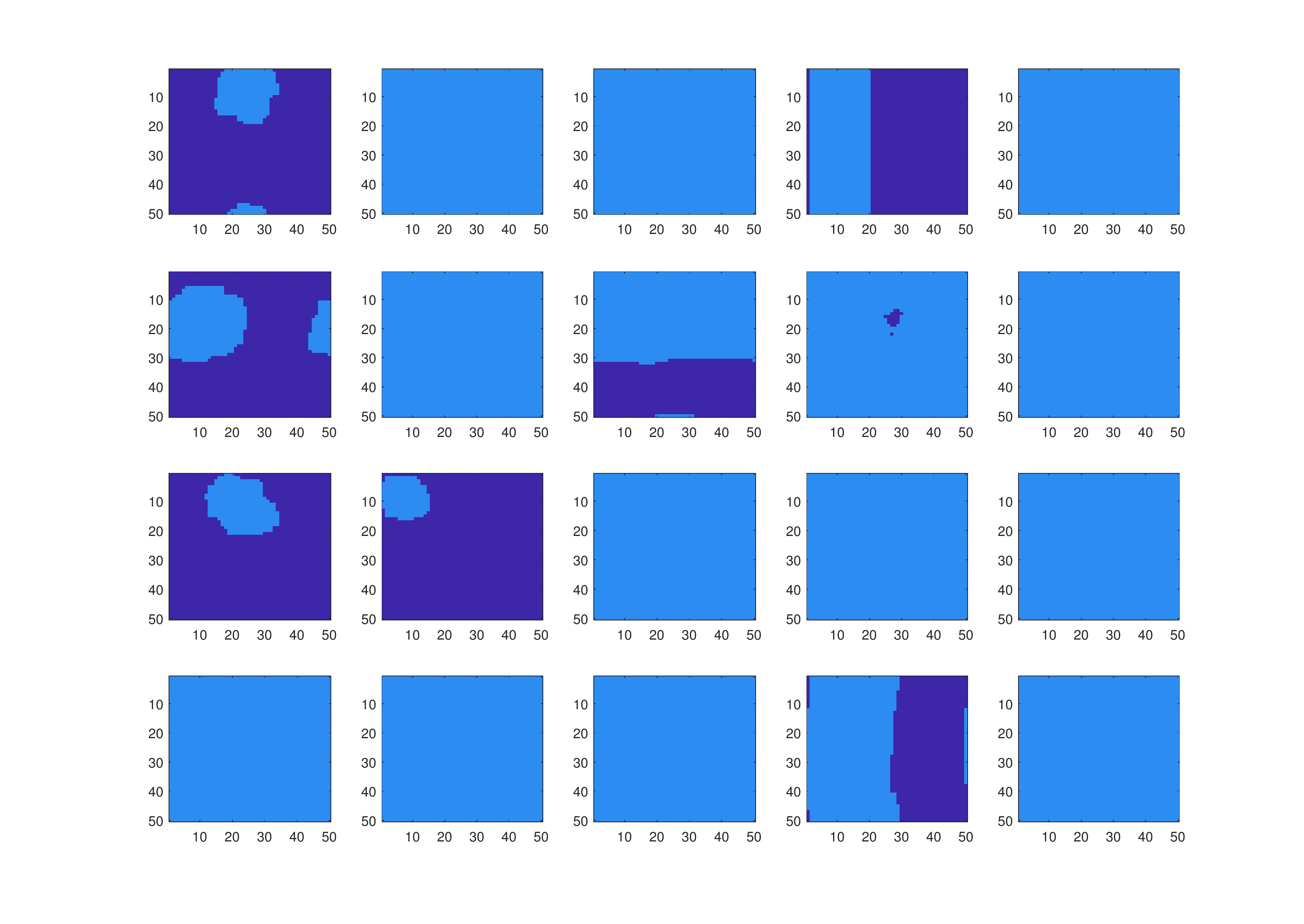,width=4cm}
 \epsfig{file=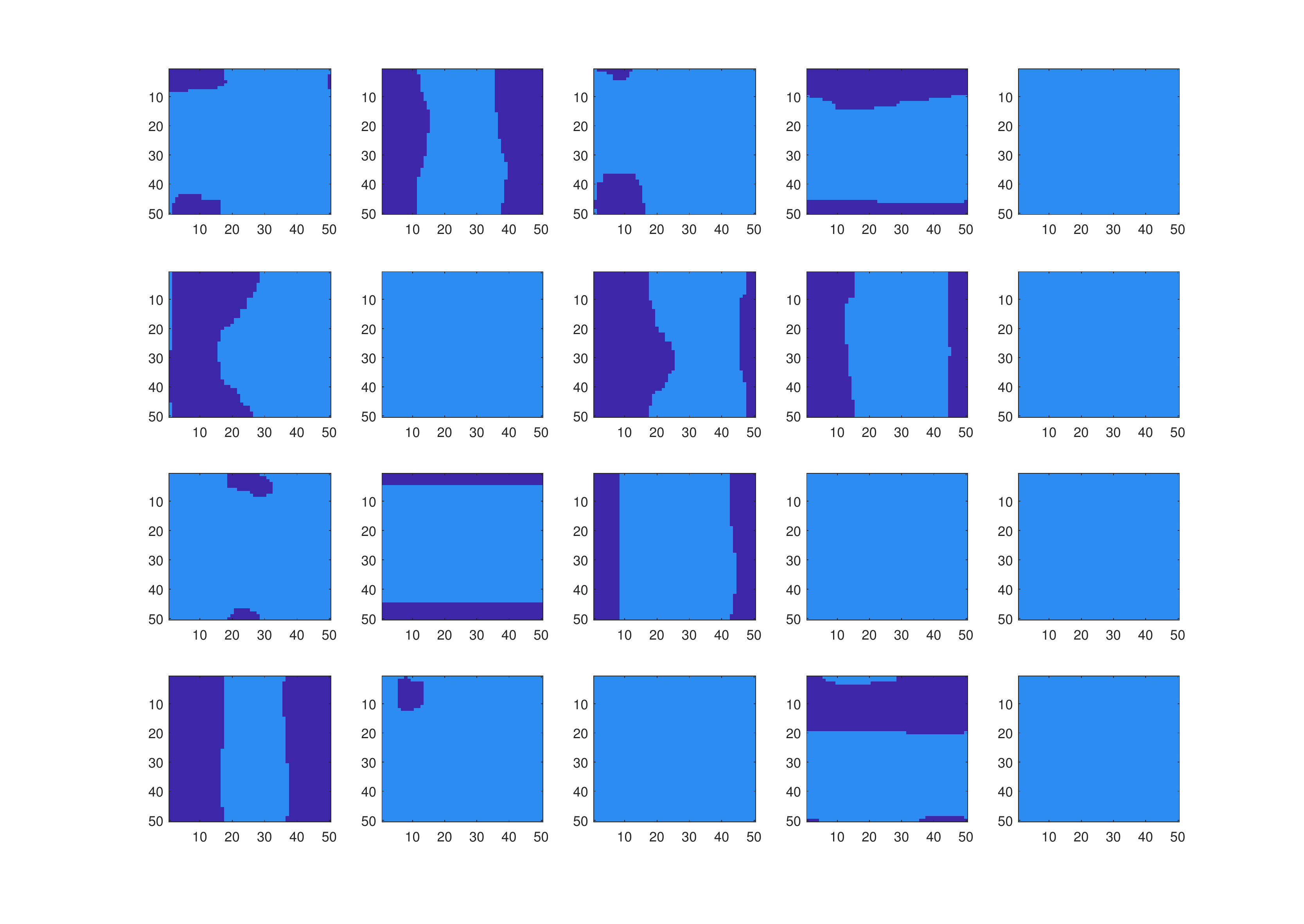,width=4cm}
 \caption{   The first image from the left side   represents the
 $FM2$ of  geneon of  $VP1$  derived by the  Ising algorithm.  The second image from the left side is $FM3$ of $VP1$.   The third image    represents  $FM2$  of  geneon of  $VP2$.
 The last image is $FM3$ of $VP2$.\label{fm_vp1}}
 \end{figure}

The image of  $FM2, FM3$ in geneon image of $VP1,VP2$ is shown in Fig. \ref{fm_vp1}.
 The Holo-Hilbert spectrum \cite{holo} of quantum biology is represented as  a spectrum of geneon wave function with different frequency mode modulations.   It will be interpreted as hidden transition states
 of geneotype  induced from evolution feedback path in protein docking system. The results can be used to detect  hidden patterns inside the  protein folding.

 \begin{figure}[!t]
 \centering
 \epsfig{file=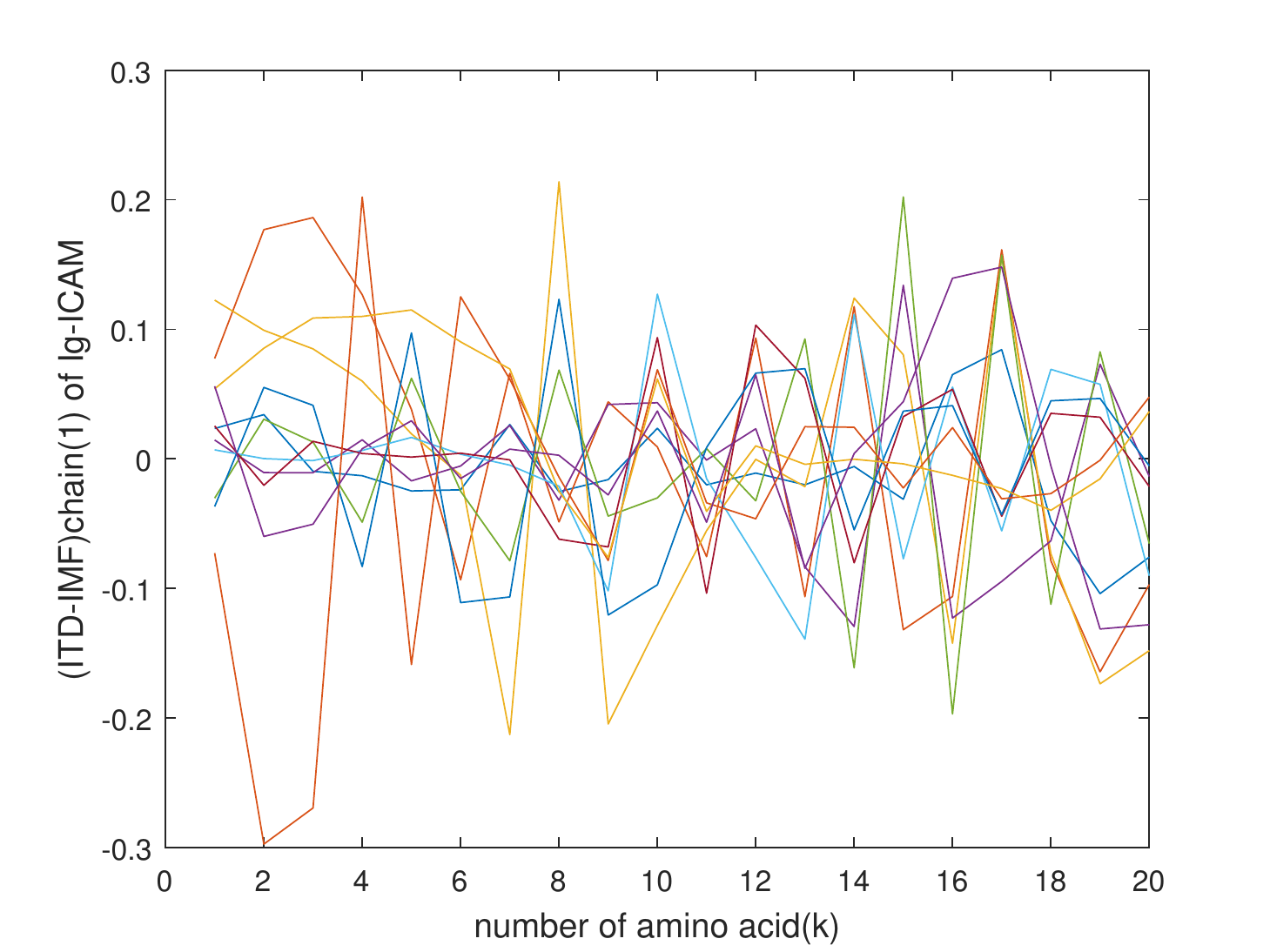,width=8cm}
\epsfig{file=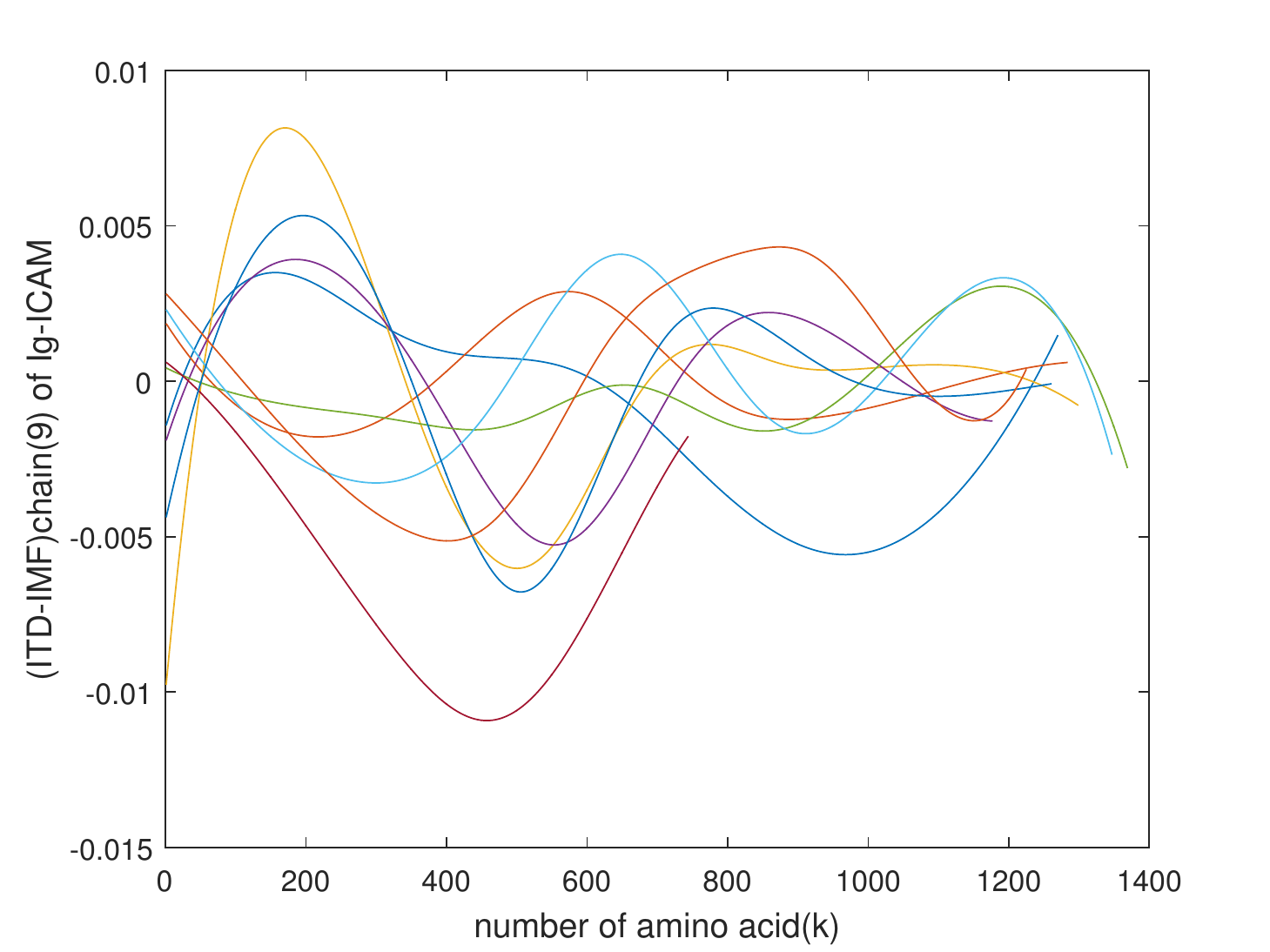,width=8cm}
\epsfig{file=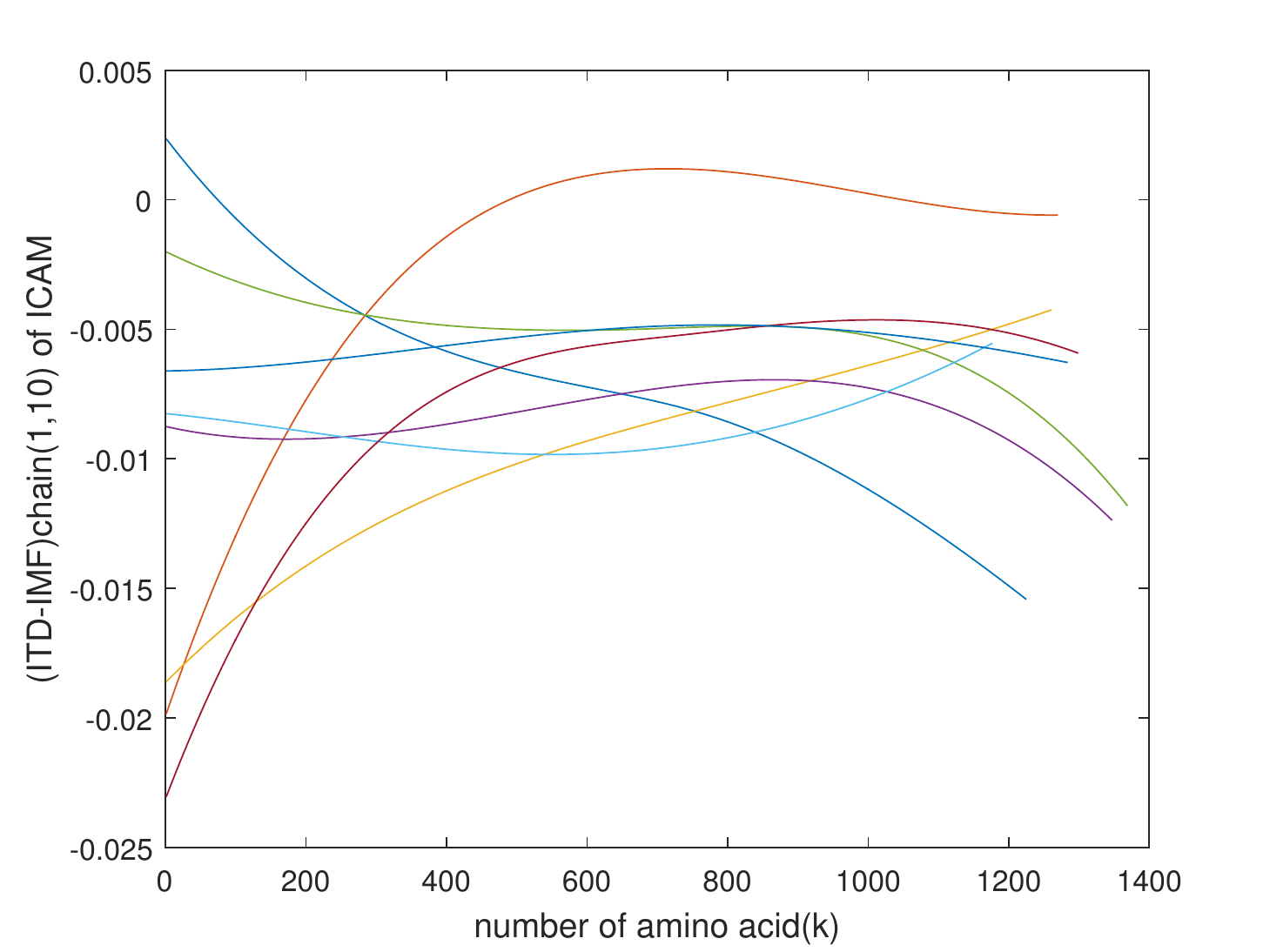,width=12cm}
  \caption{Here, we  plot the gene of  $Ig-CAM$-like protein receptor of antibody
 with 10 samples   with spinor field in time series data. On the left, we plot only the first 20 amino-acid sequence.
 The graph shows highly genetic variation compared to the  genetic variation in 
 viral glycoprotein gene.\label{result_icam}}
  \end{figure}

\begin{figure}[!t]
 \centering
 \epsfig{file=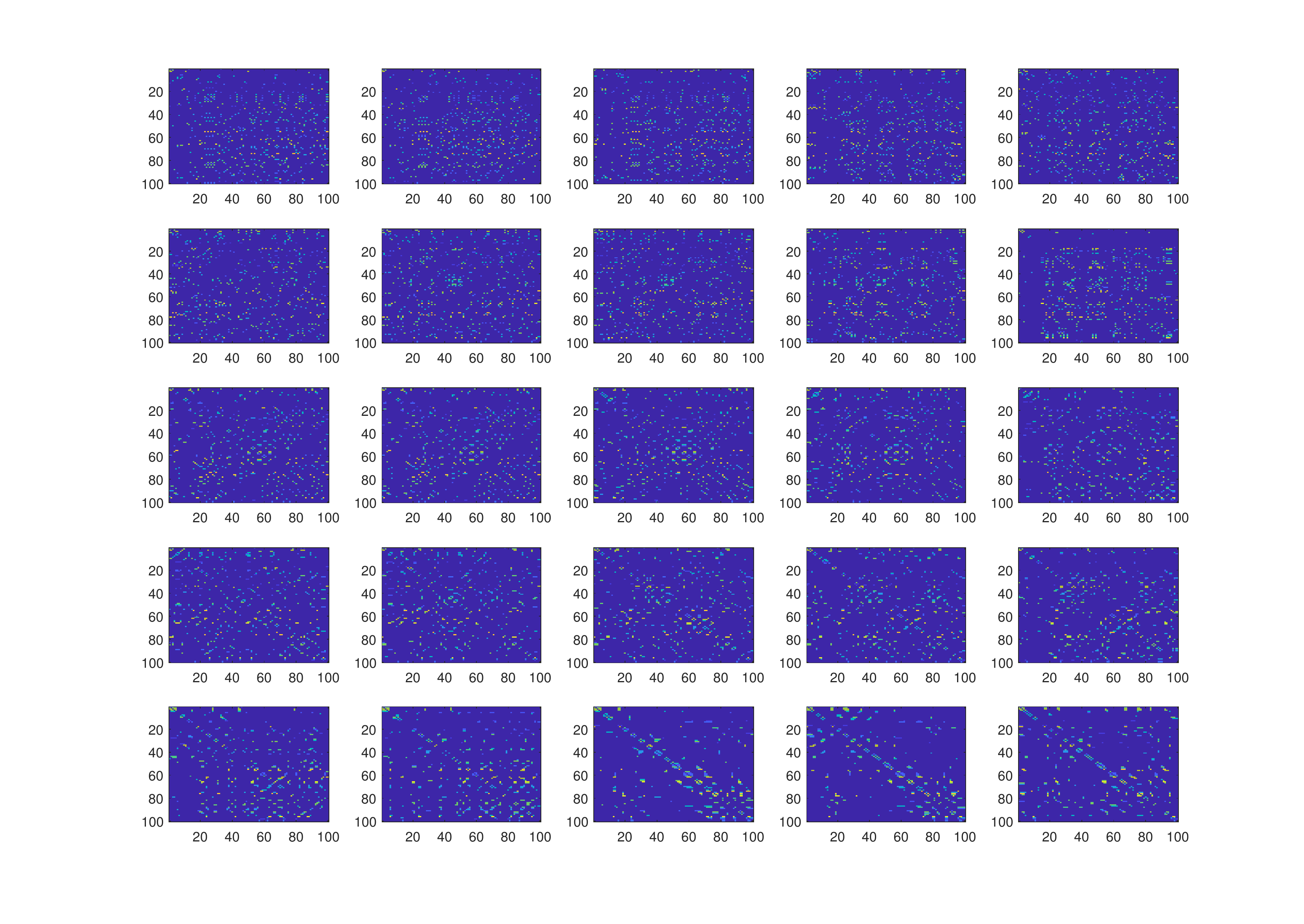,width=18cm}
   \caption{  The images  represent network of networks between the first 25 layers of  tensor network of 
correlation between geneon of all genes taken from $IgA CAM-1$, human virus receptor protein gene in antibody for  $VP1,VP2,VP3$ viral capsid glycoproteins. From the result, the image shows the  high variation without pattern compared with  image of viral capsid protein.\label{tensor_icam} }
 \end{figure}

For the  protein docking behavior, we compare results of $VP1,VP2,VP3$
with their host protein receptor genotype in antibody of $Ig-CAM$.
The results of data analysis of Chern-Simons current with $(ITD-IMF)chain(1)$ and $(ITD-IMF)chain(8)$   are shown in Fig. \ref{result_icam}. The image of tensor network analysis of $Ig-CAM$ with different layer in  network of networks is shown in Fig.\ref{tensor_icam}.

\section{Discussion and Conclusions}

A configuration space of organism can be represented as the  analogue of atomic configurations with
valence shell of cycle circle coordinates which constitute the  principal  fiber bundle of a free electron circulating in
 free energy level of a global macromolecule of protein enzyme. Due to  the metabolism of cell, it  rotates with spinor field in ribbon graph of spinor network. The
 nucleus of cell is analogue with the atomic nucleus  with the total number of proton and 
neutron as the number of chromosomes in genetic code  with a first prototype of
 knot and vortex structure. Historically,  it could be compared with the atomic model 
 first introduced by  Kelvin,  Thomson and finally Rutherford. We know that this model 
fails in  explaining the atomic structure.
Despite this failure, we can still recover the  old theory and  redefine the wave function of simple 
cell organism  introducing  new concepts based on  quantum mechanic
 of  "atomic" knotted proteins in cell. Here "atomic" means that 
 we cannot divide the  constituents of an organism more than these 
configurations. Roughly speaking, we  can separate the superspace of atomic organism in 2 parts. The first part 
is the nucleus and the second part is the cytoplasm. The orbital is the configuration of valence of  
organelle  that, as the electron,   circulates around nucleus in loop space of cell system according to theory of vortex and knot of  Kelvin. The difference between cell and atom is that the  nucleus is connected with cytoplasm by using the knotted behavior of first atomic prototype   introduced by  Kelvin and  Thomson. The nucleus is in contact to electron orbital by electrostatic force instead of genetic code.  For this construction, the  genetic code  is an artificial movement of current, i.e. the  Chern-Simons super-current of  life energy flow along coordinate of cell cycle in fiber space \cite{quantum_bio}. 

In this paper, we proposed a  superspace model of  knots and links for   time series data of proteins taking into account the  
 feedback loop from docking to undocking states of protein-protein interactions. 
 In particular, the   direction of interactions between the  hidden 
states is considered. 
A $E_{8}\times E_{8}$ unified spin model  emerges  where the  genotype, from active  
and  inactive parts of DNA  time data series, is considered.
 The approach comes  from the  loop-quantum  gravity and quantum field theory and it is  adapted to
 biology.  In this sense, we can deal with a {\it quantum field biology}. We  derive  the equations for the gene expression which describe  transitions  from ground  to excited states, and  for the 8 coupling states between geneon and anti-geneon
 transposon and retrotransposon in trash DNA.   The construction is essentially based on  
the modifications,  in view of biological applications, of the  Grothendieck   and  Khovanov cohomologies. The final result is  a 
Chern-Simons current  in  $(8+3)$ extradimensions on a given unoriented supermanifold generated by the  
  ghost and anti-ghost fields of protein structures.
 The $8$ dimensions are related to  the 8 hidden  states of spinor field of genetic code while  the extradimensions derive from the 3 types of principle fiber  bundle in the secondary protein.
 
 Specifically, we solve a central dogma paradox by using this new modified Grothendieck cohomology  to explain
 why undocking states of proteins induce docking mechanism.  The core of this mechanism is related to a new definition  of parallel transport as a connection acting  in  the quantum  gauge group of genotype in 
genetic code as the representation of gene expression. Implicitly, an  equation of gene
 expression is defined over the coupling state between the geneon and anti-geneon. In this picture,   an induced analogue 
gravitational field can be defined related to the  curvature of docking. 

From the modified  Khovanov cohomology, used for the 
construction of time series of knotted protein folding, it is possible to  define knots and links in protein 
secondary structure. Furthermore,  by using the Grothendieck topology in co-adjoint functor between 3 categories of 3 types of biological objects, i.e. DNA, RNA and proteins and 
 with Chern-Simons current, one can plot all gene in active area and in trash area.  
From a geometrical point of view, one can  give a new definition of the  superspace of living organism by a unoriented supermanifold generated by   the  sheaf cohomology. In the framework of this theory, 
the central dogma is  an adjoint functor over the group action which gives a  representation of the alphabet code with left and right symmetry. 
 
Based on the fact that  the protein docking is a non-equilibrium state, we use the 
 transition state in hidden directions of gene expression in loop-quantum  gravity for
 biology over a  modified Seiberg-Witten equation.
 The transposon induces a loop space in time series data in which unoriented states of
 knots and twistors appear from the operation of insert and delete a gene.
 The retrotransposon is an example of genotype in viral capsid protein  $VP1,VP2,VP3$.
 We give a simple model of moduli state space  for the transition of all states and hidden states of 
geneon, anti-geneon to transposon and retrotransposon. This allows us to  construct the wave function 
of all gene component  in trash  and in active part of DNA by using differential 2 forms and modified Dirac operators. 
In particular, we use the Khovanov cohomology to define the time series of knots and links in proteins and 
we give a new definition for connection on principle bundle 
of secondary protein by using the Grothendieck topology and modified Atiyah axioms. This construction says us 
how living organisms can adapt their behavior, in central dogma model, under life energy 
of ghost  and anti-ghost fields as a central unit of sheaf cohomology defined in 
 Grothendieck topology. 
The Chern-Simons current for active gene and  trash DNA can be  explained  by the 
 retroptransposon state of the  gene.
The image of tensor network  in geneon image of $VP1,VP2,VP3$ can be explained as the
 Holo-Hilbert spectrum of quantum biology wave function. It  can  be intepreted as hidden 
transition states in geneotype  induced from evolution feedback path in protein docking system.
The results of this approach    explain the evolutional band gap between $(IMF-ITD)chain(8)$ in 2 parabolic shape merging into the hyperbolic structure of  the cell manifold. The spinor field in time series of knotted protein 
is induced by the  Hopf fibration over the loop space of  transposon and retrotransponson transition in inactive states over trash DNA.
 
The information can  take off only when we give a  definition of the   Seiberg-Witten equation for
 biology by  a  Dirac operator with Chern-Simons  current  over cohomology sequence of living organism. We use model of hyperbolic knots of parasitism state in viral replication cycle  to solve this new equation which  results  a hyperbolic equation of transition states in trash DNA.
 The gene  state  of central dogma can be induced without transitive layer of protein  
as principle fiber bundle $P_{[A_{\mu}] }$. The  underlying connection (a gauge potential
 of genotype)  appears as a Yang-Mills
 field $F_{\mu\nu}$\cite{ssm}. We have cycle and co-cyle $\beta $ of  superspace of 
 living organism $X_{t}$ defining a Jacobian over the 
supermanifold fiber $g_{ij}=\frac{\partial s_{i}}{\partial s_{j}}.$  
The source of gravitational field \cite{simons}, the  Chern-Simons current in codon,
 is induced from the connection over the parallel transport of co-cycle $g^{ij}$ and cycle 
$g_{ij}$ in analogy with the definition of connection $\Gamma_{ij}^{k} :=[A_{k}]$ in General 
Relativity. 
Furthermore, we define 3 forms over alphabet code  by  Chern-Simons 3 form over alphabet 
of gauge field in  genetic code. They  represent the  protein layers as  triplet states which induce
 further 32 states from differential 3 forms with Seiberg-Witten invariant \cite{seiberg}. 
 
 Finally, we demonstrated the above construction by 
simple examples of $VP1,VP2$ and $VP3$ genes in 2 species of icosahedral virus with their
 host receptor gene, $Ig-CAM$ antibody. We used the  new algorithm
 of frequency mode modulation in Holo-Hilbert spectral algorithm \cite{holo} over a
new adaptive method of data analysis to find
 a spectrum of transition states in genotype of capsid protein of 2 viruses. 
The knotted protein time series of curvature in  genotype of secondary protein
 docking \cite{docking} state is parameterized by $k=dt^{\ast}$ the 
order number of amino-acid in form of time series data.
  
In future works, detailed applications of the present model will be developed considering other examples of genes.

\section*{Acknowledgment}
\addcontentsline{toc}{section}{Acknowledgment}

This article is based upon work from COST Action CA15117  "Cosmology and Astrophysics Network for Theoretical Advances and Training Actions" (CANTATA), supported by COST (European Cooperation in Science and Technology). S. Capozziello is supported by Istituto Nazionale di Fisica Nucleare (INFN). R. Pincak would like to thank the TH division at CERN for hospitality.   The work is partly supported by VEGA Grant No. 2/0009/16.

\end{document}